\documentclass[useAMS,usenatbib,iop,numberedappendix]{mn2e} 

\usepackage{mathptmx}
\usepackage{amsmath,amsfonts,amssymb}
\usepackage{makeidx}
\usepackage{graphicx}
\usepackage{epstopdf}
\usepackage[colorlinks=True]{hyperref}
\usepackage{multirow}
\usepackage{rotating}
\usepackage{color}
\usepackage{tablefootnote}
\usepackage{lscape}
\usepackage{subcaption}
\usepackage{tablefootnote}
\usepackage[T1]{fontenc}
\usepackage{aecompl}
\usepackage[normalem]{ulem}

\definecolor{inon}{rgb}{1.00,0.27,0.00}

\newcommand{\Munich}{$^{1}$}
\newcommand{\ExcellenceCluster}{$^{2}$}
\newcommand{\MPE}{$^{3}$}
\newcommand{\Bosch}{$^{4}$}

\author[Chiu et al.]{
I.~Chiu\Munich$^,$\ExcellenceCluster,
A.~Saro\Munich$^,$\ExcellenceCluster,
J.~Mohr\Munich$^,$\ExcellenceCluster$^,$\MPE,
S.~Desai\Munich$^,$\ExcellenceCluster,
S.~Bocquet\Munich$^,$\ExcellenceCluster,
R.~Capasso\Munich$^,$\ExcellenceCluster,
\newauthor
C.~Gangkofner\Munich$^,$\ExcellenceCluster,
N.~Gupta\Munich$^,$\ExcellenceCluster,
J.~Liu\Bosch
\\
\Munich Faculty of Physics, Ludwig-Maximilians-Universit\"at, Scheinerstr.\ 1, 81679 Munich, Germany \\
\ExcellenceCluster Excellence Cluster Universe, Boltzmannstr.\ 2, 85748 Garching, Germany \\
\MPE Max Planck Institute for Extraterrestrial Physics, Giessenbachstr.\ 85748 Garching, Germany \\
\Bosch Bosch Research and Technology Center North America,
4005 Miranda Ave \#200, Palo Alto, CA 94304, United States
}

\newcommand{\LCDM}{\ensuremath{\Lambda\textrm{CDM}}}
\newcommand{\OmegaM}{\ensuremath{\Omega_{\mathrm{M}}}}
\newcommand{\OmegaL}{\ensuremath{\Omega_{\Lambda}}}

\newcommand{\Hnow}{\ensuremath{H_{0}}}
\newcommand{\seight}{\ensuremath{\sigma_{8}}}
\newcommand{\Msun}{\ensuremath{\mathrm{M}_{\odot}}}
\newcommand{\Lsun}{\ensuremath{\mathrm{L}_{\odot}}}
\newcommand{\Rfiveoo}{\ensuremath{R_{500}}}
\newcommand{\Mfiveoo}{\ensuremath{M_{500}}}
\newcommand{\Cfiveoo}{\ensuremath{C_{500}}}
\newcommand{\redshift}{\ensuremath{z}}
\newcommand{\zd}{\ensuremath{z_{\mathrm{d}}}}
\newcommand{\dif}{\ensuremath{\mathrm{d}}}
\newcommand{\angstrom}{\textup{\AA}}

\newcommand{\XMMNEWTON}{XMM-\emph{Newton}}
\newcommand{\Spitzer}{\emph{Spitzer}}

\newcommand{\XMMBCS}{XMM-BCS}
\newcommand{\BCS}{BCS}
\newcommand{\SSDF}{SSDF}

\newcommand{\Mstar}{\ensuremath{M_{\star}}}
\newcommand{\Lstar}{\ensuremath{L_{\star}}}
\newcommand{\Mstarsat}{\ensuremath{M_{\star, \mathrm{sat}}}}
\newcommand{\Lstarsat}{\ensuremath{L_{\star, \mathrm{sat}}}}

\newcommand{\LBCG}{\ensuremath{L_{\star,\mathrm{BCG}}}}
\newcommand{\Lx}{\ensuremath{L_{\mathrm{X}}}}
\newcommand{\Lxbol}{\ensuremath{L_{\mathrm{X},~\mathrm{bol}}}}
\newcommand{\mstar}{\ensuremath{m_{\star}}}
\newcommand{\IRACone}{\ensuremath{[3.6]}}
\newcommand{\IRACtwo}{\ensuremath{[4.5]}}
\newcommand{\IRAConestar}{\ensuremath{m_{\star,[3.6]}}}
\newcommand{\IRACtwostar}{\ensuremath{m_{\star,[4.5]}}}
\newcommand{\fcom}{\ensuremath{f_{\mathrm{com}}}}
\newcommand{\ML}{\ensuremath{\Gamma_{\star}}}
\newcommand{\Ncl}{\ensuremath{N_{\mathrm{cl}}}}
\newcommand{\fblue}{\ensuremath{f_{\mathrm{blue}}}}
\newcommand{\Tx}{\ensuremath{T_{\mathrm{X}}}}

\newcommand{\MPIV}{\ensuremath{M_{\mathrm{piv}}}}
\newcommand{\ZPIV}{\ensuremath{z_{\mathrm{piv}}}}
\newcommand{\Ez}{\ensuremath{E(z)}}
\newcommand{\Ax}{\ensuremath{A_{\mathrm{X}}}}
\newcommand{\Bx}{\ensuremath{B_{\mathrm{X}}}}
\newcommand{\Cx}{\ensuremath{C_{\mathrm{X}}}}
\newcommand{\Dx}{\ensuremath{D_{\mathrm{X}}}}
\newcommand{\Dxcom}{\ensuremath{\sigma_{\ln L_{\mathrm{X}}|M_{500}}}}
\newcommand{\Astar}{\ensuremath{A_{\star}}}
\newcommand{\Bstar}{\ensuremath{B_{\star}}}
\newcommand{\Cstar}{\ensuremath{C_{\star}}}
\newcommand{\Dstar}{\ensuremath{D_{\star}}}
\newcommand{\Dstarcom}{\ensuremath{\sigma_{\ln M_{\star}|M_{500}}}}

\newcommand{\Cstat}{\ensuremath{C_{\mathrm{stat}}}}
\newcommand{\percent}{\ensuremath{\%}}

\title[Stellar Mass to Halo Mass Scaling Relation]{
Stellar Mass to Halo Mass Scaling Relation for X-ray Selected Low Mass Galaxy Clusters and Groups out to Redshift $\redshift\approx1$
}

\begin{document}
\pdfpageheight 11.7in
\pdfpagewidth 8.3in

%
%

\maketitle 

%
%

\begin{abstract}
We present the stellar mass-halo mass scaling relation for 46 X-ray selected low-mass clusters or groups detected in the \XMMBCS\ survey with masses $2\times10^{13}\Msun\lesssim\Mfiveoo\lesssim2.5\times10^{14}\Msun$ (median mass $8\times10^{13}\Msun$) at redshift $0.1\le\redshift\le1.02$ (median redshift $0.47$).  The cluster binding masses \Mfiveoo\ are inferred from the measured X-ray luminosities \Lx, while the stellar masses \Mstar\ of the galaxy populations are estimated using near-infrared imaging from the \SSDF\ survey and optical imaging from the \BCS\ survey. 
With the measured \Lx\ and stellar mass \Mstar, we determine the best fit stellar mass-halo mass relation, accounting for selection effects, measurement uncertainties and the intrinsic scatter in the scaling relation.   The resulting  mass trend is $\Mstar\propto\Mfiveoo^{0.69\pm0.15}$, the intrinsic (log-normal) scatter is $\Dstarcom=0.36^{+0.07}_{-0.06}$, and there is no significant redshift trend $\Mstar\propto(1+z)^{-0.04\pm0.47}$, although the uncertainties are still large.  We also examine \Mstar\ within a fixed projected radius of $0.5$~Mpc, showing that it provides a cluster binding mass proxy with intrinsic scatter of $\approx93$\percent\ (1$\sigma$ in \Mfiveoo).
We compare our $\Mstar=\Mstar(\Mfiveoo, \redshift)$ scaling relation from the \XMMBCS\ clusters with samples of massive, SZE-selected clusters ($\Mfiveoo\approx6\times10^{14}\Msun$) and low mass NIR-selected clusters ($\Mfiveoo\approx10^{14}\Msun$) at redshift $0.6\lesssim\redshift\lesssim1.3$.  After correcting for  the known mass measurement systematics in the compared samples, we find that the scaling relation is in good agreement with the high redshift samples, suggesting that for both groups and clusters the stellar content of the galaxy populations within \Rfiveoo\ depends strongly on mass but only weakly on redshift out to $z\approx1$.
\end{abstract}

%
%


%
%

\section{INTRODUCTION}
\label{sec:introduction}

Over the past decade and a half extensive galaxy cluster surveys have been undertaken in the X-ray \citep[e.g.][]{boehringer04,pierre04,mantz08,vikhlinin09b}, at mm wavelengths \citep[e.g.][]{staniszewski09,planck11-5.1a,hasselfield13} employing the Sunyaev-Zel'dovich Effect \citep[SZE;][]{sunyaev70b,sunyaev72}, and in the optical \citep[e.g.][]{gladders05,koester07a} and near infrared \citep[NIR;][]{lacy05,stanford14}.  While the primary goal of many of these surveys has been to use galaxy clusters to study cosmology and, in particular, the dark energy or cosmic acceleration \citep{haiman01,weller03}, the buildup of large regions of the sky with overlapping, multiwavelength surveys provides not only data for cluster cosmological studies, but also data that enable the study of the clusters themselves.

Of particular importance to cluster studies is the need to account for the impact of the cluster sample selection.  A uniform selection can simplify the interpretation of the results.  Another element of critical importance is that one needs to have precise mass estimates from low scatter mass proxies where the remaining systematic uncertainties are quantified.  Because most properties of the cluster vary with cluster-centric distance, a precise mass is crucial for making it possible to study the same portion of the cluster virial region at all redshifts.  

Recently, the stellar and intracluster medium mass trends with cluster binding mass \Mfiveoo\ and redshift have been studied in a sample of massive ($\Mfiveoo\gtrsim3\times10^{14}\Msun$), SZE selected clusters at redshfit $\redshift\approx0.9$ with X-ray, optical and NIR followup data \citep{chiu16}.  Cluster binding masses were determined using the SZE signatures of the clusters, and the X-ray and optical/NIR observations were used to study the intracluster medium and galaxy populations, respectively.  This sample exhibits a trend for the stellar mass fraction to fall with cluster binding mass, which has been noted previously \citep[e.g.][]{lin04a,andreon10}.  In addition, this sample is at relatively high redshift and thus-- in combination with results from previous studies of low redshift clusters-- these clusters indicate that there is no significant redshift trend in the stellar mass fraction out to redshift $\redshift\approx1.32$ \citep{chiu16}.  Such a result is troubling at first glance, because it suggests that massive halos exhibit different stellar mass fractions than their building blocks, which include the low mass halos.

It must be noted that the current constraints on the trends in stellar mass fraction with redshift suffer from the systematic uncertainties induced from the joint analysis of heterogeneous samples using different mass measurement techniques.  Moreover, these constraints are only available for the massive systems, and of course the progenitors of the local high mass clusters are low mass clusters at higher redshift.  It is therefore imperative to extend the scaling relation studies to include lower mass halos and to use consistent mass measurement techniques over as wide a range of redshift as possible to improve constraints on the matter assembly history of both groups and galaxy clusters.  

In this work, we aim to measure the relationship between stellar and binding mass for the low mass clusters and groups detected in a 6~deg$^2$ region of the XMM-\textit{Newton}-Blanco Cosmology Survey \citep[\XMMBCS;][]{suhada12}.   The \XMMBCS\ survey employs the \XMMNEWTON\ telescope (Proposal Id 050538, PI H. Boehringer) to survey a total sky area of 14~$\deg^{2}$ within a region fully covered by the optical $griz$ Blanco Cosmology Survey \citep[BCS;][]{desai12} and the NIR \Spitzer-South Pole Telescope Deep Field survey \citep[\SSDF;][]{ashby13a}.  This sky region has also been imaged in the SZE by 
the South Pole Telescope \cite[SPT,][]{carlstrom02}, and the \XMMBCS\ cluster sample has already been used to study the SZE signature-halo mass scaling relation \citep[][L15 hereafter]{liu15a}.  By combining the SPT-SZ maps and the \XMMNEWTON\ sample, it was possible to study the relationship between SZE signature and halo binding mass to a mass threshold  $\approx3$ times lower than the masses of the SPT-SZ selected clusters \citep{bleem15}.

This paper is organized as follows. The cluster sample and data are described in Section~\ref{sec:cluster_and_data}, while the analysis method is given in Section~\ref{sec:method}.
We present the results and discussion in Section~\ref{sec:result_discussion}.  The potential systematics are quantified in Section~\ref{sec:sys}, and the conclusions are presented in Section~\ref{sec:conclusion}.  Throughout this paper, we adopt the \LCDM\ cosmology with the fiducial cosmological parameters $(\OmegaM, \OmegaL, \Hnow, \seight) = (0.3, 0.7, 70~\mathrm{km}~\mathrm{s}^{-1}~\mathrm{Mpc}^{-1}, 0.8)$.  Unless otherwise stated, the uncertainties indicate the $1\sigma$ confidence level, the halo binding mass \Mfiveoo\ is estimated at the overdensity of 500 with respect to the critical density at the cluster redshift, and the photometry is in the AB magnitude system.

%
%

\section{CLUSTER SAMPLE AND OBSERVATIONS}
\label{sec:cluster_and_data}

In this section we briefly introduce the cluster sample in \XMMBCS\ catalog in Section~\ref{sec:xmmbcs_catalog} and the \SSDF\ catalog used to derive the stellar masses in Section~\ref{sec:ssdf_catalog}.

\begin{figure}
\centering
\includegraphics[scale=0.5]{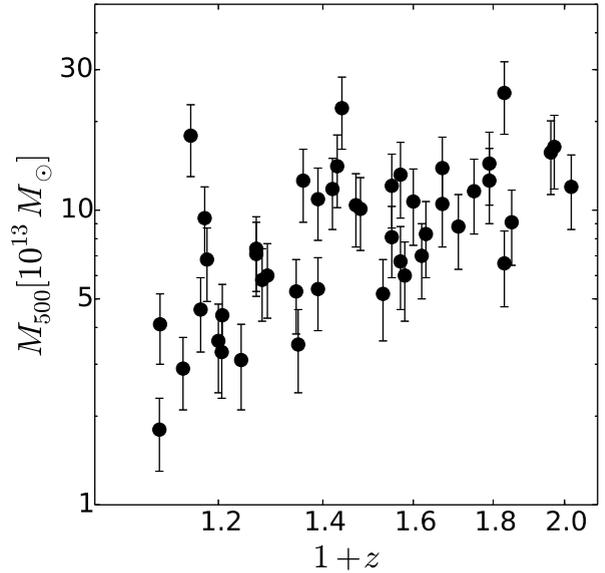}
\vskip-0.1in
\caption{
The \XMMBCS\ sample plotted in mass versus redshift (similar to Figure~1 in L15).
The cluster masses \Mfiveoo\ are derived via the X-ray luminosity \Lx\ to mass scaling relation, and the redshifts are estimated using the red sequence overdensity technique.  The median mass is $\Mfiveoo=8\times10^{13}\Msun$, and the median redshift is $\redshift=0.47$.
}
\label{fig:sample}
\end{figure}

\subsection{\XMMBCS\ catalog}
\label{sec:xmmbcs_catalog}

We use the galaxy clusters detected in the \XMMBCS\ survey (S12), which is also the same sample used in L15 to study the SZE signature-halo mass scaling relations.  The \XMMBCS\ sample consists of 46 clusters with the median \Mfiveoo\ of $8\times10^{13}\Msun$ and redshift range from $0.1$ to $1.02$ with a median of $0.47$ (see FIgure~\ref{fig:sample}).  A full description of the X-ray data reduction, source detection and mass calibration is given in S12, we briefly summarize the \XMMBCS\ catalog in the following.  Each galaxy cluster is detected by their X-ray emission in the energy range $0.5-2~\mathrm{keV}$ in the central $6~\deg^{2}$ footprint of the \XMMBCS\ survey, this results in a flux-limited sample with the limiting flux $f_{\mathrm{lim}}=10^{-14}\mathrm{erg}\mathrm{s}^{-1}\mathrm{cm}^{-2}$.  After the optical confirmation and redshift estimation (see below), the X-ray luminosity in the energy range $0.5-2~\mathrm{keV}$ ($L_{\mathrm{X},[0.5-2~\mathrm{keV}]}$) for each cluster is iteratively measured within \Rfiveoo, which is defined as the radius corresponding to the enclosed mass \Mfiveoo, through the X-ray luminosity-halo mass relation \citep{pratt09}.  The measured X-ray luminosity is then converted into the bolometric luminosity \Lxbol\ using the characteristic temperature \Tx\ and the redshift for the cluster.  The \Tx\ is derived from a scaling relation with the observed luminosity.  Following L15, we use \Lxbol\ (hereafter abbreviated as \Lx) as the mass proxy for \XMMBCS\ sample through the \Lx-\Mfiveoo\ relation \citep{pratt09}:
\begin{equation}
\label{eq:lxbol2mass}
\Lx = \Ax \left( \frac{\Mfiveoo}{2\times10^{14}\Msun} \right)^{\Bx} \Ez^{\Cx} \, ,
\end{equation}
with 
$\Ez\equiv\sqrt{\OmegaM(1+\redshift)^3 + \OmegaL}$, 
$\Ax = 1.38\pm0.12\times10^{44}~\mathrm{erg}~\mathrm{s}^{-1}$, 
$\Bx = 2.08 \pm 0.13$ and 
$\Cx = 7/3$. 
The intrinsic log-normal scatter of \Lx\ for a given mass in equation~(\ref{eq:lxbol2mass}) is $\Dx\equiv\Dxcom=0.383\pm0.061$.  Note that the Malmquist and Eddington biases are both taken into account and corrected in fitting equation~(\ref{eq:lxbol2mass}).

The redshifts of the majority of the \XMMBCS\ sample are determined by the \BCS\ photometry except for a few exceptions where the spectroscopic redshifts are available (S12).  The photometric redshift for each cluster is estimated by modeling the excess of the red sequence (RS) galaxies within $0.8$~Mpc centered on the X-ray center (S12).  
The RS model is constructed, using the python package \texttt{EZGAL} \citep{mancone12b}, by a Composite Stellar Population (CSP) of \citet[][hereafter BC03]{bruzual03} model with the formation redshift $\redshift_{\mathrm{f}}=3$ and an exponentially decaying $e$-folding timescale $\tau=0.4$~Gyr.  The \cite{chabrier03} initial mass function is used in the model construction.  The color-magnitude relation of the RS is determined by using six different metallicities, which are calibrated by the metallicity-luminosity relation of the RS of Coma cluster \citep[for more details see][]{song12a}.  This model has been used to successfully measure the photometric redshifts of SPT clusters out to redshift $\redshift>1$ with the root-mean-square of the redshift uncertainties at level of 0.017 \citep{song12b}.  Calibrating  the \XMMBCS\ clusters with the available spectroscopic redshifts, the photometric redshift estimations for \XMMBCS\ groups result in the root-mean-square of the redshift uncertainties $\Delta\redshift/(1+\redshift)$ of 0.023 (S12), which is in  good agreement with that for SPT clusters.

\subsection{\SSDF\ catalog}
\label{sec:ssdf_catalog}

The major goal of the \SSDF\ survey is to enable  study the evolution and structure of baryons in the distant Universe by observing the regions overlapping with the multi-wavelength surveys (e.g., SPT, \XMMNEWTON\ and \BCS).  The whole survey consists of a sky area of $94\deg^{2}$ and is the largest wide field \Spitzer\ extragalactic survey to date.
The \SSDF\ survey was completed in 2013 and the data reduction, the photometry calibration and the source extraction is fully described in \citet{ashby13a}.  We briefly summarize the survey below.  

Two IRAC channels of $3.6~\micron$ and $4.5~\micron$ are imaged in \SSDF\ survey to depths which result in $5\sigma$ limiting magnitudes for $4\arcsec$ diameter apertures of $21.79$~mag ($7.0$~$\mu$Jy) and $21.47$~mag ($9.4$~$\mu$Jy) for $3.6~\micron$ (\IRACone) and $4.5~\micron$ (\IRACtwo), respectively.  Source detection is performed by running \texttt{SExtractor} \citep{bertin96} in dual-image mode.  The $3.6~\micron$ and $4.5~\micron$ mosaic images are used in turn as the detection image, resulting in $3.6~\micron$-selected or $4.5~\micron$-selected source catalogs with \texttt{MAG\_AUTO} mesurements for each object.  
The completeness of the source catalogs as a function of magnitude $\fcom(m)$ is derived through simulation.   A vast number of simulated objects with a wide range of magnitudes are injected into the mosaics and the same detection pipeline is used to extract those objects and derive the catalog completeness \citep{ashby13a}.  The resulting $90\percent$ ($50\percent$) completeness of the source detection is at $19.60$~mag ($21.45$~mag) and $19.72$~mag ($21.47$~mag) for $3.6~\micron$ and $4.5~\micron$, respectively. The completeness function $\fcom(m)$ in detail is given in \citet{ashby13a}.

\begin{figure}
\vskip-0.2in
\centering
\includegraphics[scale=0.6]{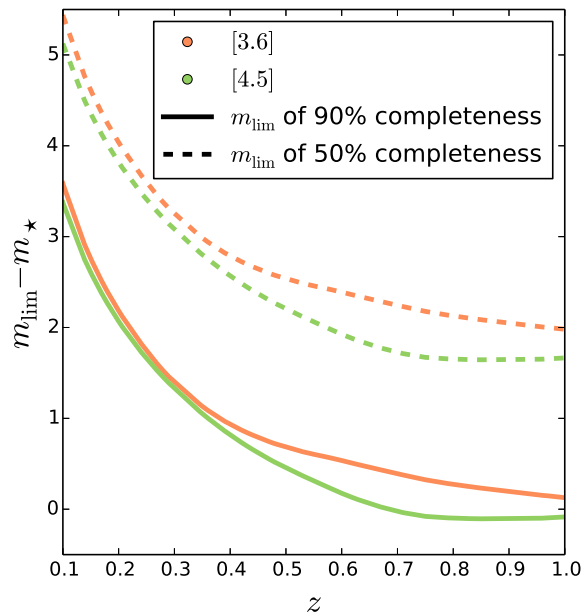}
\vskip-0.2in
\caption{
The offset of the \SSDF\ survey limiting magnitudes relative to the cluster galaxy population characteristic magnitudes (\IRAConestar\  (red) and \IRACtwostar\ (green)) from our CSP model out to redshift $\redshift=1$.  The $90\percent$ ($50\percent$) completeness magnitudes appear as solid (dashed) lines.   The \SSDF\ at 50\% completeness has adequate depth to allow us to estimate stellar masses for the \XMMBCS\ cluster sample.
}
\label{fig:depth}
\end{figure}

We compare the characteristic magnitudes of \XMMBCS\ clusters and the \SSDF\ limiting magnitudes in \IRACone\ and \IRACtwo\ in Figure~\ref{fig:depth}.  The characteristic magnitude of each cluster in $3.6~\micron$ and $4.5~\micron$ (\IRAConestar\ and \IRACtwostar) is estimated using the same CSP model used to estimate the cluster redshift (see Section~\ref{sec:xmmbcs_catalog}).  The 50\percent\ completeness limit is deeper than the characteristic magnitudes of \IRAConestar\ and \IRACtwostar\ by $\gtrsim1.8$~mag out to redshift $\redshift\approx1$, thus ensuring that about 70\percent\ (80\percent) of the light emitted from the cluster galaxies out to $\redshift\approx1$, assuming a \cite{schechter76} luminosity function (LF) with a faint end power law index of $-1.1$ \citep{lin04a} (-0.9; see Section~\ref{sec:stellarmass_estimation}), is directly detected after suitable completeness corrections with the \SSDF\ survey.  That is, the depth of the \SSDF\ survey is adequate to enable us to measure the stellar masses of the \XMMBCS\ clusters.
%

%
%

%
\begin{figure*}
\centering
\includegraphics[scale=0.5]{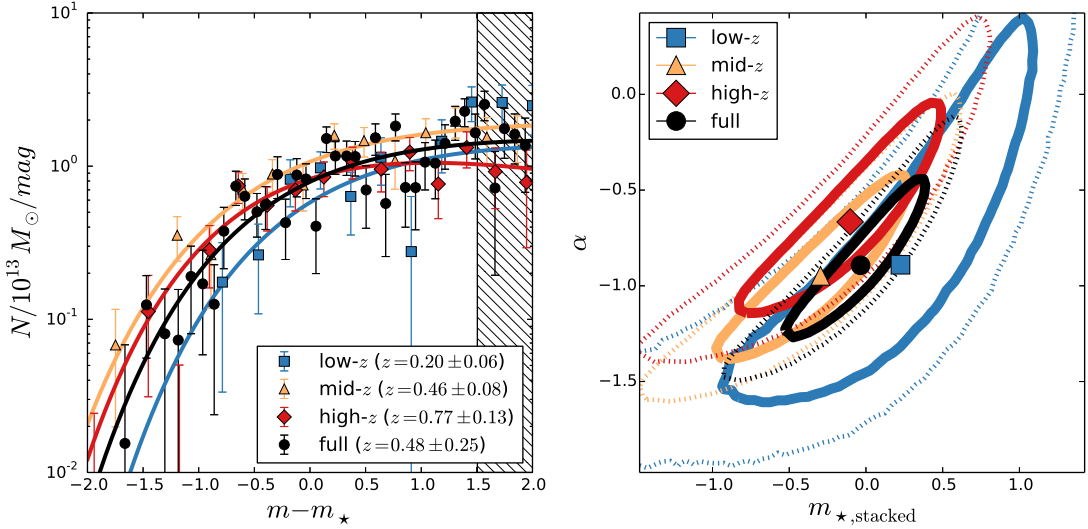}
\vskip-0.2in
\caption{
The stacked luminosity functions (left) in various redshift ranges together with the best fit LF parameters (right).  On the left the x-axis shows the magnitudes with respect to the \mstar\ predicted by our CSP model, while the y-axis shows the number density of galaxies normalized to per magnitude and per cluster mass of $10^{13}\Msun$.  The stacked profiles of the full, low-\redshift, mid-\redshift\ and high-\redshift\ samples are shown in black circles, blue squares, orange triangles and red diamonds, respectively.  The best-fit profiles are in the solid lines with the same color. The shaded region indicates the magnitudes which are fainter than $\mstar+1.5$, which are not used in the fitting. The mean and the standard deviation of the redshift distribution for the stacked samples are shown in the lower right corner. The joint constraints of ${\mstar}_{,~\mathrm{stacked}}$ and $\alpha$ appear on the right for the different luminosity functions using the same color coding.  The LFs in all redshift ranges are in good agreement with the CSP model, and there is little evidence for a redshift dependence in $\alpha$.
 }
\label{fig:stacked_lf}
\end{figure*}

\section{METHODS}
\label{sec:method}

For each cluster, we derive the total stellar masses of the galaxies which are photometrically identified in the \SSDF\ $3.6~\micron$-selected source catalog.  The most robust stellar mass estimates would come from spectral energy distribution (SED) template fitting on a single galaxy basis, and that would require photometry in multiple bands (e.g., \BCS\ plus \Spitzer) with depths that are comparable to the \SSDF\ imaging.  However, the \BCS\ optical survey is too shallow to be used for this purpose for all \XMMBCS\ clusters in a uniform manner out to $\redshift\approx1$.  Therefore, we model the NIR Luminosity Function (LF) for each cluster and then use a derived mass-to-light ratio for a stellar population that includes a group intracluster medium temperature and redshift dependent blue fraction to convert the total stellar luminosity into mass.

We first describe the LF modeling in Section~\ref{sec:lf_xmmbcs} and then the mass-to-light ratio, which varies depending upon the measured blue fraction, in Section~\ref{sec:m2l_method}.  We describe the measurement of the total stellar mass, which leverages parameters from stacked LFs,  in Section~\ref{sec:stellarmass_estimation}.  The Bayesian method for fitting the stellar mass-halo mass scaling relation is presented in Section~\ref{sec:likelihood}.

\subsection{Stacked Luminosity Function of \XMMBCS\ sample}
\label{sec:lf_xmmbcs}

Because the LF parameters are ill-constrained in the case of a single group or low mass cluster, we use information from the stacked profiles to constrain the faint end slope $\alpha$ and to test a model for the characteristic magnitude as a function of redshift $\mstar(z)$ \citep[e.g.,][]{lin04a}. To derive the stacked LF of the \XMMBCS\ sample, we first construct the observed LF of each individual system.

Using the \IRACone\ selected catalogs described in Section~\ref{sec:ssdf_catalog} above, we first discard the point sources with $\IRACone<17.75$~mag identified in the stellar branch of the \texttt{FLUX\_RADIUS}-\IRACone\ relation; the \texttt{FLUX\_RADIUS} for the stellar branch is between $1.75$ and $2.30$ pixels (corresponding to $1.05\arcsec$ and $1.38\arcsec$, respectively).  We further discard the non-extended objects which have \texttt{FLUX\_RADIUS} smaller than $1.75$ pixels.  Note that this removal of the non-extended sources will allow some stars to leak into the analysis sample but few if any galaxies will be excluded.  The remaining stellar contamination in the extended source list is subtracted during the analysis by statistical foreground and background subtraction (we refer to it as  the background subtraction hereafter).  The $6~\deg^{2}$ \XMMBCS\ footprint excluding the cluster fields, which are the $3$~Mpc diameter apertures centered on each cluster, is defined as the blank sky used to measure the background for our sample.

We then identify the Brightest Cluster Galaxy (BCG) within \Rfiveoo\ for each cluster in the pseudo-color image reconstructed from the mosaics of $3.6~\micron$, $4.5~\micron$ and \BCS-$z$ filter.  For each cluster, we derive the \IRACone\ magnitude distribution with a bin width of 0.25~mag for the galaxies that lie projected within \Rfiveoo\ and are fainter than the BCG.  The background magnitude distribution is obtained by randomly drawing 25 non-overlapping apertures with the radii of cluster \Rfiveoo\ from the blank sky.  The mean of the 25 background magnitude distributions is used in making a statistical background subtraction.  Finally, we apply the completeness correction to the observed overdensity of galaxies within the cluster.

We derive the stacked profiles for the full sample of the \XMMBCS\ clusters and compare it to the samples which are divided according to the cluster redshifts-- namely, the low-\redshift\ ($0.1\le\redshift<0.33$), mid-\redshift\ ($0.33\le\redshift<0.58$) and the high-\redshift\ ($0.58\le\redshift\le1.02$) subsamples.
To construct the stacked LF of the subsample, we first normalize the observed LF of each cluster by dividing by the total mass \Mfiveoo\ inferred from the X-ray luminosity \Lx.  In this way we mostly remove the dependence of the LF's normalization on the cluster mass \citep{lin04a}.  Second, we shift the LF of each individual cluster along the magnitude-axis by subtracting the characteristic magnitude \mstar\ predicted by our CSP model at the cluster redshift.  That is, if one fits the \citet{schechter76} LF model to the normalized LFs, then any deviation of the best-fit \mstar\ from zero implies a deviation of the cluster galaxy population from our CSP model prediction.  In the end, we derive the stacked LFs by taking the inverse-variance weighted average of the stacked clusters at each magnitude bin.

After constructing the stacked LF of the samples, we fit the \citet{schechter76} LF model-- with the normalization $\phi_{0}$, characteristic magnitude ${\mstar}_{,~\mathrm{stacked}}$ and the faint end power law index $\alpha$ varied-- to the stacked profiles via $\chi^{2}$ minimization.  We restrict the fitting to  magnitudes with $m - \mstar \le 1.5$~mag, which ensures that all the magnitude bins of the stacked clusters are above $50\percent$ completeness (see Section~\ref{sec:ssdf_catalog}) where the LF modeling is not dominated by the incompleteness.  The resulting stacked LF, the best-fit model and the constraints of ${\mstar}_{,~\mathrm{stacked}}$ and $\alpha$ are shown in Figure~\ref{fig:stacked_lf}.

As shown in the right panel of Figure~\ref{fig:stacked_lf}, the best fit LF
for the full sample stack shows no significant deviation from the CSP model with  ${\mstar}_{,~\mathrm{stacked}} = -0.035\pm0.31$~mag, and provides a faint end constraint $\alpha=-0.89\pm0.29$.  This implies that the cluster galaxy population of the \XMMBCS\ sample can be well described by our CSP model.  This value for the faint end slope is consistent with the values $\alpha=-0.84\pm0.08$ \citep{muzzin07a}  and $\alpha=-0.84\pm0.02$ \citep{lin04a} published in two previous analyses.

Moreover, the LF parameters derived from the different subsamples exhibit no significant discrepancy among themselves or with the full sample (see Figure~\ref{fig:stacked_lf}).  All four samples are consistent (within $\lesssim1\sigma$).  This implies that the evolution of the characteristic magnitude of the cluster galaxy populations of the \XMMBCS\ are consistent with the CSP model and that there is no significant redshift trend in $\alpha$.  As a result, for the individual cluster LF fits we use the characteristic magnitude \mstar\ predicted by the CSP model and the faint end power law index $\alpha=-0.89$ (see Section~\ref{sec:stellarmass_estimation}). 

\subsection{Mass-to-light ratio \ML\ of \XMMBCS\ sample}
\label{sec:m2l_method}

The stellar mass of each cluster is estimated by multiplying the total NIR light by the mass-to-light ratio \ML.  In general, \ML\ varies among the types of galaxies. Evidence tends to show that the cluster galaxy population is dominated by early type galaxies \citep[e.g.][]{dressler80},
but with a component of late type galaxies that vary with cluster mass \citep{desai07, jeltema07, mei09}.  Thus, using a constant \ML\ appropriate for a passively evolving population would unavoidably bias the  stellar mass of the cluster galaxies.  In this work, we use the synthetic \ML\ in the \IRACone\ filter for each cluster derived by mixing the mass-to-light ratio of a passive red and a star forming blue galaxy population, using the estimated blue fractions \fblue\ of the \XMMBCS\ clusters.

\begin{figure}
\vskip-0.2in
\centering
\includegraphics[scale=0.6]{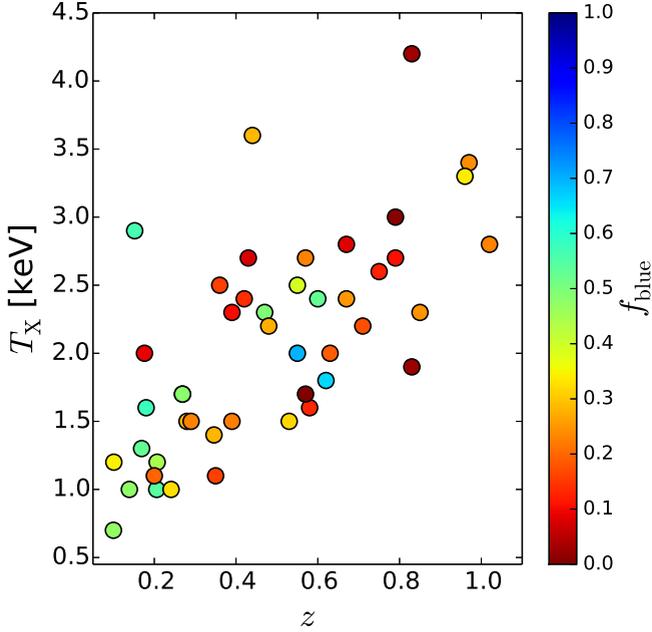}
\caption{
A plot of the blue fraction \fblue\ of \XMMBCS\ clusters as a function of X-ray temperature \Tx\ and redshift $z$.  The value \fblue\ for each cluster is color coded according to the scaling given in the colorbar.  The uncertainties for \Tx, $z$ and \fblue\ are omitted for clarity (see discussion in Appendix~\ref{app:fblue_xmmbcs}).
}
\label{fig:fblue_scatter}
\end{figure}

We estimate \fblue\ for the 46 \XMMBCS\ clusters by exploiting the \BCS\ optical catalog (see discussion in Appendix~\ref{app:fblue_xmmbcs}). The estimated \fblue\ for all \XMMBCS\ clusters is presented in Table~\ref{tab:measurement}, and \fblue\ is plotted as a function of the \XMMBCS\ cluster X-ray temperature (\Tx) and redshift \redshift\ in Figure~\ref{fig:fblue_scatter}.  We find no significant redshift trend of \fblue\ for $\Tx\gtrsim2~\mathrm{keV}$ in the \XMMBCS\ cluster sample, while the mass trend that \fblue\ increases toward low mass (or low \Tx) clusters is apparent.  
This weak mass trend of \fblue\ is consistent with the recent result based on the sample of clusters at low redshift $\redshift\lesssim0.05$ \citep{shan15}.
We extract the \Tx\ and redshift trends of \fblue\ from the observed \fblue\ of the \XMMBCS\ clusters by fitting a model to the estimated \fblue.  Specifically, we fit a function $\fblue(\Tx, \redshift)$, finding
\begin{equation}
\label{eq:fblue_in_main_texts}
\fblue(\Tx,\redshift) = \frac{(0.21\pm0.40) \redshift+(0.31\pm0.15)}{\Tx} \, .
\end{equation}
Appendix~\ref{app:fblue_xmmbcs} contains more details of the fitting of the function $\fblue(\Tx, \redshift)$.

To derive the mass-to-light ratio (${\ML}_{,\mathrm{blue}}$) in \IRACone\ for the blue population, we adopt an extreme star formation history ($\tau=10$~Gyr) at $\redshift_{\mathrm{f}}=3$ with solar metallicity (see Section~\ref{sec:xmmbcs_catalog} for the model configuration); the resulting ${\ML}_{,\mathrm{blue}}$ is $\approx(52\mathrm{-}59)\percent$ of the mass-to-light ratio ${\ML}_{,\mathrm{CSP}}$ derived from the CSP model for $0.1\le \redshift \le1.0$.

We derive the synthetic mass-to-light ratio \ML\ for each cluster using the best-fit \fblue\ (equation~\ref{eq:fblue_in_main_texts}), ${\ML}_{,\mathrm{blue}}$ and the mass-to-light ratio ${\ML}_{,\mathrm{CSP}}$ estimated from the CSP model, i.e.,
\begin{equation}
\label{eq:derive_m2l}
\ML = (1 - \fblue(\Tx,\redshift))~{\ML}_{,\mathrm{CSP}} + \fblue(\Tx,\redshift)~{\ML}_{,\mathrm{blue}} \, .
\end{equation}
In this way we use a mass-to-light ratio that accounts for the trends in blue fraction variation with temperature and redshift, allowing us to avoid introducing stellar mass biases over the range of mass and redshift probed by our sample.  Using the ensemble fit results provides a way of avoiding the use of the noisy, single cluster blue fraction measurements directly in the mass-to-light ratio calculation.

\subsection{Stellar Mass Estimations}
\label{sec:stellarmass_estimation}

We derive the total stellar mass of each of the 46 \XMMBCS\ cluster by multiplying the total luminosity, estimated from the satellite galaxies that lie projected within cluster \Rfiveoo\ and the BCG,  by the derived mass-to-light ratios.  To estimate the luminosity of the satellite galaxies (BCG excluded) in $3.6~\micron$, we fit a model to the observed magnitude distribution of each cluster.  Specifically, the magnitude distribution model $M(m)$ is constructed using the observed background magnitude distribution
$B(m)$, a \citet{schechter76} LF $\phi(m)$ that represents the cluster galaxies and the \SSDF\ completeness function $\fcom(m)$ as a function of magnitude $m$:
\begin{equation}
\label{eq:md_model}
M(m) =  \phi(m, \phi_{0}, \mstar, \alpha) \fcom(m)  + B(m) \, ,
\end{equation}
where $\phi_{0}$, $\mstar$ and $\alpha$ are the normalization, characteristic apparent magnitude and the faint end power law index of the LF, respectively.  The background $B(m)$ is fixed in the model because the uncertainties of $B(m)$ are small due to its being drawn from an area that is 25 times the area of the cluster.  That is, the uncertainties of the cluster LF are dominated by the cluster field.  We use the \citet{cash79} statistic with a maximum likelihood estimator \Cstat\ to estimate the parameters in the fitting.
\begin{equation}
\label{eq:cstat}
\Cstat = 2\sum_{j} \left( M(m_{j}) - N(m_{j}) + N(m_{j}) \ln\left( \frac{N(m_{j})}{M(m_{j})} \right) \right) \, ,
\end{equation}
where $j$ runs over all the magnitude bins in the fit and $N$ is the observed magnitude distribution for the cluster.
Using the estimator \Cstat\ allows us to estimate the goodness of fit (GOF) for the data following the Poisson distribution in the same way as a $\chi^2$-distribution.  The GOF of the LF fitting is defined by the ratio of the best-fit \Cstat\ to the degrees of freedom $(\mathrm{d.o.f})$ (i.e., $\mathrm{GOF} = \Cstat/\mathrm{d.o.f}$) and has a corresponding probability to exceed that provides information about tension between the best fit model and the data.

We find that the magnitude distribution is generally too noisy to constrain the three parameters of the LF for the individual clusters.  Therefore, we fix \mstar\ to the \IRAConestar\ predicted by the CSP model and the faint end slope to $\alpha=-0.89$, which is measured in the stacked profile (see discussion in Section~\ref{sec:lf_xmmbcs}).  Essentially, we fit for only one parameter $\phi_{0}$ on a single cluster basis.  The fit is done using the magnitude range extending from the magnitude of the BCG to the 50\percent\ completeness limit.

We convert the BCG magnitude to the rest-frame luminosity (\LBCG) at the cluster redshift with the $k$-correction-- estimated from our CSP model \citep{mancone12b}.  In this work we do not attempt to correct for the contribution from the intracluster light around the BCG.  The total luminosity of the cluster galaxies \Lstar\ is the sum of the BCG luminosity \LBCG\ and the luminosity of the satellite galaxies \Lstarsat\ within the \Rfiveoo\ sphere, where  \Lstarsat\ is obtained as follows.
\begin{equation}
\label{eq:totl}
\Lstarsat = D_{\mathrm{prj}}\int^{\LBCG}_{L_{\mathrm{min}}} \phi(L)~L~\dif L \, ,
\end{equation}
where $D_{\mathrm{prj}}$ is the deprojection correction from cylinder to sphere, and $L_{\mathrm{min}}$ is the lower threshold of the integrated interval. We use $D_{\mathrm{prj}}=0.69$, which is derived by assuming an NFW \citep[][]{navarro96} distribution with concentration $\Cfiveoo=1.8$ with respect to \Rfiveoo\ for the distribution of cluster galaxies \citep{lin04a}.  We set $L_{\mathrm{min}}$ to be the luminosity corresponding to $\IRAConestar+2$, ensuring that \Lstar\ of each cluster is estimated to a consistent depth.  The luminosity is converted from the magnitude at the cluster redshift with the $k$-correction applied.

The uncertainty of \Lstarsat\ for each cluster is derived by bootstrapping the galaxies in the observed magnitude distribution of the cluster field and repeating the whole process described above. For the photometric uncertainty of the BCG we would like to be able to perform \textit{repeatability tests} \citep{desai12,liu15b} on the multiple single epoch images of a particular BCG. In that process  one calculates the scatter of the photometric measurements of the same objects and uses that to characterize the uncertainty.  However, we do not have the required data products to carry out this test, and besides this test would likely not include uncertainties due to the extended halo of light surrounding them.  So for this analysis we have scaled up the  \texttt{MAG\_AUTO} uncertainties of the BCG using the method in \cite{barmby08}, which leads to the typical uncertainty of the BCG at the level of $\approx9\percent$.  Given that the BCG typically contributes $\approx10\mathrm{-}40\percent$ of the total luminosity of the \XMMBCS\ clusters, this implies that the photometric uncertainty of the BCG is at the level $\approx1\mathrm{-}4\percent$ of the total luminosity and therefore is not the dominant source of uncertainty in our analysis.  

In the end, the stellar mass for each cluster is obtained by
\begin{eqnarray}
\label{eq:light2mass}
\Mstarsat        &= &\ML~\Lstarsat \, \nonumber\\
\Mstar            &= &{\ML}_{,\mathrm{CSP}}~\LBCG\ + \ML~\Lstarsat \, ,
\end{eqnarray}
where \Mstar\ is the total stellar mass of the cluster, \Mstarsat\ is the total stellar mass of the satellite galaxies and \ML\ is the synthetic mass-to-light ratio in \IRACone\ derived in Section~\ref{sec:m2l_method}.   We use the mass-to-light ratio estimated from the CSP model ${\ML}_{,\mathrm{CSP}}$ for the BCG, given the evidence that the BCG can typically be well described by a passively evolving model out to redshift $\approx1.5$ \citep{lidman12,wylezalek14}.

\subsection{Stellar Mass to Halo Mass Scaling Relations}
\label{sec:likelihood}

The scaling relation of the stellar mass-halo mass is defined as:
\begin{equation}
\label{eq:sr}
\Mstar = \Astar 
\left( \frac{\Mfiveoo}{\MPIV   } \right)^{\Bstar} 
\left(\frac{1+z}{1+\ZPIV} \right)^{\Cstar} \, ,
\end{equation}
with the intrinsic, log-normal scatter $\Dstar\equiv\Dstarcom$ of the observed \Mstar\ for a given cluster mass \Mfiveoo. The normalization, the mass power index and the redshift power law index of the scaling relations are denoted by \Astar, \Bstar\ and \Cstar, respectively.
The \MPIV\ and \ZPIV\ are fixed to the median values of the cluster sample: $\MPIV\equiv0.8\times10^{14}\Msun$ and $\ZPIV\equiv0.47$.

\begin{table}
\centering
\caption{
The measurements of \XMMBCS\ clusters.
Column~1: the unique ID of the \XMMBCS\ clusters.
Column~2: the cluster redshift.
Column~3: the stellar mass estimate of the cluster in units of $10^{12}\Msun$.
Column~4: the normalization $\phi_{0}$ of the best-fit LF in units of ${\Lsun}^{-1}$.
Column~5: the $p$-value of consistency between the LF model and the data.
Column~6: the measured blue fraction \fblue.
}
\label{tab:measurement}
\resizebox{\columnwidth}{!}{
\begin{tabular}{cccccc}
\hline
ID & Redshift 
 & \Mstar\ 
 & $\phi_{0}$ 
 & $p$-value
 & \fblue   \\  
 & & [$10^{12}\Msun$] & [${\Lsun}^{-1}$] & & \\
\hline \hline
     $ 11 $ &$ 0.970 $ &$ 0.65  \pm  0.40 $ &$ 5.83  \pm  6.22 $ &$ 0.181 $ &$ 0.241  \pm  0.338 $   \\  
     $ 18 $ &$ 0.390 $ &$ 2.57  \pm  0.50 $ &$ 33.5  \pm  8.16 $ &$ 0.368 $ &$ 0.104  \pm  0.082 $   \\  
     $ 32 $ &$ 0.830 $ &$ 3.48  \pm  0.64 $ &$ 40.6  \pm  9.83 $ &$ 0.189 $ &$ 0.030  \pm  0.077 $   \\  
     $ 33 $ &$ 0.790 $ &$ 3.43  \pm  0.54 $ &$ 30.5  \pm  8.34 $ &$ 0.505 $ &$ 0.009  \pm  0.036 $   \\  
     $ 34 $ &$ 0.280 $ &$ 2.15  \pm  0.41 $ &$ 30.0  \pm  6.87 $ &$ 0.805 $ &$ 0.289  \pm  0.162 $   \\  
     $ 35 $ &$ 0.670 $ &$ 2.33  \pm  0.49 $ &$ 29.5  \pm  7.90 $ &$ 0.392 $ &$ 0.253  \pm  0.199 $   \\  
     $ 38 $ &$ 0.390 $ &$ 1.05  \pm  0.33 $ &$ 13.4  \pm  5.66 $ &$ 0.752 $ &$ 0.223  \pm  0.298 $   \\  
     $ 39 $ &$ 0.180 $ &$ 0.80  \pm  0.24 $ &$ 6.20  \pm  3.90 $ &$ 0.983 $ &$ 0.575  \pm  0.230 $   \\  
     $ 44 $ &$ 0.440 $ &$ 7.81  \pm  0.82 $ &$ 111.  \pm  12.8 $ &$ 0.019 $ &$ 0.287  \pm  0.052 $   \\  
     $ 69 $ &$ 0.750 $ &$ 2.64  \pm  0.48 $ &$ 31.7  \pm  7.73 $ &$ 0.057 $ &$ 0.125  \pm  0.241 $   \\  
     $ 70 $ &$ 0.152 $ &$ 2.67  \pm  0.48 $ &$ 37.7  \pm  7.69 $ &$ 0.056 $ &$ 0.563  \pm  0.233 $   \\  
     $ 81 $ &$ 0.850 $ &$ 1.56  \pm  0.47 $ &$ 17.1  \pm  7.72 $ &$ 0.154 $ &$ 0.248  \pm  0.336 $   \\  
     $ 82 $ &$ 0.630 $ &$ 2.94  \pm  0.54 $ &$ 39.5  \pm  8.84 $ &$ 0.336 $ &$ 0.190  \pm  0.124 $   \\  
     $ 88 $ &$ 0.430 $ &$ 2.44  \pm  0.55 $ &$ 32.4  \pm  8.82 $ &$ 0.135 $ &$ 0.076  \pm  0.097 $   \\  
     $ 90 $ &$ 0.580 $ &$ 1.99  \pm  0.41 $ &$ 23.5  \pm  6.87 $ &$ 0.182 $ &$ 0.131  \pm  0.172 $   \\  
     $ 94 $ &$ 0.269 $ &$ 0.59  \pm  0.28 $ &$ 4.65  \pm  4.72 $ &$ 0.451 $ &$ 0.225  \pm  0.341 $   \\  
     $ 109 $ &$ 1.020 $ &$ 2.98  \pm  0.55 $ &$ 36.4  \pm  8.73 $ &$ 0.587 $ &$ 0.227  \pm  0.231 $   \\  
     $ 110 $ &$ 0.470 $ &$ 3.03  \pm  0.48 $ &$ 28.3  \pm  7.55 $ &$ 0.073 $ &$ 0.496  \pm  0.094 $   \\  
     $ 126 $ &$ 0.420 $ &$ 1.99  \pm  0.46 $ &$ 24.0  \pm  7.43 $ &$ 0.568 $ &$ 0.137  \pm  0.070 $   \\  
     $ 127 $ &$ 0.207 $ &$ 0.81  \pm  0.28 $ &$ 7.27  \pm  4.86 $ &$ 0.057 $ &$ 0.443  \pm  0.201 $   \\  
     $ 132 $ &$ 0.960 $ &$ 2.01  \pm  0.56 $ &$ 25.1  \pm  8.86 $ &$ 0.075 $ &$ 0.340  \pm  0.253 $   \\  
     $ 136 $ &$ 0.360 $ &$ 3.15  \pm  0.54 $ &$ 45.2  \pm  8.65 $ &$ 0.469 $ &$ 0.160  \pm  0.080 $   \\  
     $ 139 $ &$ 0.169 $ &$ 2.01  \pm  0.34 $ &$ 29.9  \pm  5.76 $ &$ 0.956 $ &$ 0.531  \pm  0.123 $   \\  
     $ 150 $ &$ 0.176 $ &$ 2.36  \pm  0.41 $ &$ 27.1  \pm  6.65 $ &$ 0.148 $ &$ 0.090  \pm  0.086 $   \\  
     $ 152 $ &$ 0.139 $ &$ 0.26  \pm  0.14 $ &$ 0.0  \pm  2.48 $ &$ 0.273 $ &$ 0.473  \pm  0.341 $   \\  
     $ 156 $ &$ 0.670 $ &$ 1.30  \pm  0.44 $ &$ 9.51  \pm  6.96 $ &$ 0.316 $ &$ 0.083  \pm  0.221 $   \\  
     $ 158 $ &$ 0.550 $ &$ 3.11  \pm  0.54 $ &$ 40.7  \pm  8.75 $ &$ 0.974 $ &$ 0.391  \pm  0.110 $   \\  
     $ 210 $ &$ 0.830 $ &$ 1.44  \pm  0.41 $ &$ 15.7  \pm  6.89 $ &$ 0.137 $ &$ 0.026  \pm  0.098 $   \\  
     $ 227 $ &$ 0.346 $ &$ 0.98  \pm  0.34 $ &$ 6.79  \pm  5.78 $ &$ 0.726 $ &$ 0.283  \pm  0.311 $   \\  
     $ 245 $ &$ 0.620 $ &$ 1.71  \pm  0.41 $ &$ 19.8  \pm  6.77 $ &$ 0.255 $ &$ 0.663  \pm  0.159 $   \\  
     $ 275 $ &$ 0.290 $ &$ 0.65  \pm  0.30 $ &$ 8.36  \pm  5.06 $ &$ 0.852 $ &$ 0.230  \pm  0.153 $   \\  
     $ 287 $ &$ 0.570 $ &$ 1.65  \pm  0.40 $ &$ 21.8  \pm  6.73 $ &$ 0.001 $ &$ 0.000  \pm  0.000 $   \\  
     $ 288 $ &$ 0.600 $ &$ 2.69  \pm  0.50 $ &$ 29.1  \pm  8.04 $ &$ 0.484 $ &$ 0.529  \pm  0.109 $   \\  
     $ 357 $ &$ 0.480 $ &$ 3.21  \pm  0.53 $ &$ 38.3  \pm  8.48 $ &$ 0.463 $ &$ 0.277  \pm  0.170 $   \\  
     $ 386 $ &$ 0.530 $ &$ 2.83  \pm  0.45 $ &$ 39.8  \pm  7.71 $ &$ 0.051 $ &$ 0.317  \pm  0.136 $   \\  
     $ 430 $ &$ 0.206 $ &$ 0.83  \pm  0.22 $ &$ 4.24  \pm  3.79 $ &$ 0.627 $ &$ 0.543  \pm  0.297 $   \\  
     $ 444 $ &$ 0.710 $ &$ 2.13  \pm  0.45 $ &$ 27.9  \pm  7.36 $ &$ 0.556 $ &$ 0.175  \pm  0.286 $   \\  
     $ 457 $ &$ 0.100 $ &$ 1.05  \pm  0.26 $ &$ 16.5  \pm  5.04 $ &$ 0.000 $ &$ 0.471  \pm  0.321 $   \\  
     $ 476 $ &$ 0.101 $ &$ 1.78  \pm  0.31 $ &$ 25.2  \pm  5.29 $ &$ 0.105 $ &$ 0.347  \pm  0.151 $   \\  
     $ 502 $ &$ 0.550 $ &$ 1.88  \pm  0.46 $ &$ 25.8  \pm  7.55 $ &$ 0.035 $ &$ 0.698  \pm  0.208 $   \\  
     $ 511 $ &$ 0.269 $ &$ 2.91  \pm  0.45 $ &$ 42.2  \pm  7.48 $ &$ 0.236 $ &$ 0.482  \pm  0.115 $   \\  
     $ 527 $ &$ 0.790 $ &$ 2.60  \pm  0.52 $ &$ 27.1  \pm  8.28 $ &$ 0.316 $ &$ 0.107  \pm  0.214 $   \\  
     $ 528 $ &$ 0.350 $ &$ 0.96  \pm  0.28 $ &$ 11.3  \pm  5.07 $ &$ 0.382 $ &$ 0.160  \pm  0.156 $   \\  
     $ 538 $ &$ 0.200 $ &$ 0.91  \pm  0.29 $ &$ 13.2  \pm  5.13 $ &$ 0.073 $ &$ 0.196  \pm  0.230 $   \\  
     $ 543 $ &$ 0.570 $ &$ 1.79  \pm  0.51 $ &$ 21.3  \pm  8.16 $ &$ 0.591 $ &$ 0.228  \pm  0.161 $   \\  
     $ 547 $ &$ 0.241 $ &$ 0.97  \pm  0.28 $ &$ 10.9  \pm  5.06 $ &$ 0.198 $ &$ 0.326  \pm  0.229 $   \\  

\hline
\end{tabular}
}
\end{table}

We use the same likelihood as in L15 to estimate the best-fit parameters $\mathbf{r}_{\star}\equiv(\Astar,~\Bstar,~\Cstar,~\Dstar)$ of equation~(\ref{eq:sr}).  Namely, the scaling relation parameters $\mathbf{r}_{\star}$ are estimated by evaluating the likelihood 
\begin{equation}
\label{eq:L14likelihood}
P(\mathbf{r}_{\star}) = \sum_{i}^{\Ncl}\frac{
\int \dif \Mfiveoo\ 
P({\Mstar}_{i}, {\Lx}_{i}| \Mfiveoo, \redshift_{i}, \mathbf{r}_{\star}, \mathbf{r}_{\mathrm{X}}) 
n(\Mfiveoo, \redshift_{i})
}
{
\int \dif \Mfiveoo\ 
P({\Lx}_{i}| \Mfiveoo, \redshift_{i}, \mathbf{r}_{\mathrm{X}}) 
n(\Mfiveoo, \redshift_{i})
} \, ,
\end{equation}
where \Ncl\ is the total number of \XMMBCS\ clusters, ${\Mstar}_{i}$ and  ${\Lx}_{i}$ indicate respectively the stellar mass and the X-ray luminosity of the cluster $i$ at redshift $\redshift_{i}$,  and $\mathbf{r}_{\mathrm{X}}$ denotes the parameters of the \Lx-\Mfiveoo\ scaling relation (equation~(\ref{eq:lxbol2mass})) in the same form as for $\mathbf{r}_{\star}$.  
The probability of observing the cluster $i$ with ${\Lx}_{i}$, given the mass \Mfiveoo, redshift $z$ and the scaling relation $\mathbf{r}_{\mathrm{X}}$ is $P({\Lx}_{i}| \Mfiveoo, \redshift_{i}, \mathbf{r}_{\mathrm{X}})$, and the probability of observing the cluster $i$ with ${\Mstar}_{i}$ and ${\Lx}_{i}$ given the mass \Mfiveoo\ and the scaling relations $\mathbf{r}_{\star}$ and $\mathbf{r} _{\mathrm{X}}$ is denoted as $P({\Mstar}_{i}, {\Lx}_{i}| \Mfiveoo, \redshift_{i}, \mathbf{r}_{\star}, \mathbf{r}_{\mathrm{X}})$. The probabilities in both the numerator and the denominator are weighted by the mass function $n(\Mfiveoo,\redshift_{i})$, derived using the \cite{tinker08} mass function at cluster redshift $\redshift_{i}$ within the cosmology framework used in this analysis.  

The full derivation of the likelihood is provided in L15, to which we refer the readers.  There it is shown that the selection of the sample (in this case the clusters were selected using their X-ray flux) does not impact the derivation of an unbiased scaling relation $\mathbf{r}_{\star}$.  

In L15 it was only possible to constrain the normalization and mass trend of the SZE signal-to-noise mass relation, because of the weak SZE signatures of these low mass systems.  In the current analysis the relatively higher signal to noise measurements of the stellar mass for \XMMBCS\ clusters allow us to constrain all four of the parameters of the scaling relation $\mathbf{r}_{\star}$ by maximizing the likelihood of equation~(\ref{eq:L14likelihood}).  The parameter space is explored by using \texttt{emcee} \citep{foreman13}, the Python package to search for the maximum likelihood using the Affine Invariant Markov Chain Monte Carlo  (MCMC) algorithm.  The cosmology parameters and the redshifts of the clusters are fixed in the MCMC likelihood maximization.  The measurement uncertainty of \Lx\ for each cluster and the intrinsic scatter of the scaling relation $\mathbf{r}_{\mathrm{X}}$ (equation~(\ref{eq:lxbol2mass})), which is assumed to be uncorrelated with the intrinsic scatter of the scaling relation $\mathbf{r}_{\star}$ (equation~(\ref{eq:sr})), are taken into account in deriving the stellar mass scaling relation $\mathbf{r}_{\star}$. We study both the total stellar mass to mass scaling relation $\Mstar(\Mfiveoo, \redshift)$ and the equivalent scaling relation when excluding the BCG mass $\Mstarsat (\Mfiveoo, \redshift)$. We apply uniform priors as shown in Table~\ref{tab:prior_and_fit}.  The widths of these priors are chosen to be larger than the recovered probability distributions.

\begin{figure}
\centering
\includegraphics[width=0.5\textwidth]{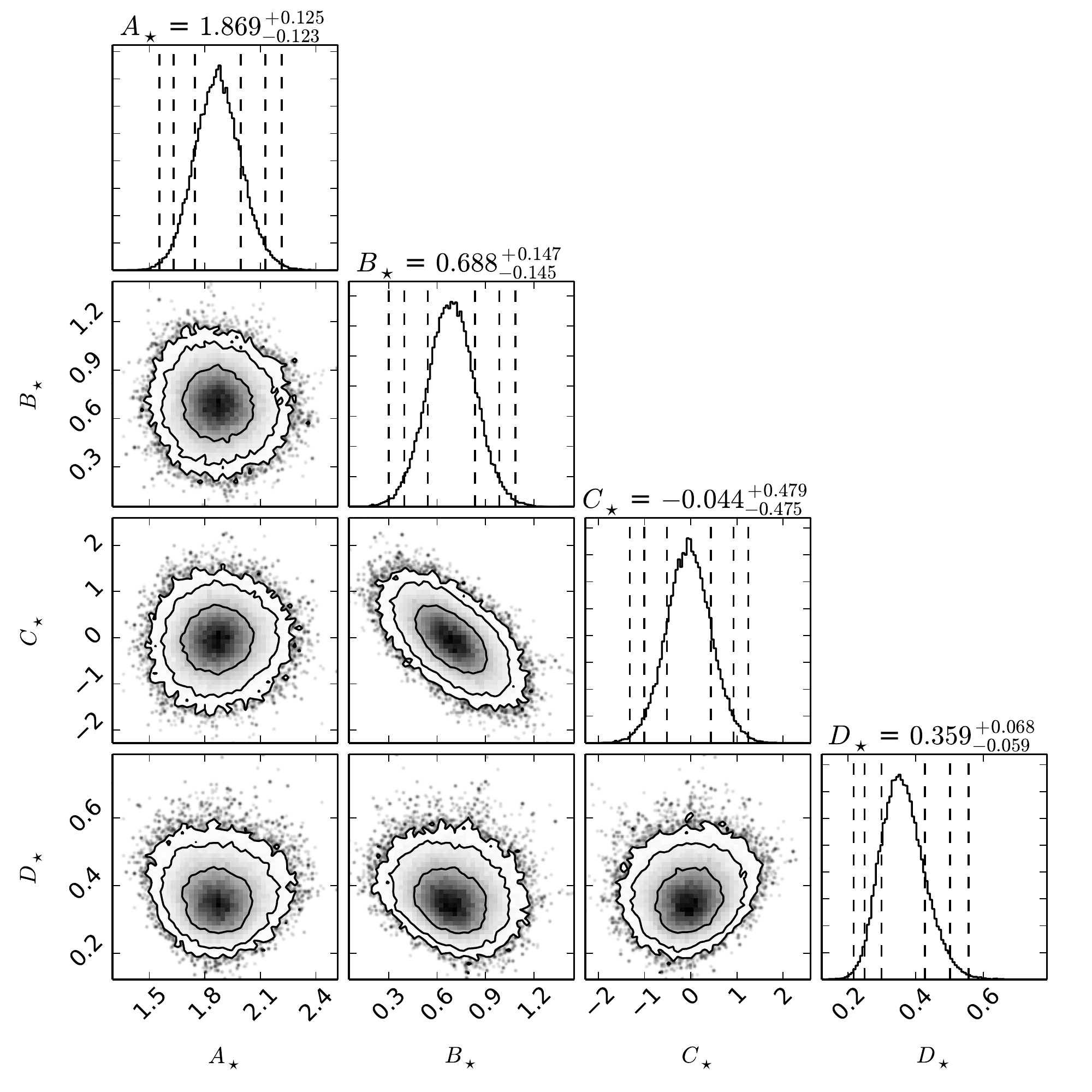}
\caption{Scaling relation parameter $\mathbf{r_{\star}}$ constraints for $\Mstar(\Mfiveoo, \redshift)$. The parameters are the normalization \Astar, power law index in mass \Bstar, power law index in redshift \Cstar\ and the intrinsic log-normal scatter \Dstar.  Both joint and fully marginalized constraints are shown.  The numerical values of the best-fit parameters and $1\sigma$ uncertainties are quoted at the top of each column, and the off-diagonal plots show joint constraints with $1\sigma$, $2\sigma$ and $3\sigma$ confidence contours.
}
\label{fig:triangle_full}
\end{figure}
%

%
%

%
\begin{table}
\centering
\caption{
Stellar mass to halo mass scaling relation parameter constraints and priors.  The columns contain
the normalization in units of $10^{12}\Msun$, the mass and redshift power law indices and the intrinsic, log-normal scatter in the observable at fixed mass.
}
\label{tab:prior_and_fit}
\begin{tabular}{ccccc}
\hline\hline
                           &$\Astar[10^{12}\Msun]$   &\Bstar\        &\Cstar\        &\Dstar\        \\
\hline
priors                  &$[0.1,~50]$       &$[-3,~3]$        &$[-6.5,~6.5]$        &$[0,~1.5]$        \\ [4pt]
\Mstar\               &$1.87^{+0.13}_{-0.12}$       
                           &$0.69^{+0.15}_{-0.15}$        
                           &$-0.04^{+0.48}_{-0.48}$        
                           &$0.36^{+0.07}_{-0.06}$        \\ [4pt]
\Mstarsat\          &$1.37^{+0.11}_{-0.11}$       
                           &$0.80^{+0.18}_{-0.18}$        
                           &$-0.26^{+0.58}_{-0.58}$        
                           &$0.43^{+0.08}_{-0.07}$        \\ [4pt]
\hline
\end{tabular}
\end{table}
\begin{figure*}
\centering
    \begin{subfigure}[b]{0.475\textwidth} \includegraphics[width=\textwidth]{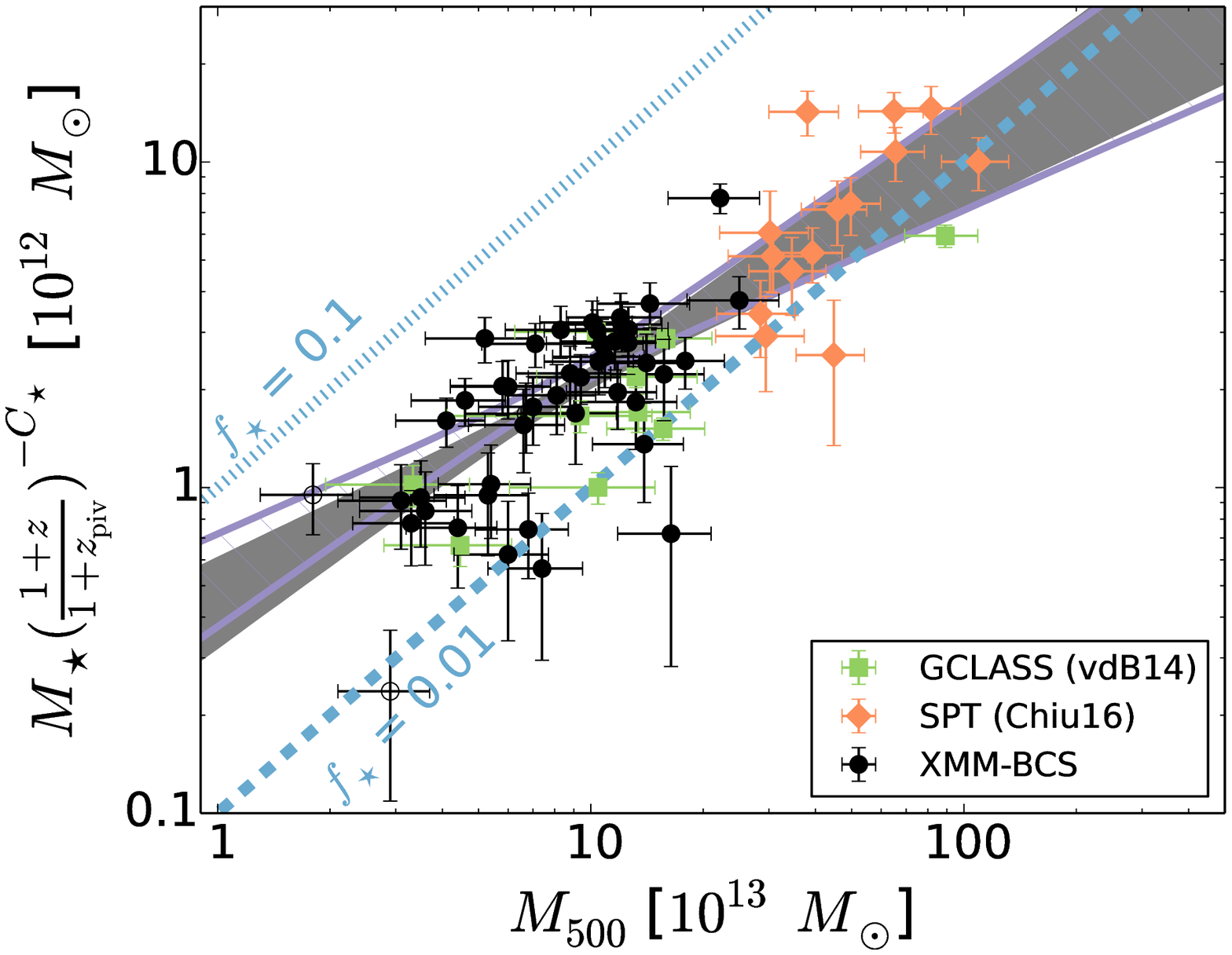} \end{subfigure}    
    \begin{subfigure}[b]{0.475\textwidth} \includegraphics[width=\textwidth]{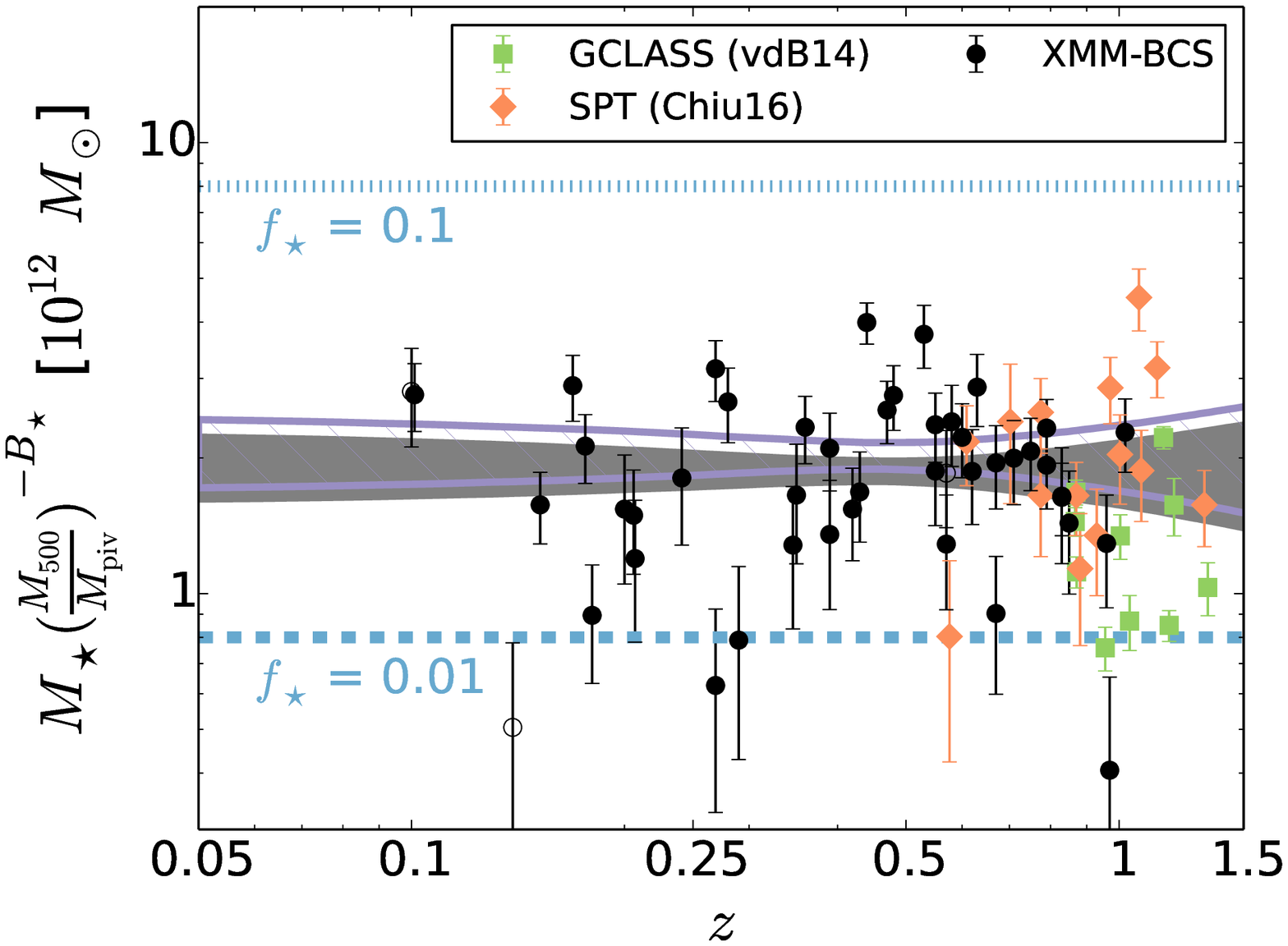}\end{subfigure}  
    \vskip-0.1in
    \caption{
The \Mstar\ of \XMMBCS\ clusters (black) and the comparison samples from SPT (red) and GCLASS (green) along with the best-fit scaling relation obtained in this work.  On the left, the total stellar masses corrected to the characteristic redshift $\ZPIV=0.47$ with the best fit redshift evolution are plotted as a function of cluster mass \Mfiveoo.  On the right, the total stellar masses corrected to the pivot mass $\MPIV=8\times10^{13}\Msun$ using the best fit mass trend are plotted as a function of cluster redshift.  Three problematic clusters (\XMMBCS152, \XMMBCS287 and \XMMBCS457) are shown with open circles.  The $1\sigma$ confidence region (see Table~\ref{tab:prior_and_fit}) of the best fit scaling relation is shaded. The $1\sigma$ confidence region of the best-fit relation assuming $\ML={\ML}_{,\mathrm{CSP}}$ (see the text in Section~\ref{sec:sys_m2l}) is enclosed by the thick purple lines.  For reference, the stellar mass fractions of 0.1 and 0.01 are shown using dotted and dashed lines, respectively.  The SPT and GCLASS samples have been corrected for estimated binding mass systematic offsets with respect to the \XMMBCS\ sample as described in the text. 
    }
    \label{fig:sr_mz}
\end{figure*}

\begin{figure*}
    \centering
    \begin{subfigure}[b]{0.20\textwidth} \includegraphics[width=\textwidth]{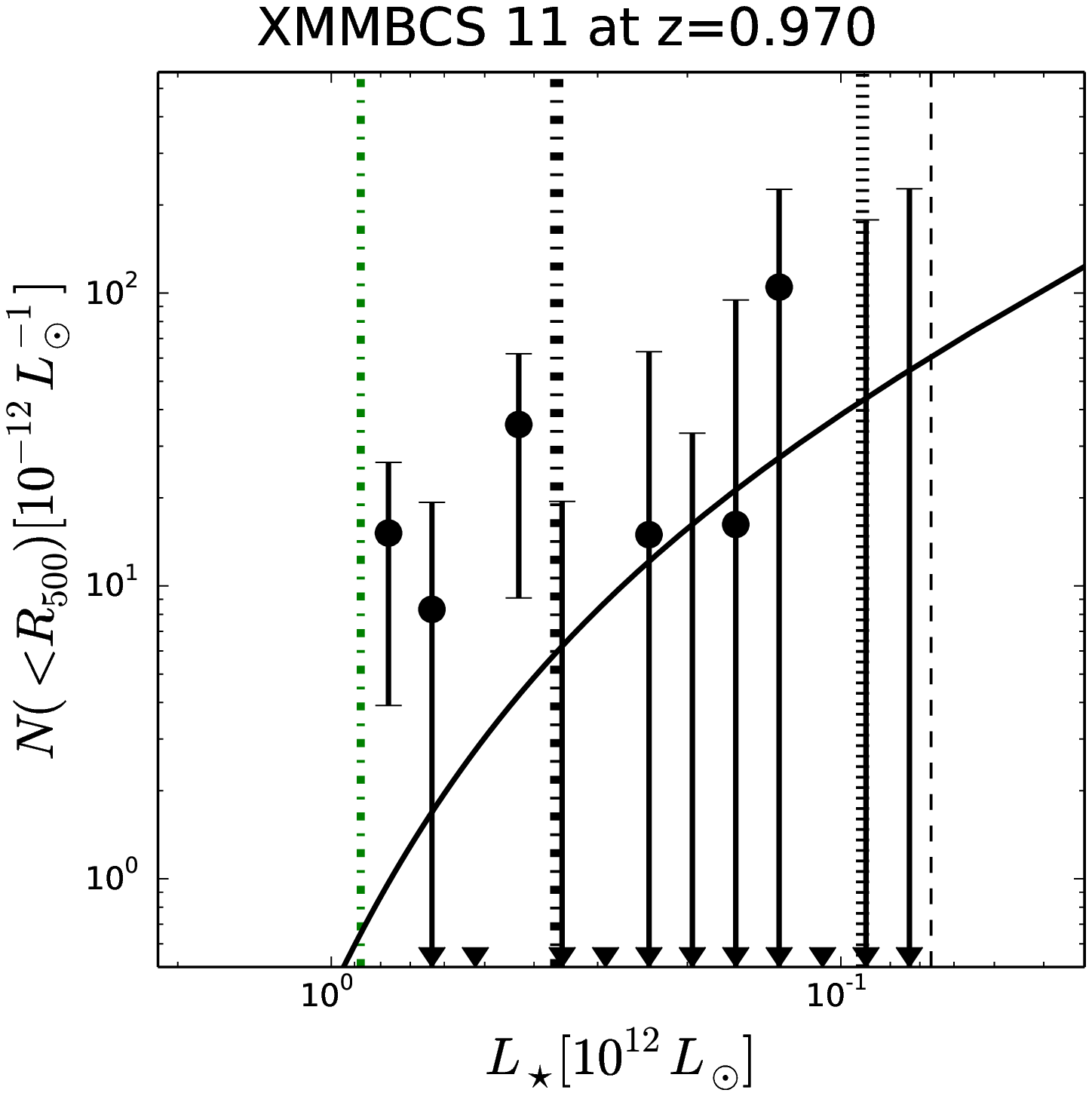} \end{subfigure}    
    \begin{subfigure}[b]{0.20\textwidth} \includegraphics[width=\textwidth]{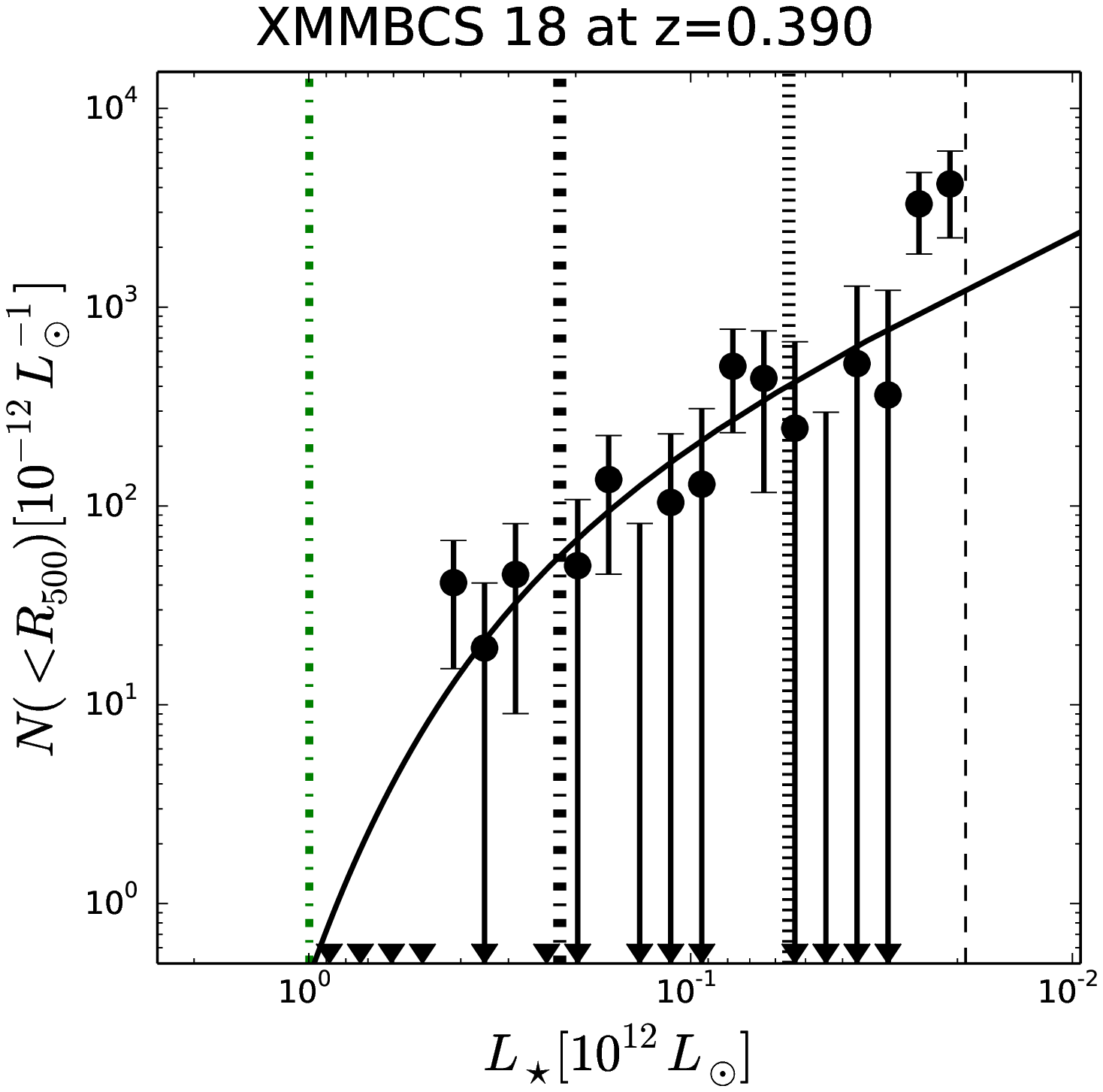} \end{subfigure}    
    \begin{subfigure}[b]{0.20\textwidth} \includegraphics[width=\textwidth]{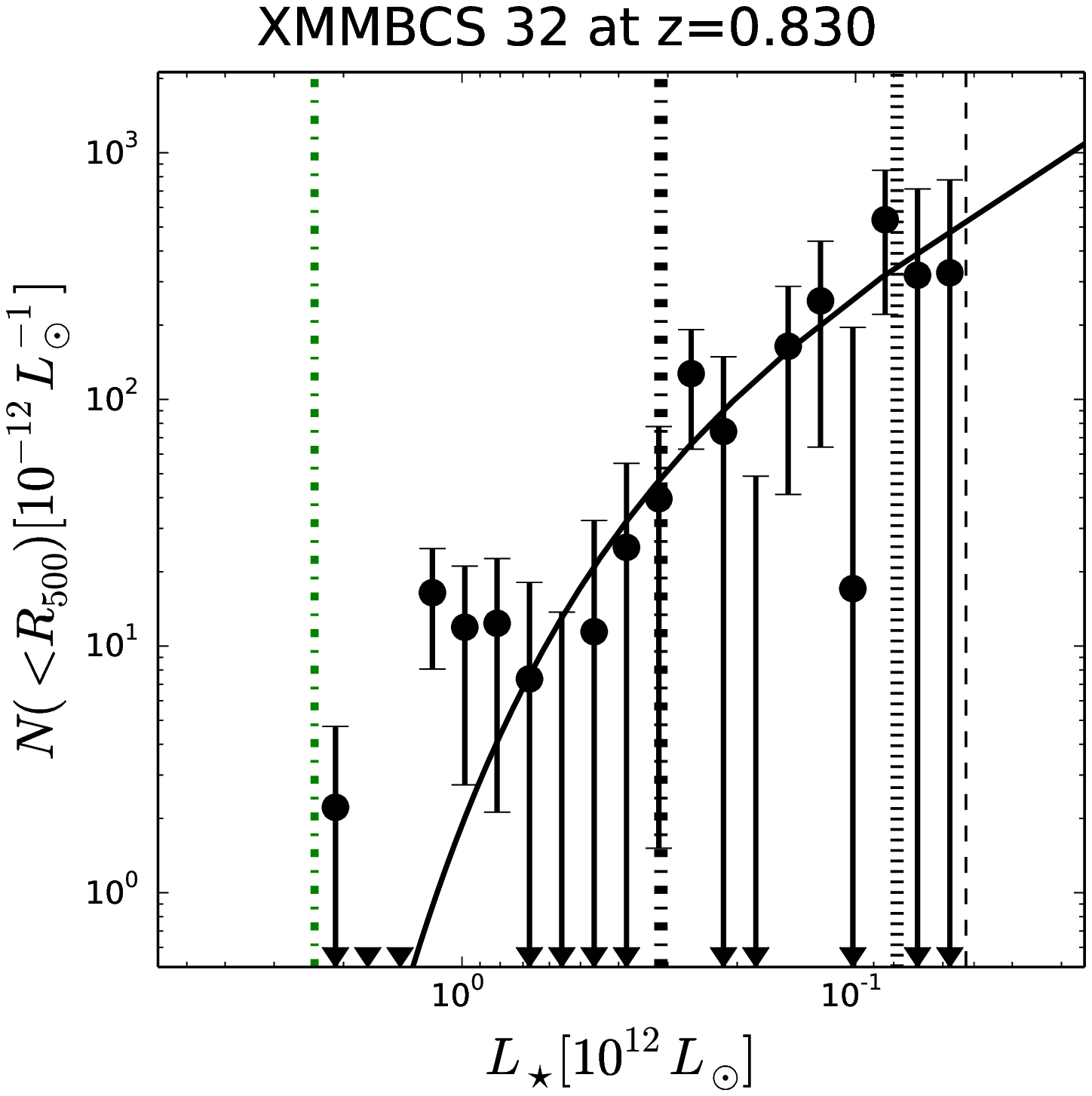} \end{subfigure}        
    \begin{subfigure}[b]{0.20\textwidth} \includegraphics[width=\textwidth]{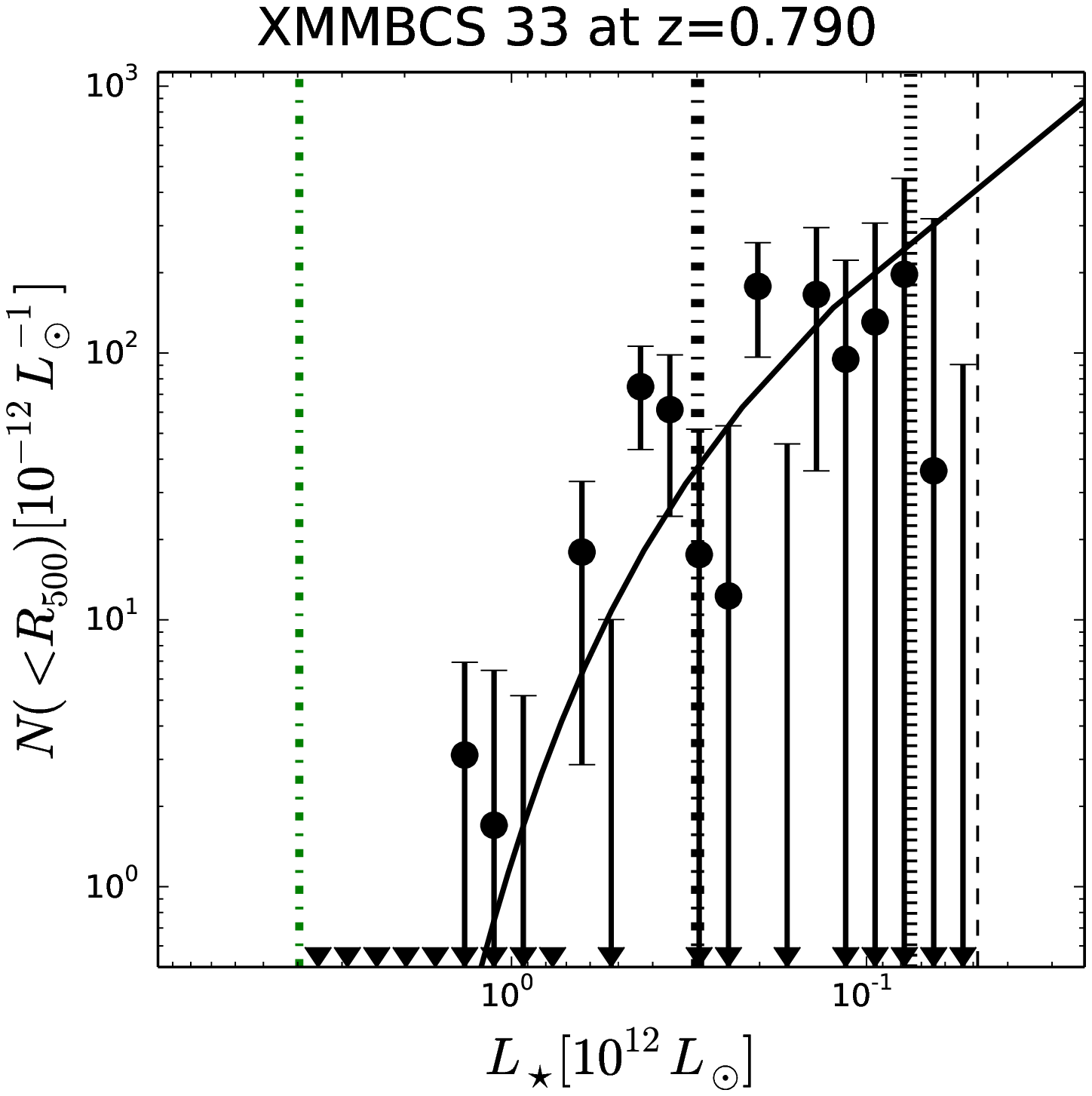} \end{subfigure}    
    
    \begin{subfigure}[b]{0.20\textwidth} \includegraphics[width=\textwidth]{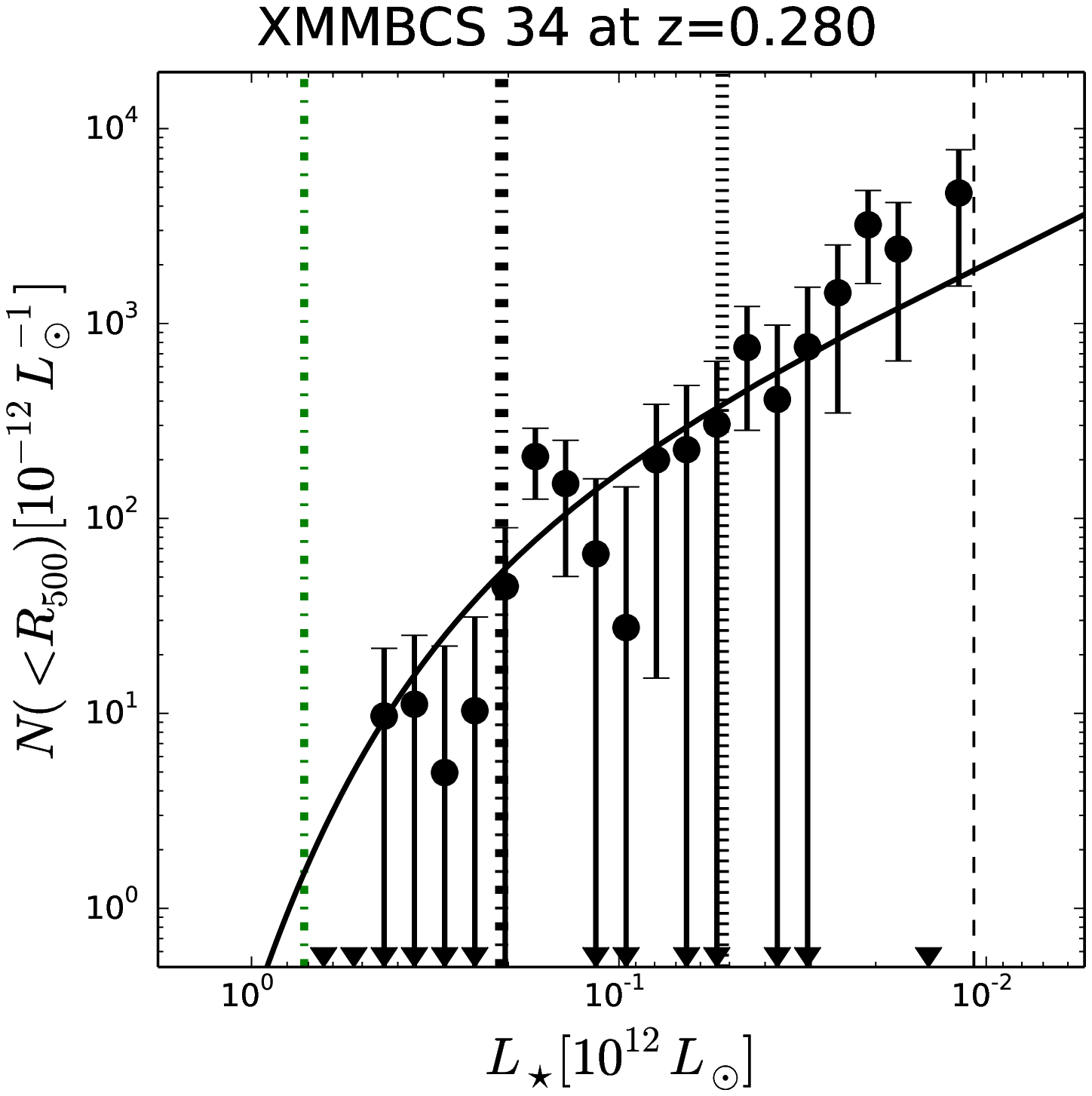} \end{subfigure}    
    \begin{subfigure}[b]{0.20\textwidth} \includegraphics[width=\textwidth]{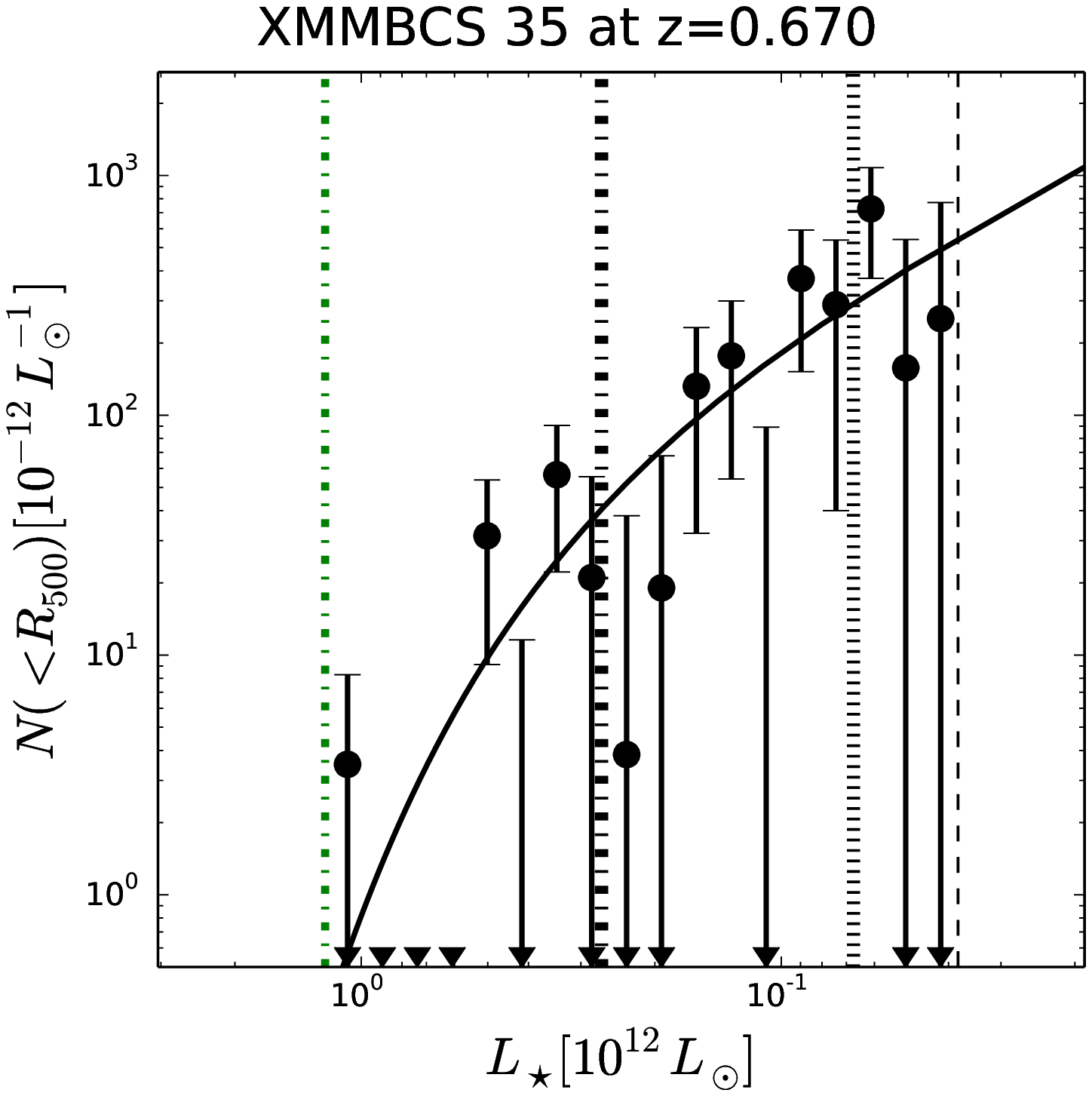} \end{subfigure}      
    \begin{subfigure}[b]{0.20\textwidth} \includegraphics[width=\textwidth]{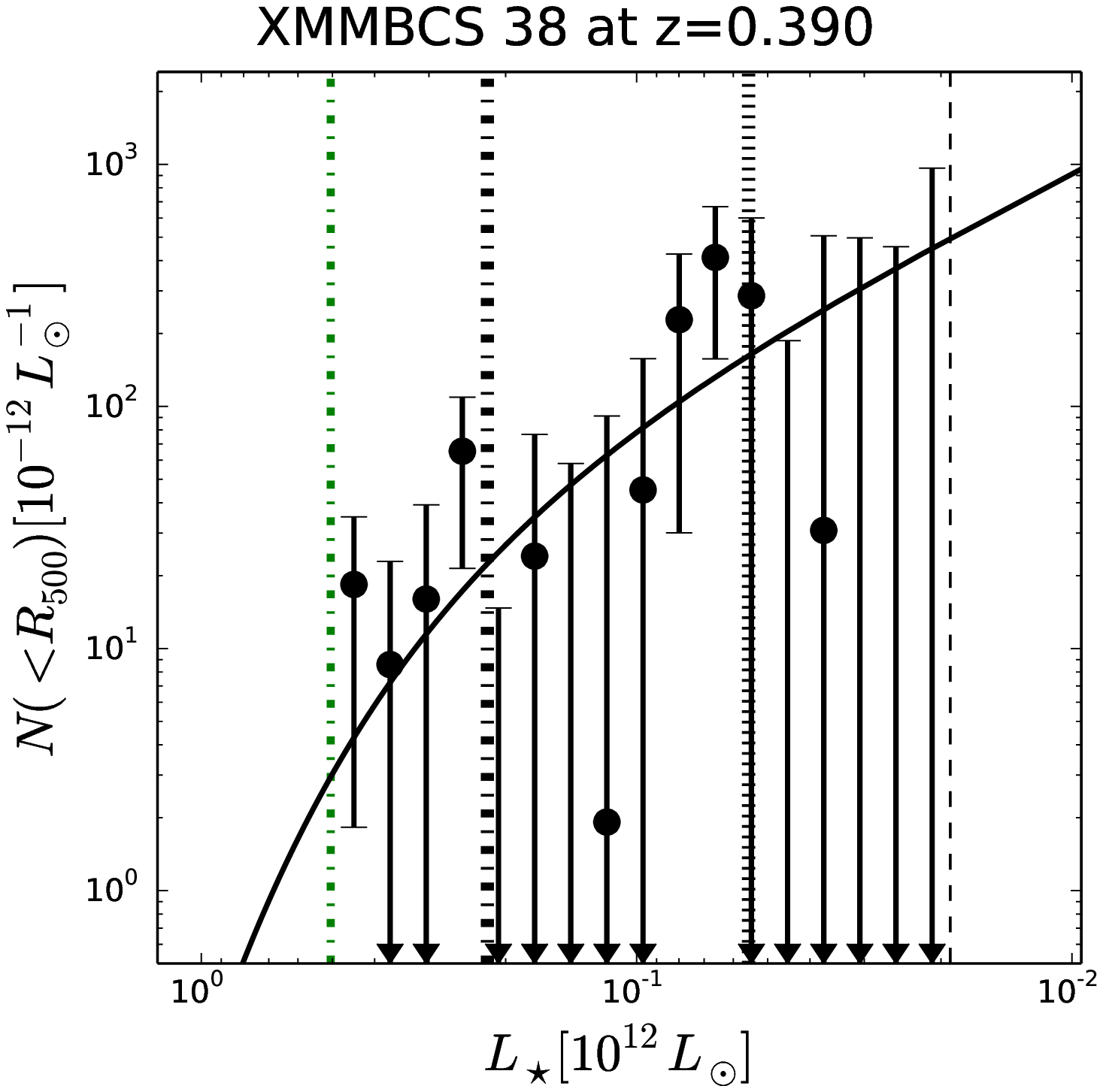} \end{subfigure}    
    \begin{subfigure}[b]{0.20\textwidth} \includegraphics[width=\textwidth]{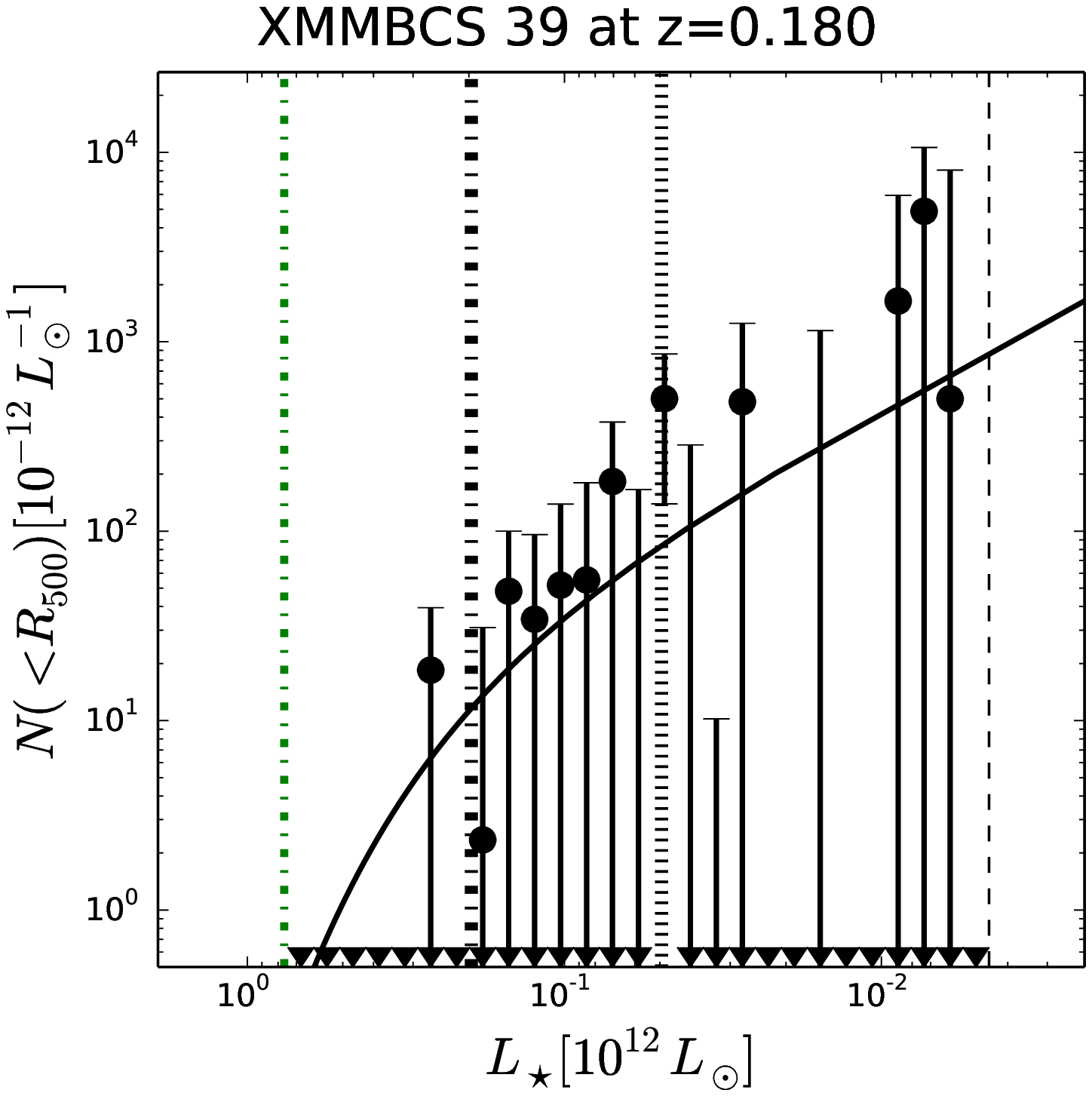} \end{subfigure}    

    \begin{subfigure}[b]{0.20\textwidth} \includegraphics[width=\textwidth]{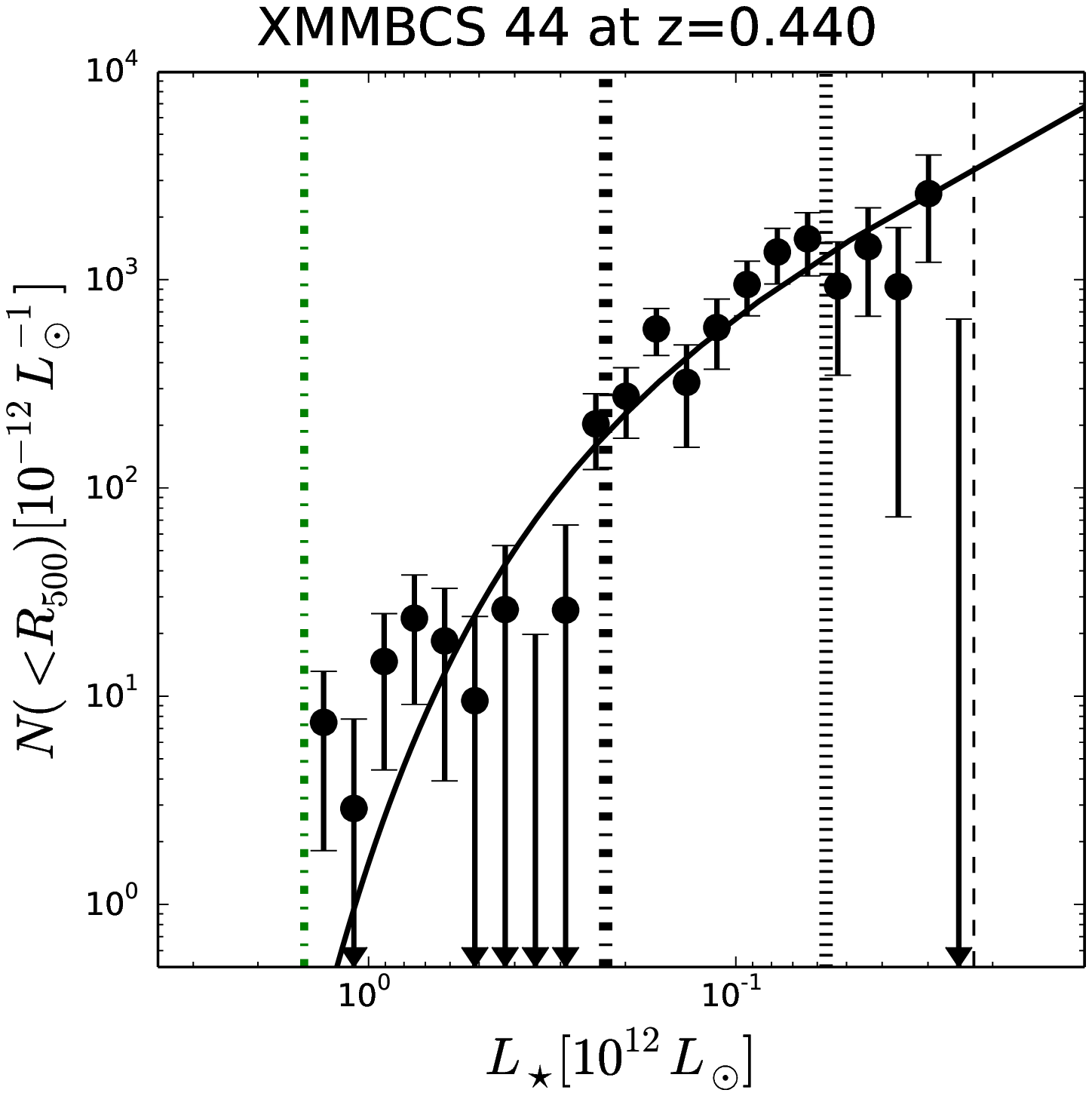} \end{subfigure}      
    \begin{subfigure}[b]{0.20\textwidth} \includegraphics[width=\textwidth]{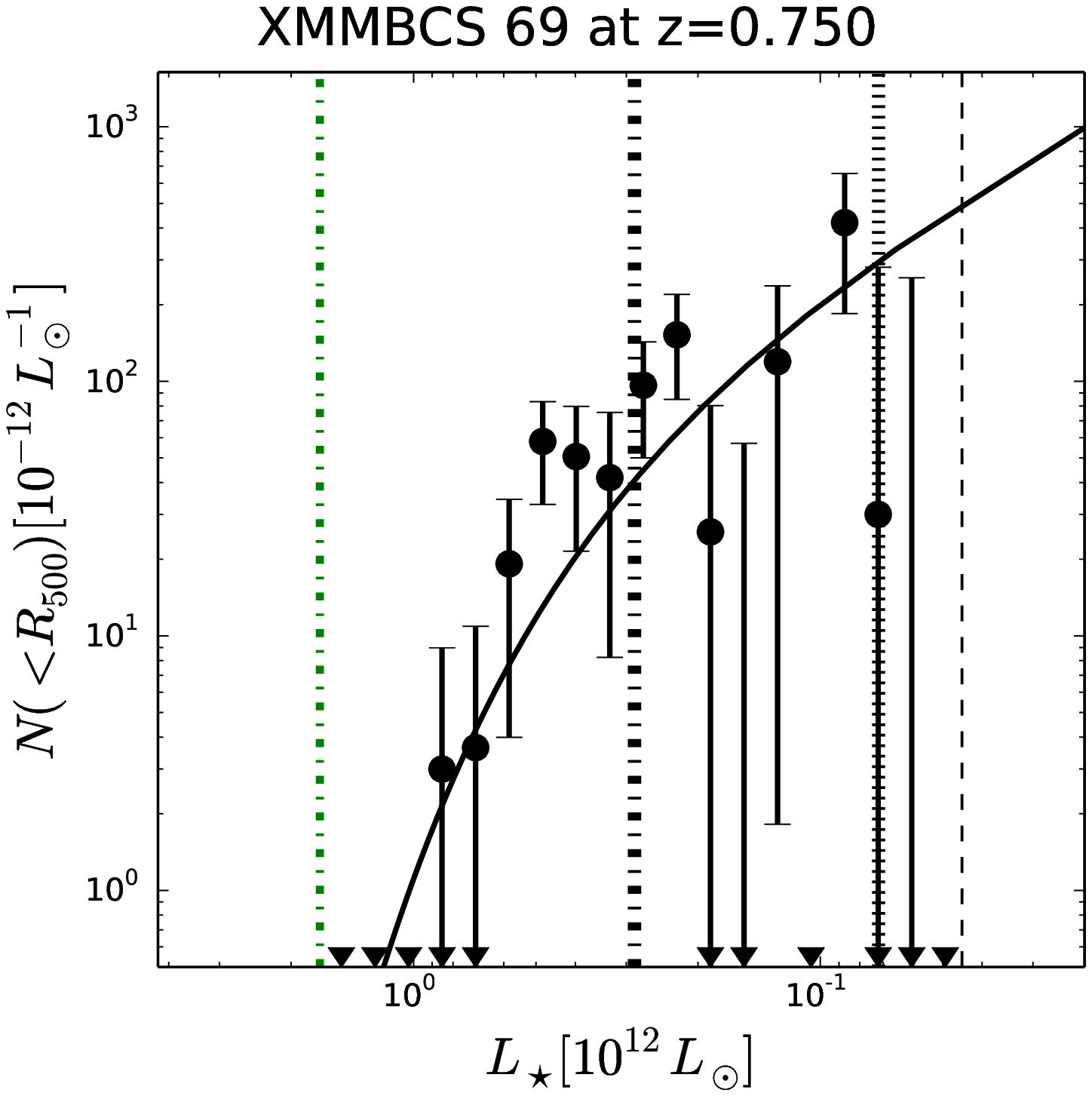} \end{subfigure}    
    \begin{subfigure}[b]{0.20\textwidth} \includegraphics[width=\textwidth]{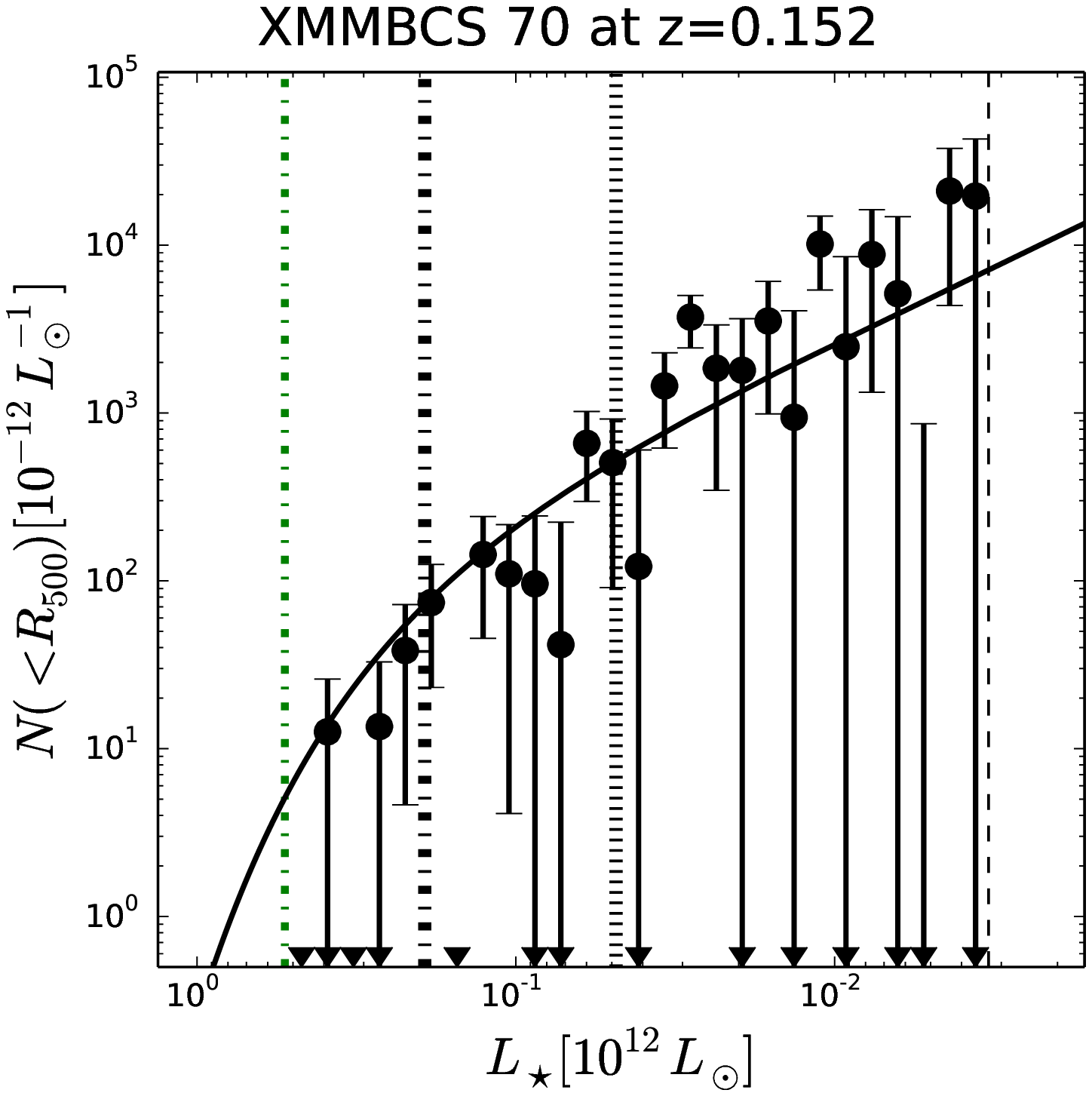} \end{subfigure}    
    \begin{subfigure}[b]{0.20\textwidth} \includegraphics[width=\textwidth]{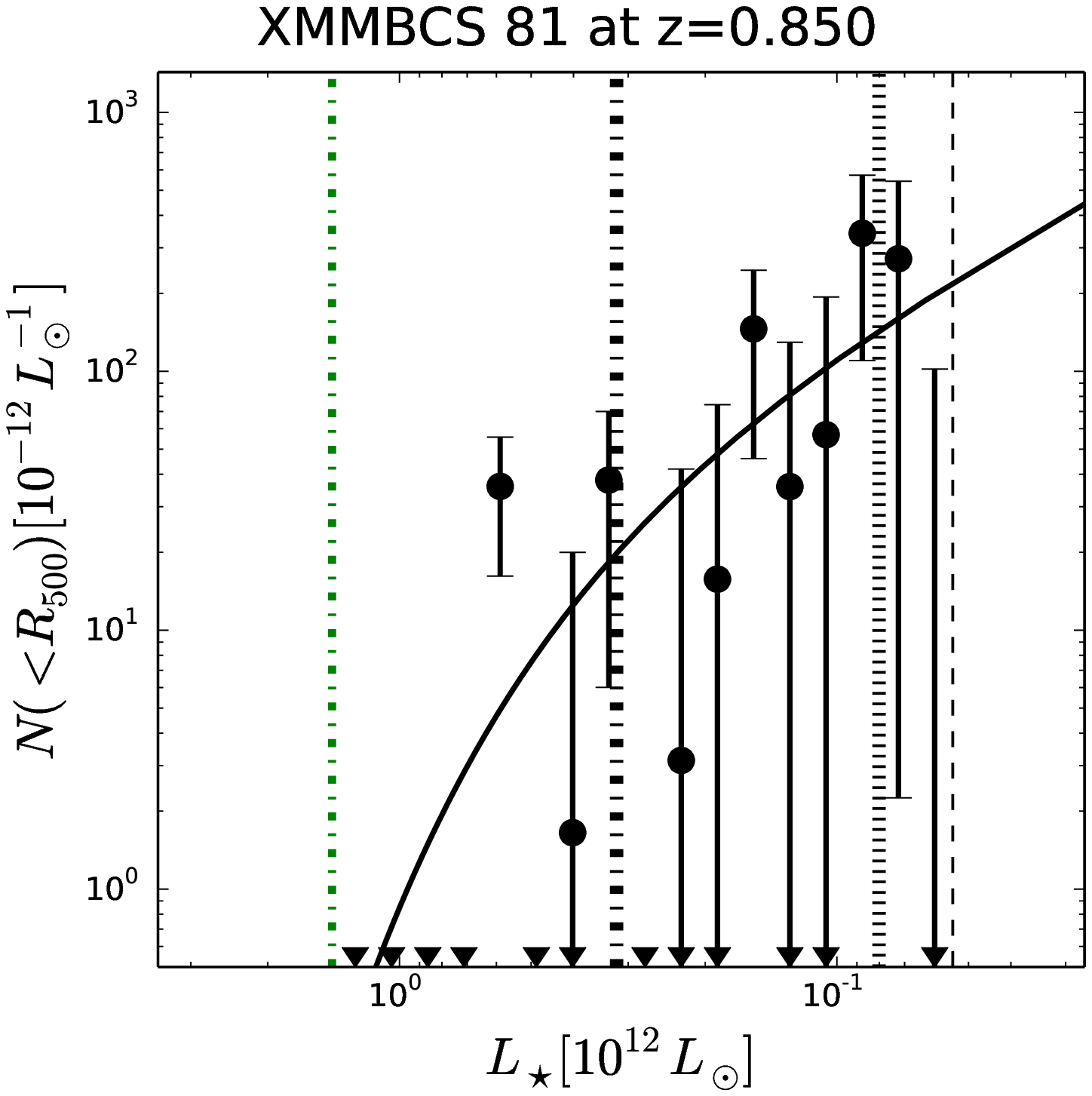} \end{subfigure}      
  
    \begin{subfigure}[b]{0.20\textwidth} \includegraphics[width=\textwidth]{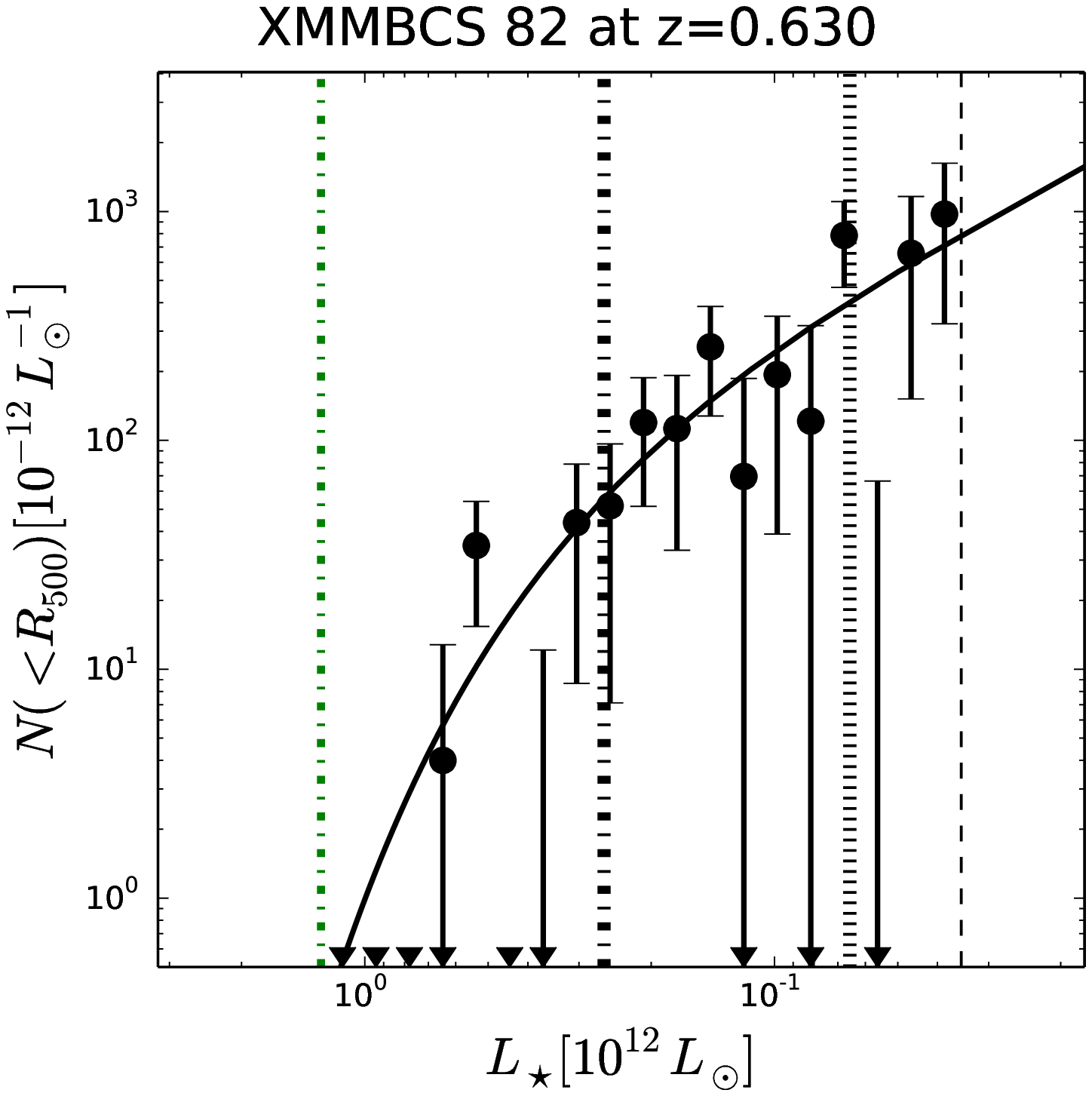} \end{subfigure}    
    \begin{subfigure}[b]{0.20\textwidth} \includegraphics[width=\textwidth]{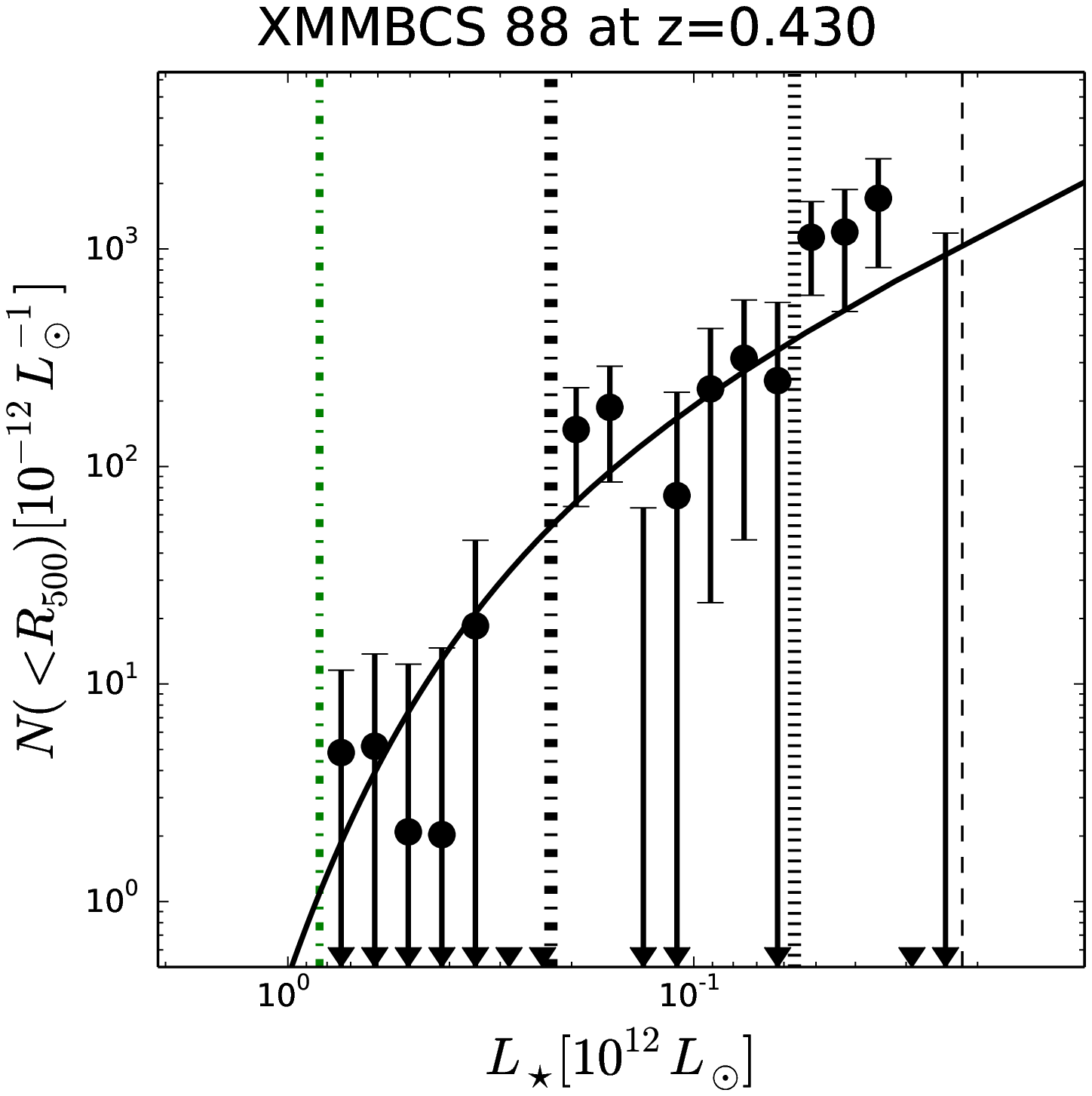} \end{subfigure}    
    \begin{subfigure}[b]{0.20\textwidth} \includegraphics[width=\textwidth]{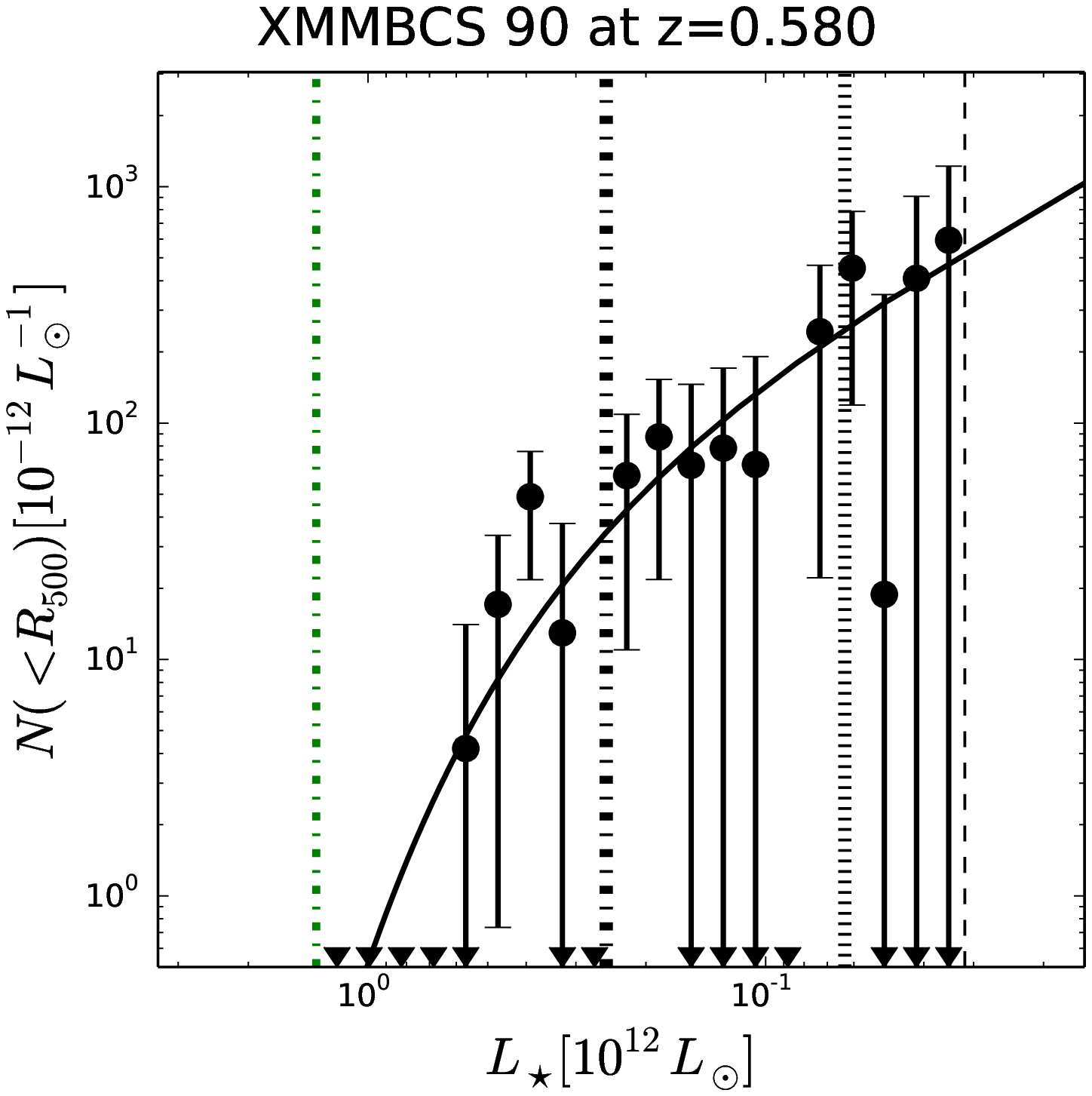} \end{subfigure}        
    \begin{subfigure}[b]{0.20\textwidth} \includegraphics[width=\textwidth]{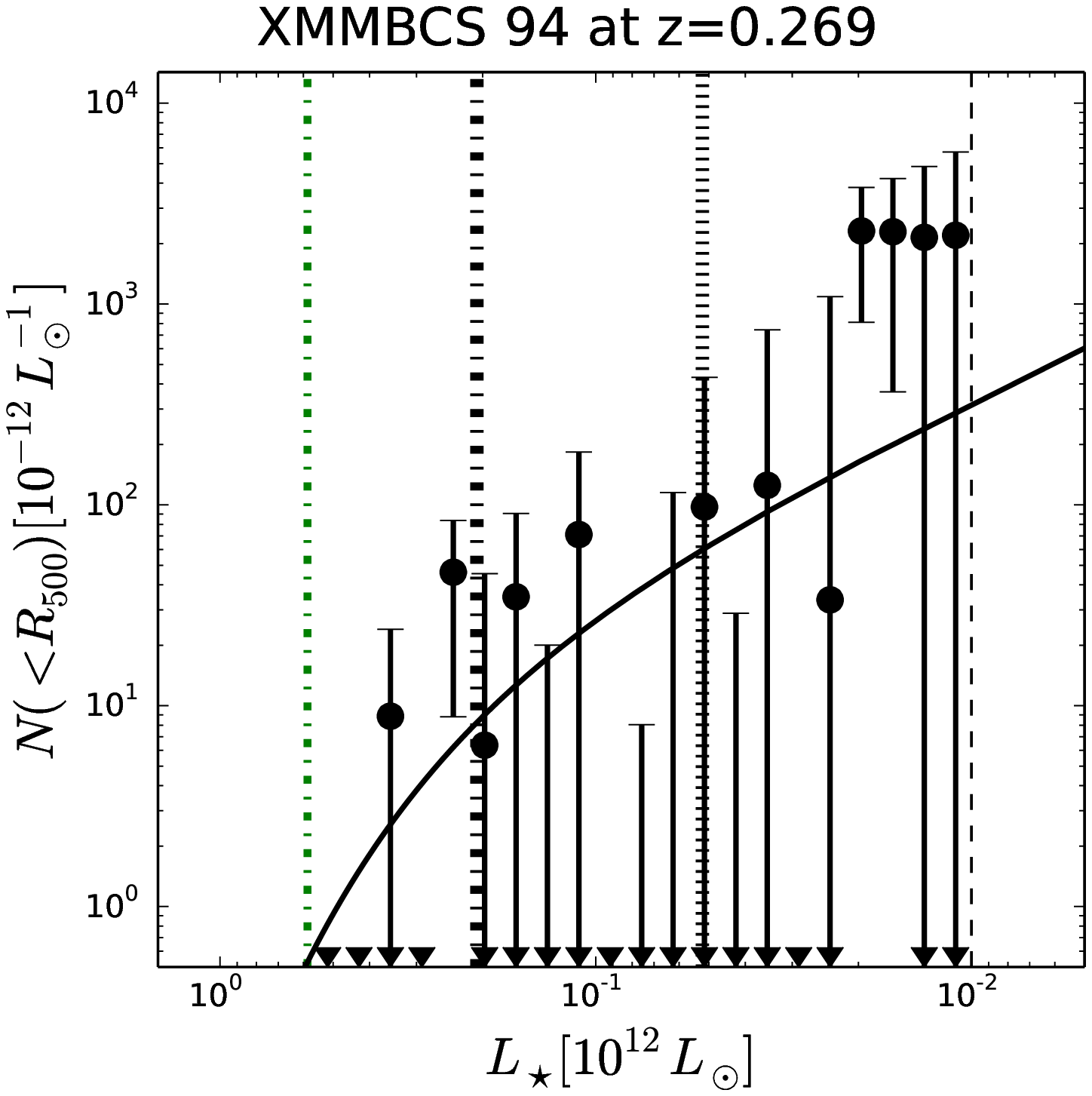} \end{subfigure}    

    \begin{subfigure}[b]{0.20\textwidth} \includegraphics[width=\textwidth]{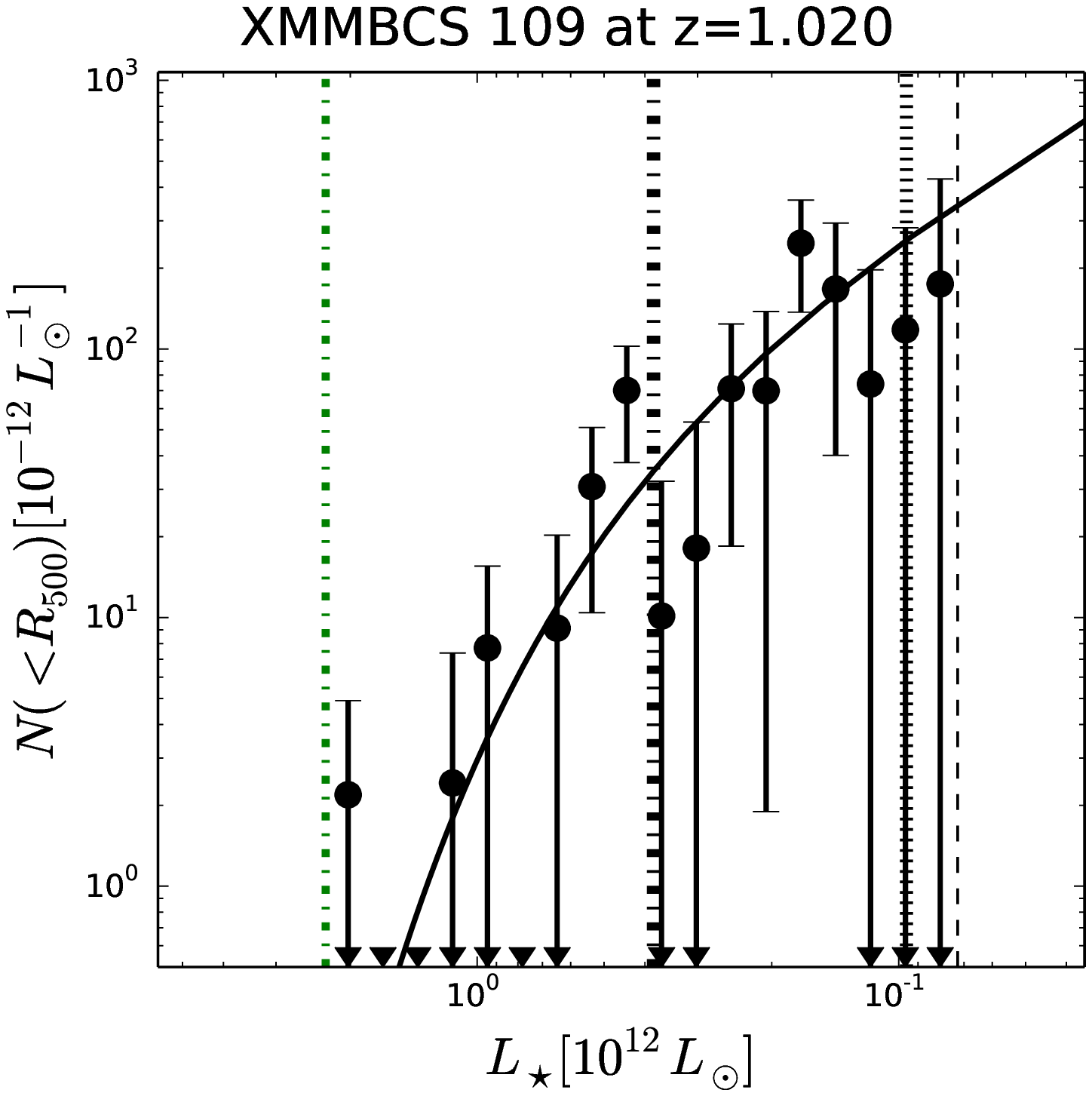} \end{subfigure}    
    \begin{subfigure}[b]{0.20\textwidth} \includegraphics[width=\textwidth]{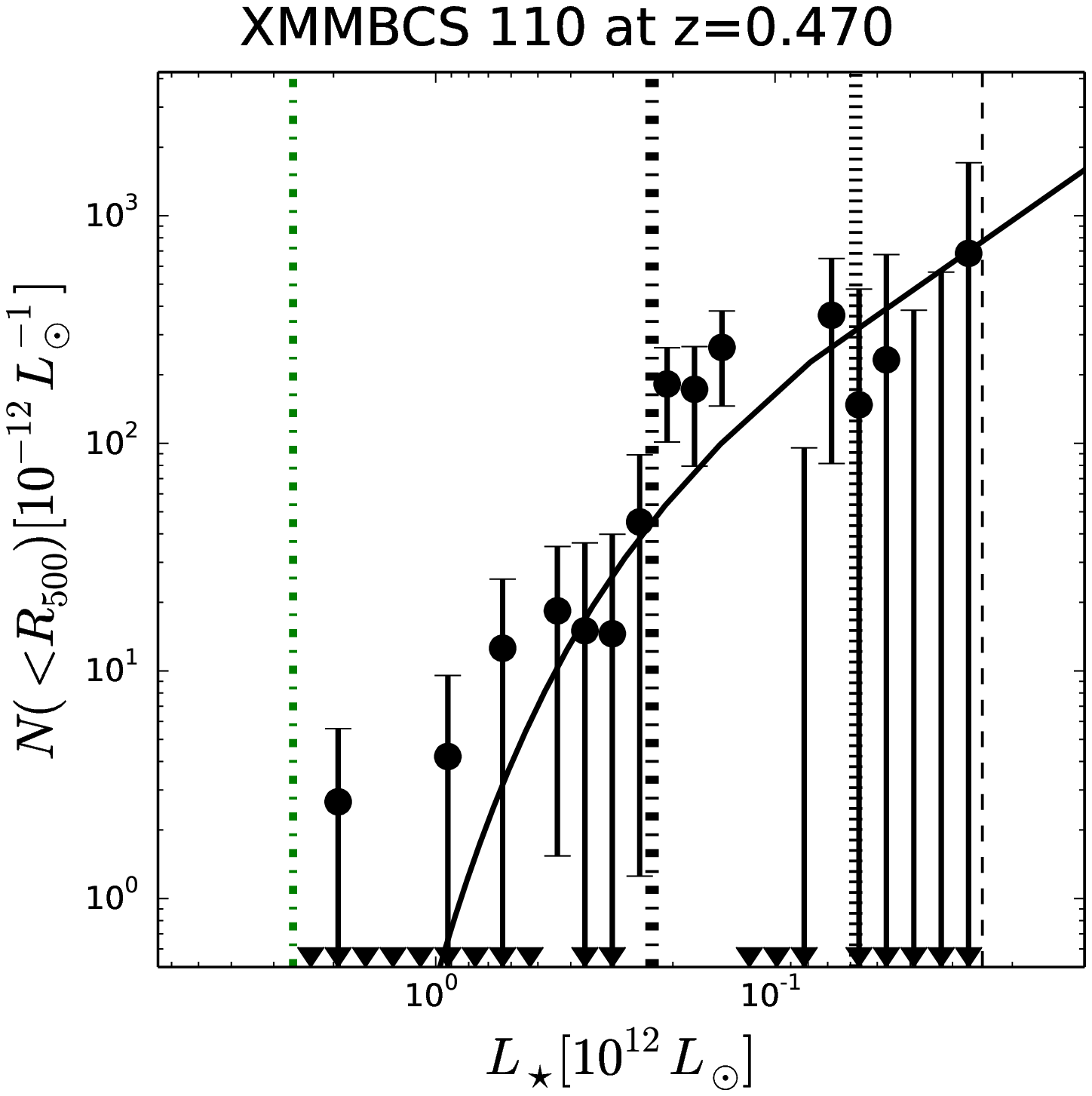} \end{subfigure}      
    \begin{subfigure}[b]{0.20\textwidth} \includegraphics[width=\textwidth]{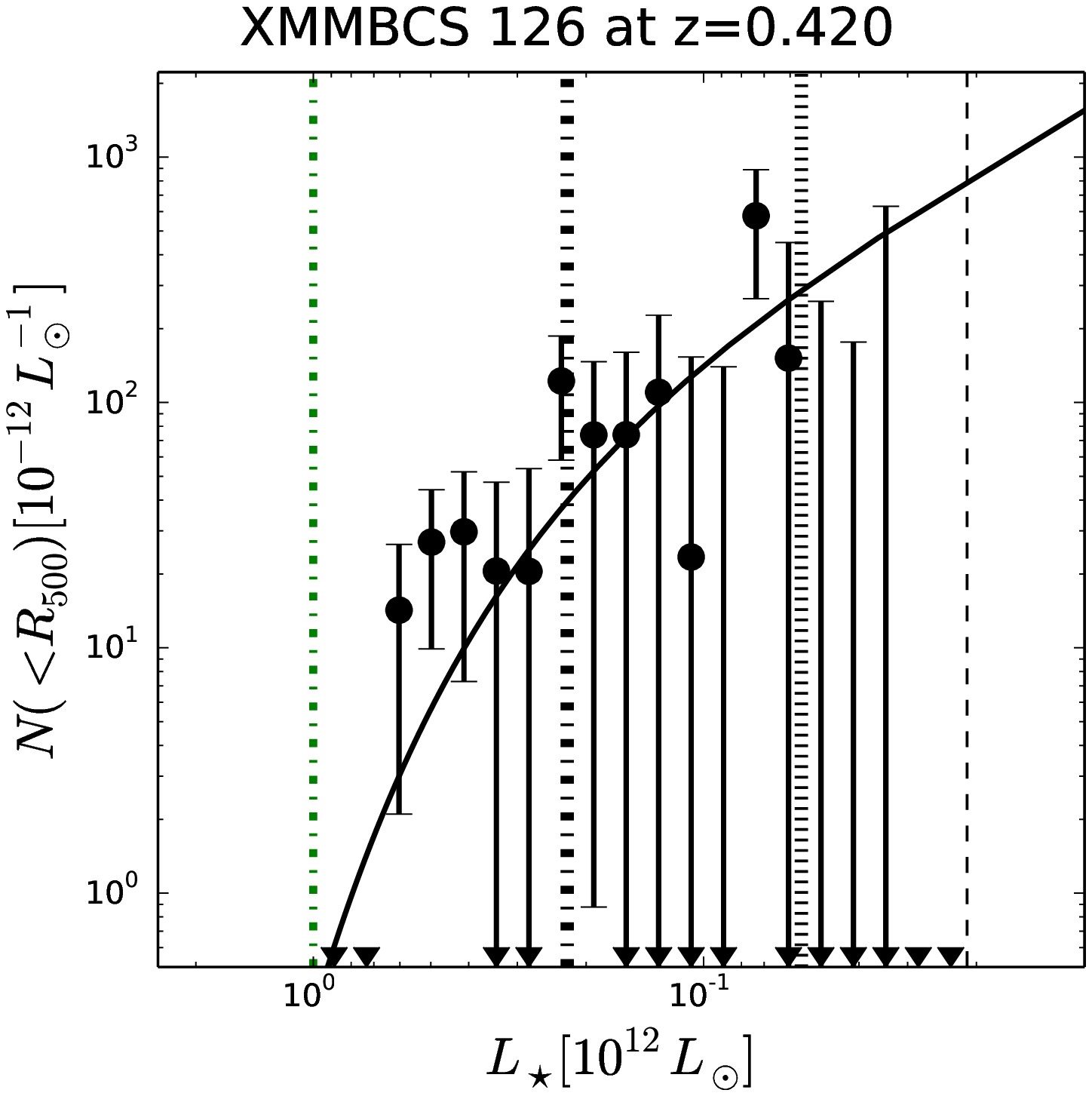} \end{subfigure}    
    \begin{subfigure}[b]{0.20\textwidth} \includegraphics[width=\textwidth]{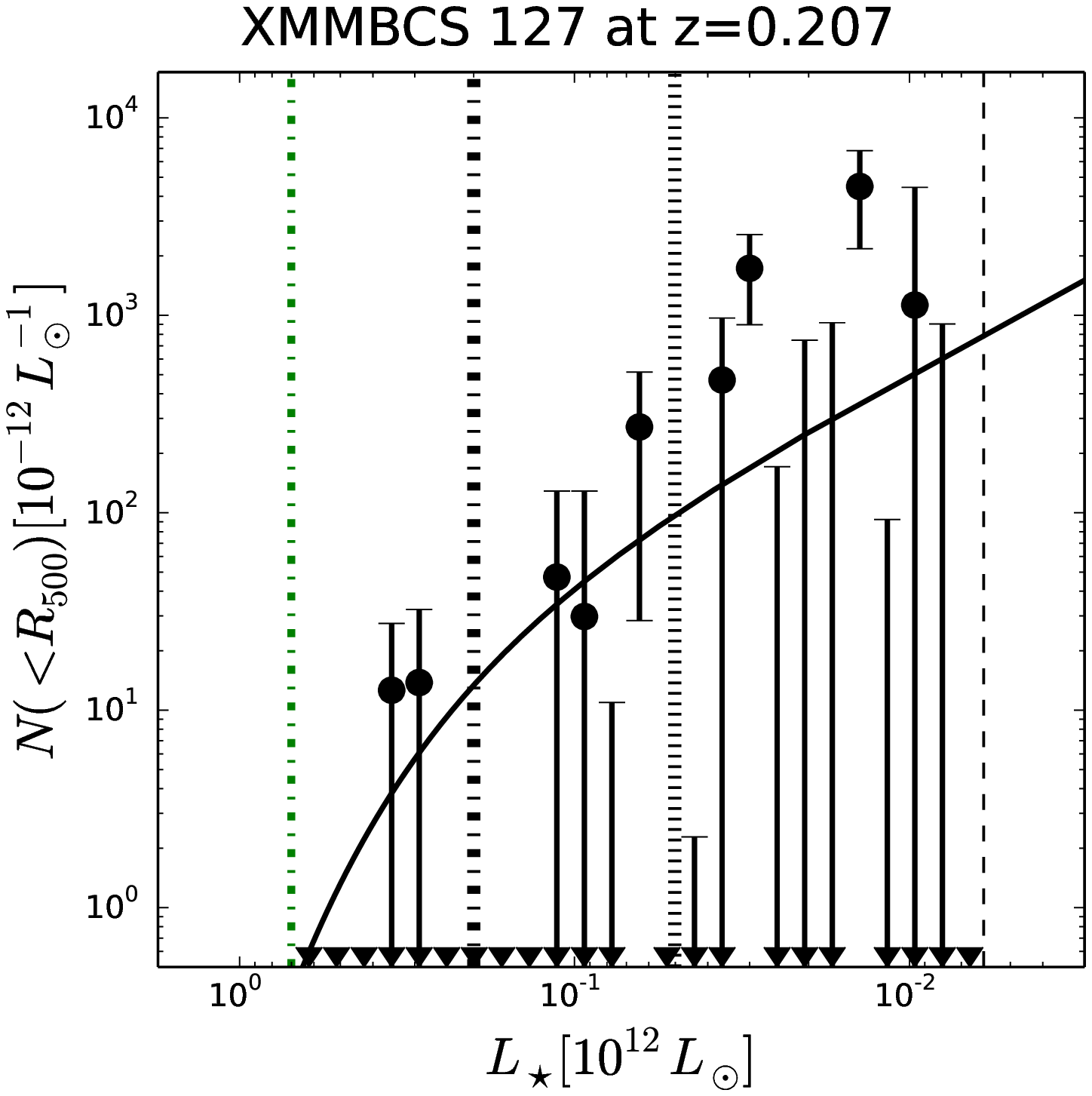} \end{subfigure}    

    \begin{subfigure}[b]{0.20\textwidth} \includegraphics[width=\textwidth]{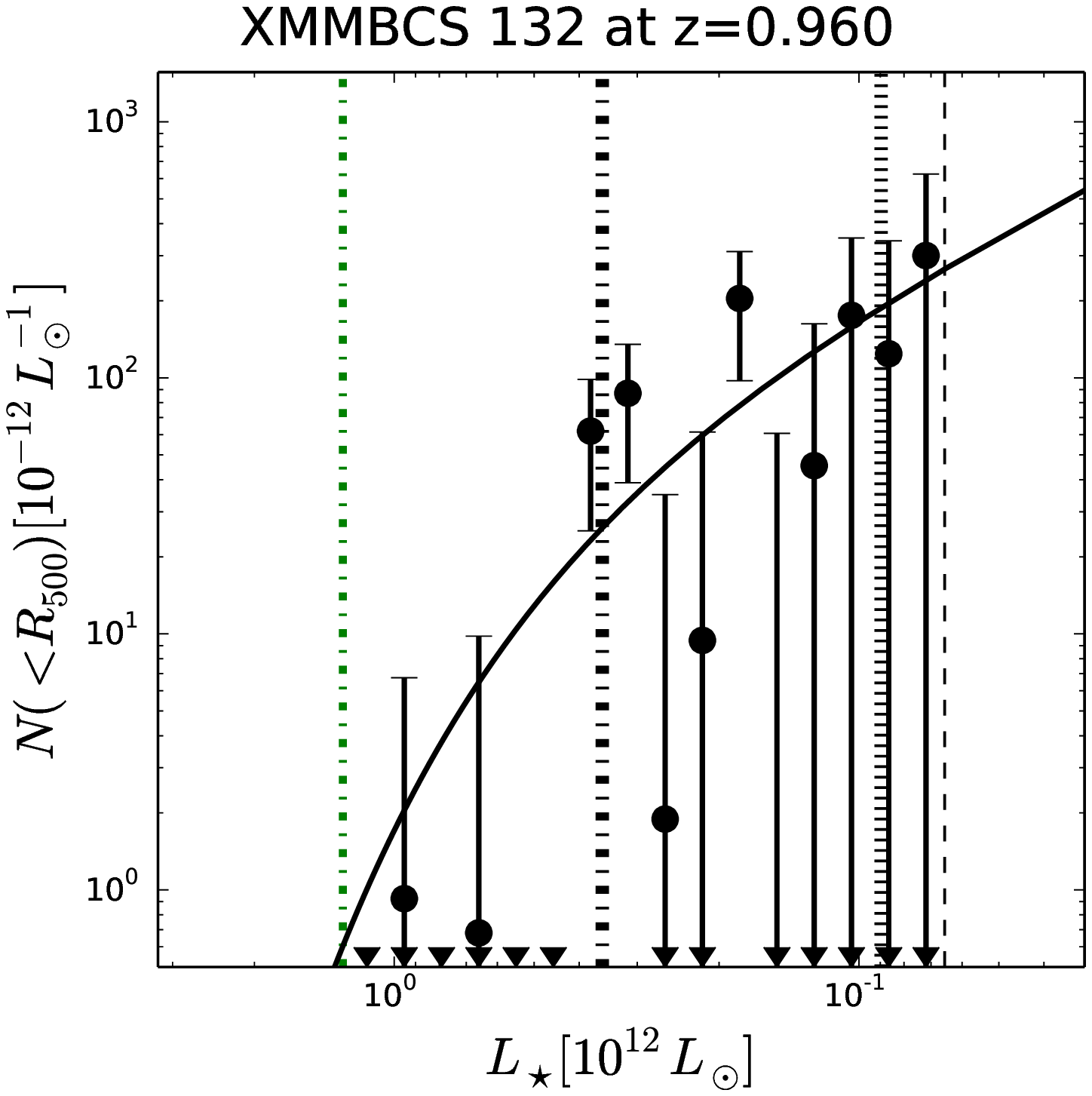} \end{subfigure}  
    \begin{subfigure}[b]{0.20\textwidth} \includegraphics[width=\textwidth]{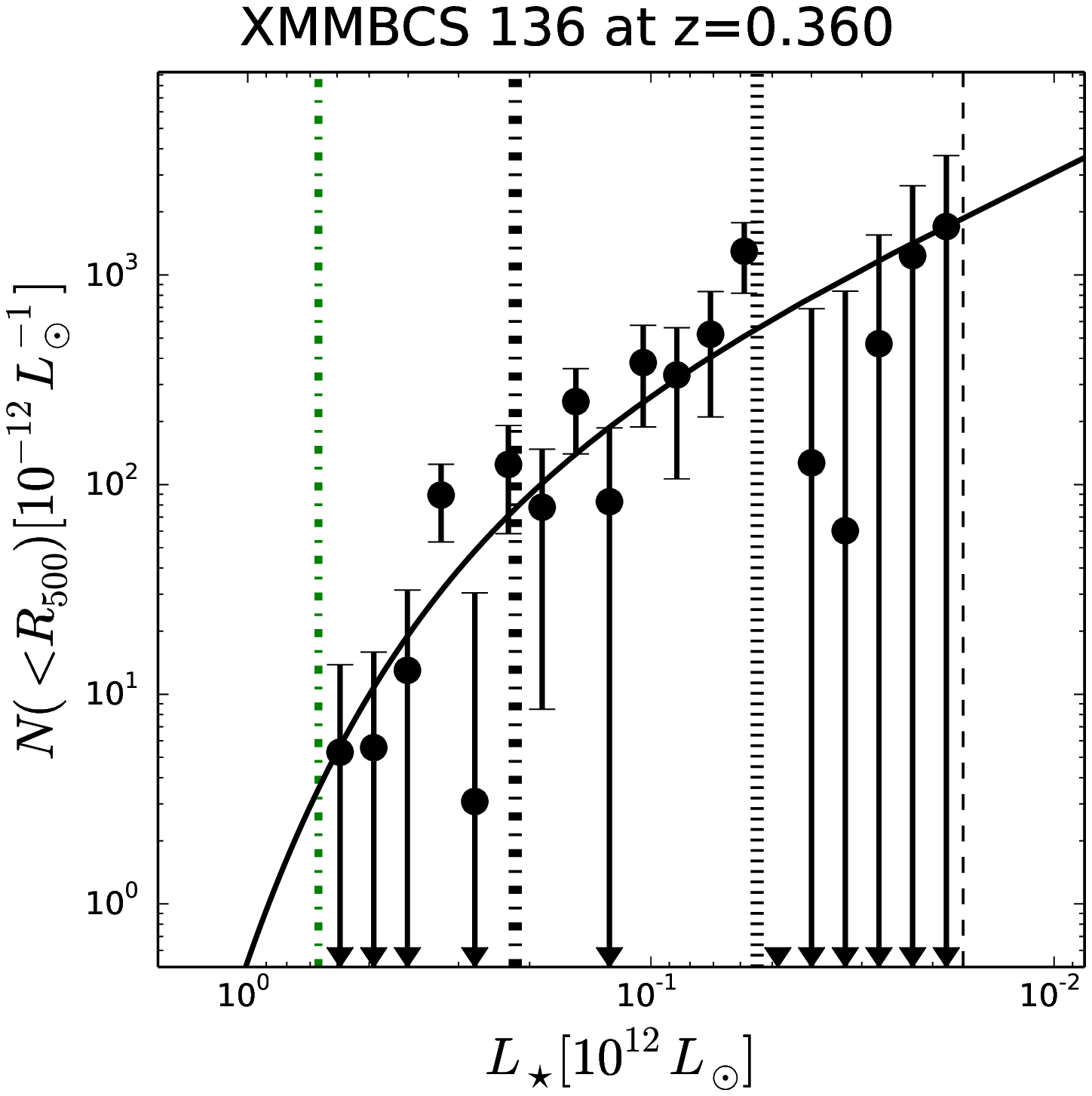} \end{subfigure}    
    \begin{subfigure}[b]{0.20\textwidth} \includegraphics[width=\textwidth]{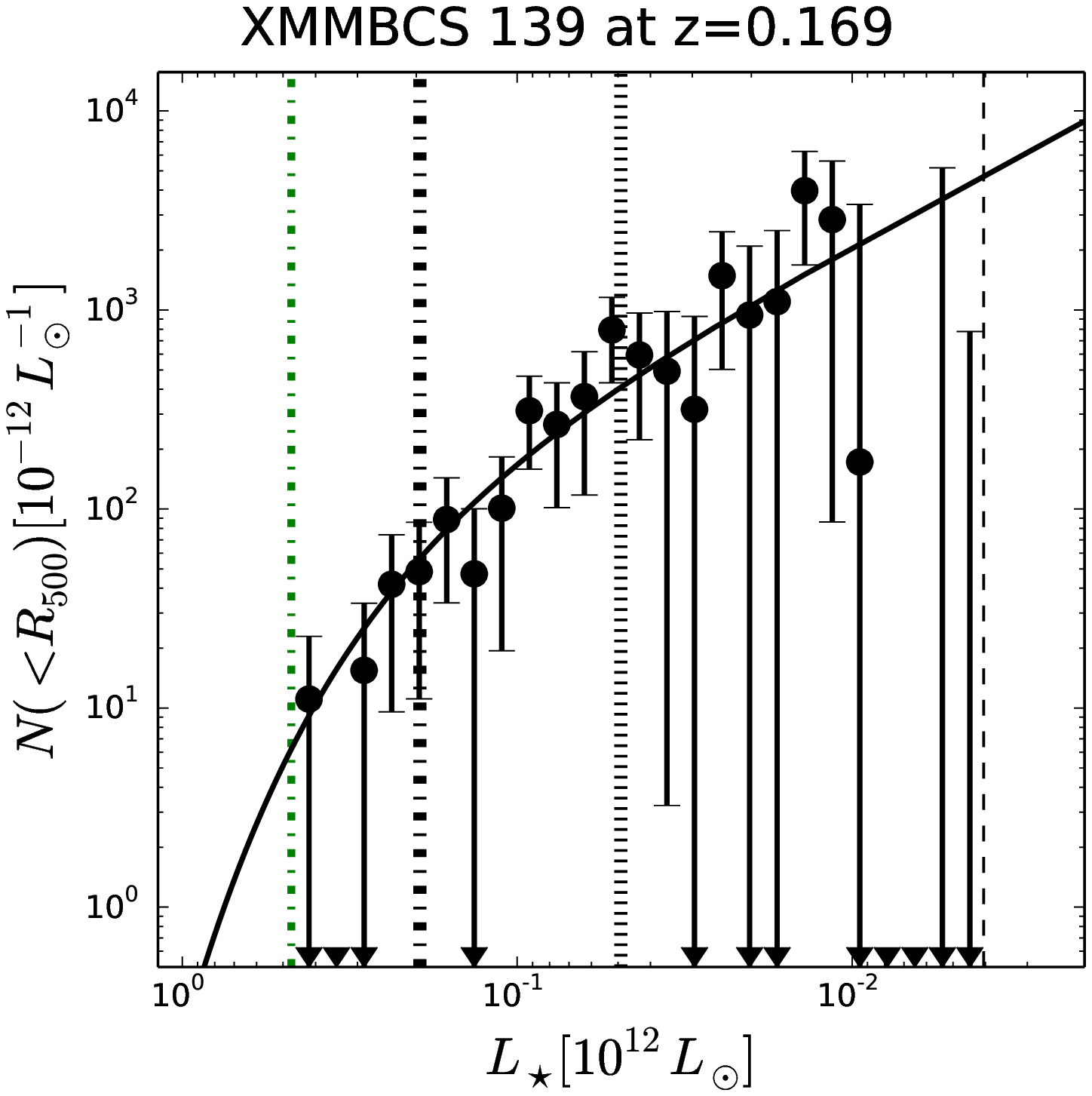} \end{subfigure}    
    \begin{subfigure}[b]{0.20\textwidth} \includegraphics[width=\textwidth]{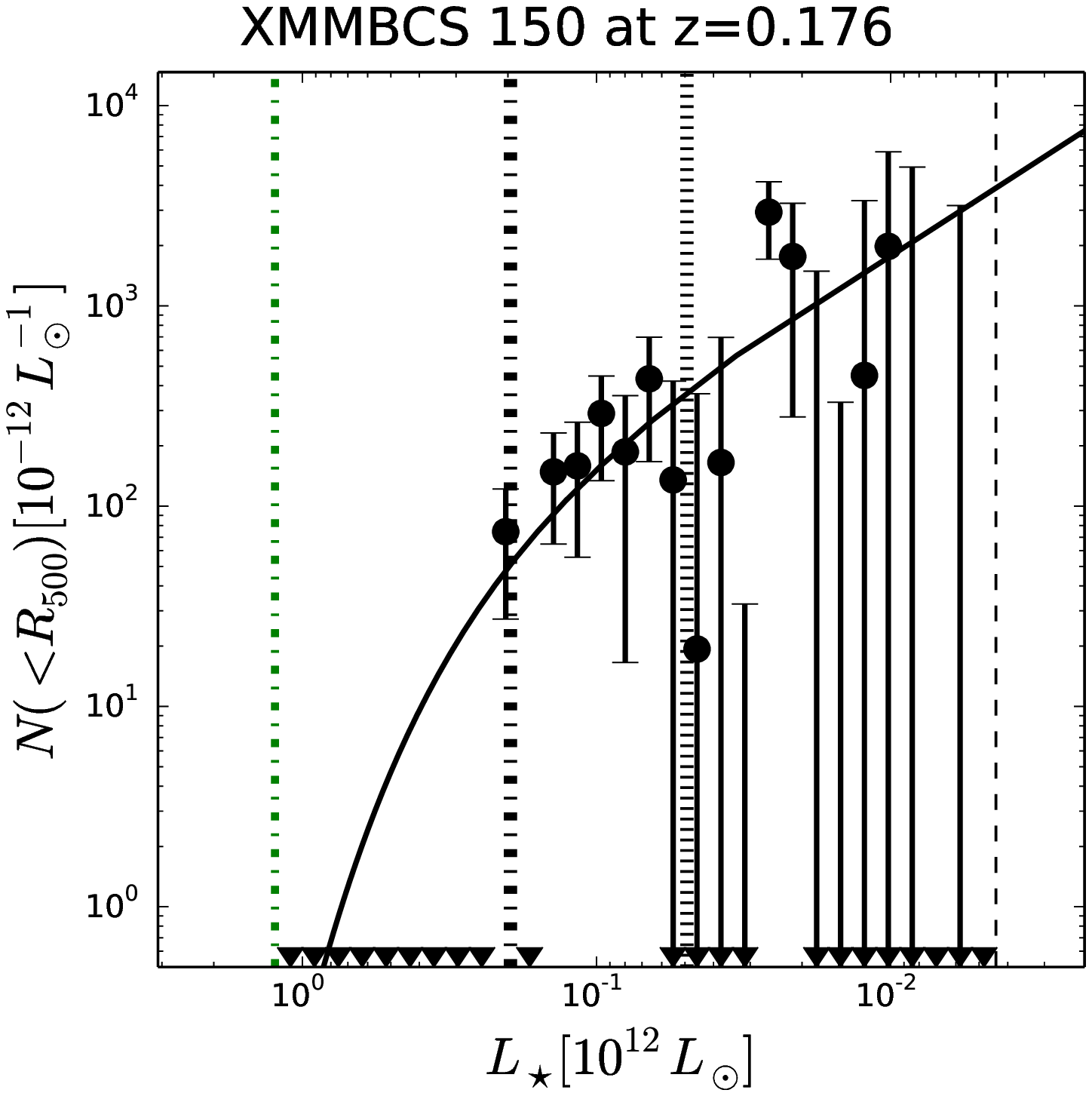} \end{subfigure}  
    \caption{
    The observed LF (points with error bars) and the best-fit model (solid line) for the \XMMBCS\ clusters.
    The unique ID and cluster redshift are listed in the title of each plot.
    The LFs are presented in units of solar luminosity in the rest-frame at the cluster redshift.
    The green dot-dashed line shows the luminosity of the BCG. 
    The black dot-dashed (dotted) line indicates the luminosity corresponding to the characteristic magnitude \mstar\ ($\mstar+1.5$) predicted by our CSP model, while the black dashed line is the luminosity corresponding to the $50\percent$ completeness limit in the \SSDF\ survey.
    }
    \label{fig:lf_cm_1}    
\end{figure*}
\begin{figure*}
\centering
    \begin{subfigure}[b]{0.20\textwidth} \includegraphics[width=\textwidth]{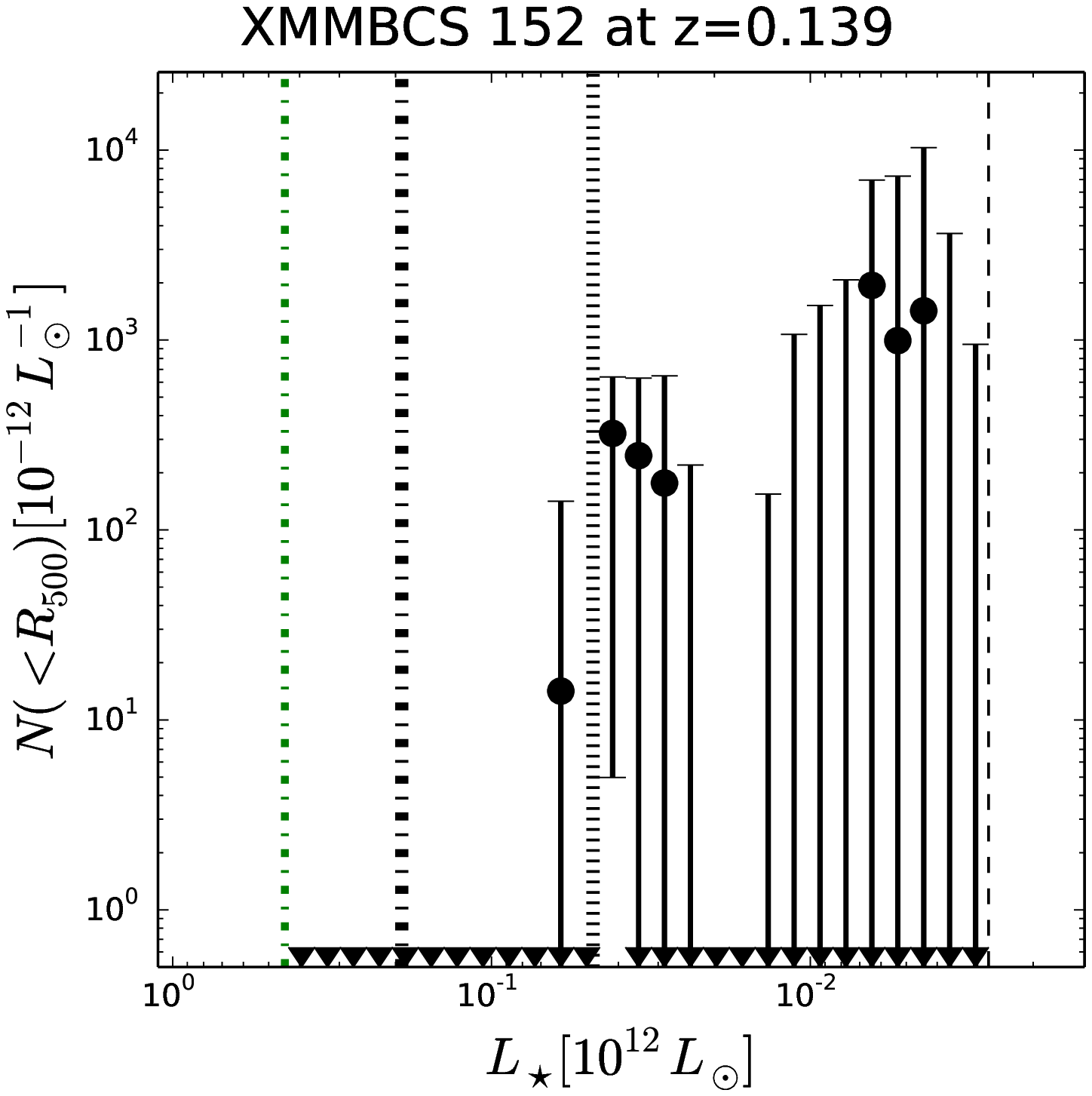} \end{subfigure}    
    \begin{subfigure}[b]{0.20\textwidth} \includegraphics[width=\textwidth]{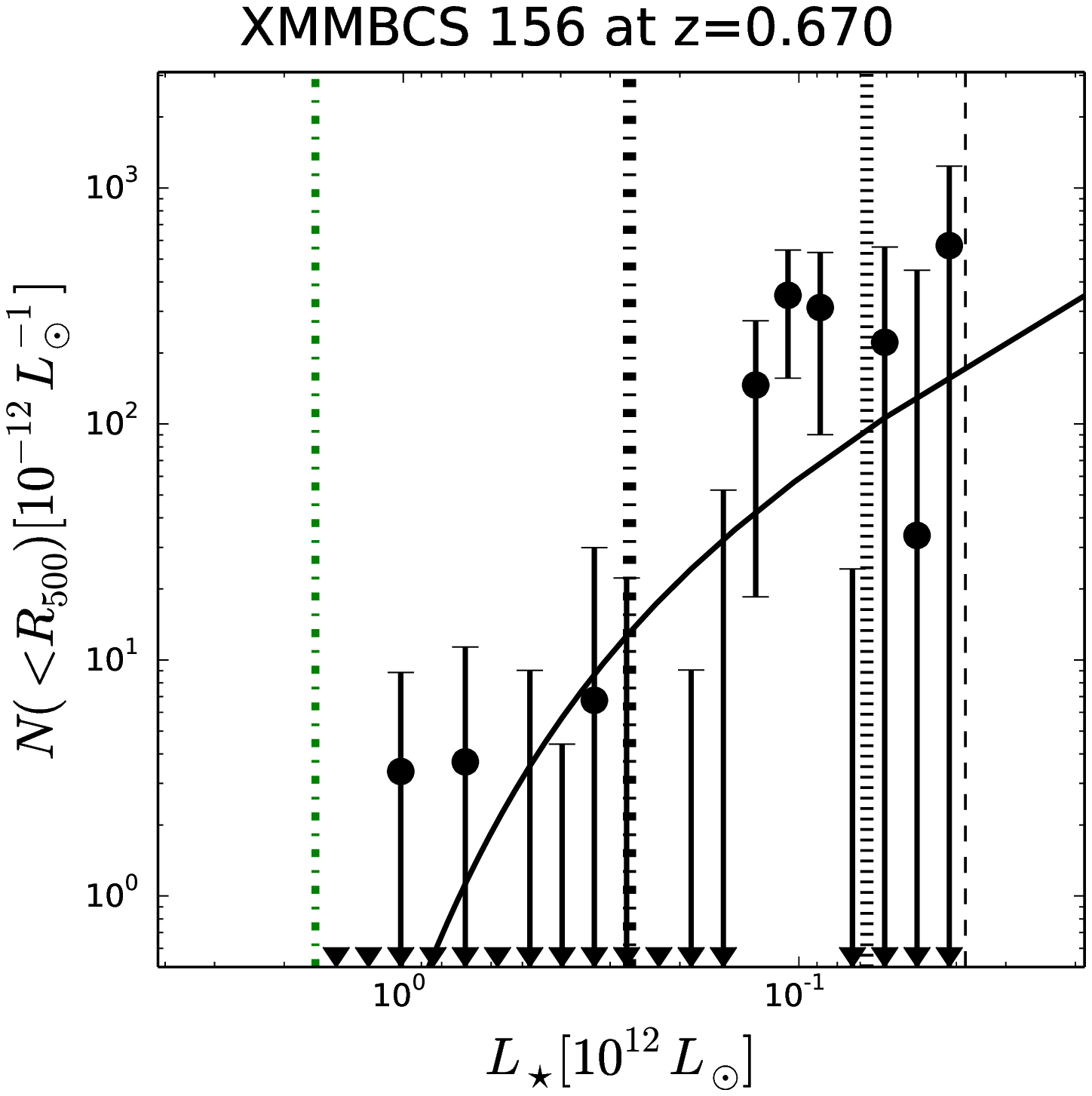} \end{subfigure}    
    \begin{subfigure}[b]{0.20\textwidth} \includegraphics[width=\textwidth]{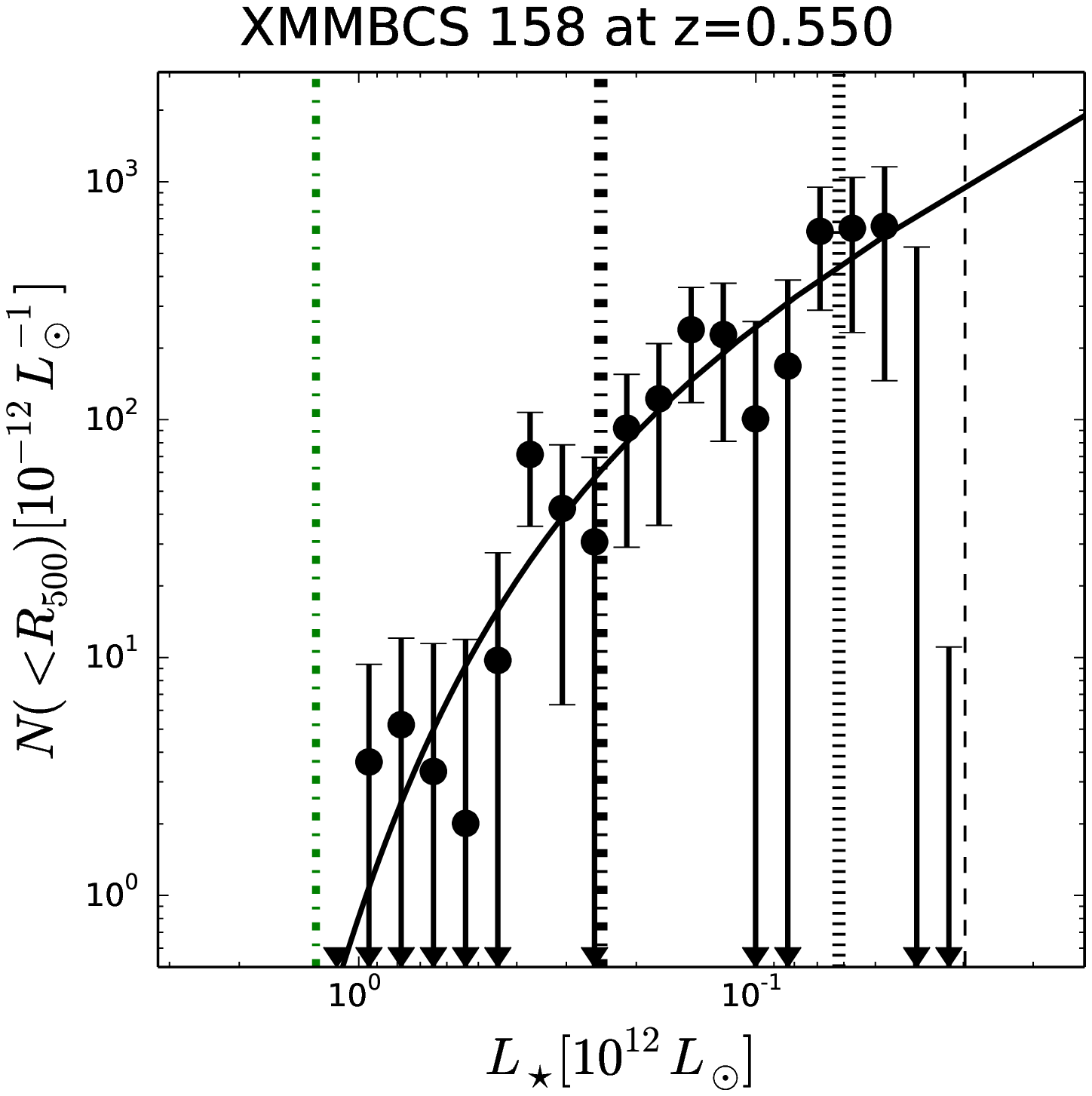} \end{subfigure}      
    \begin{subfigure}[b]{0.20\textwidth} \includegraphics[width=\textwidth]{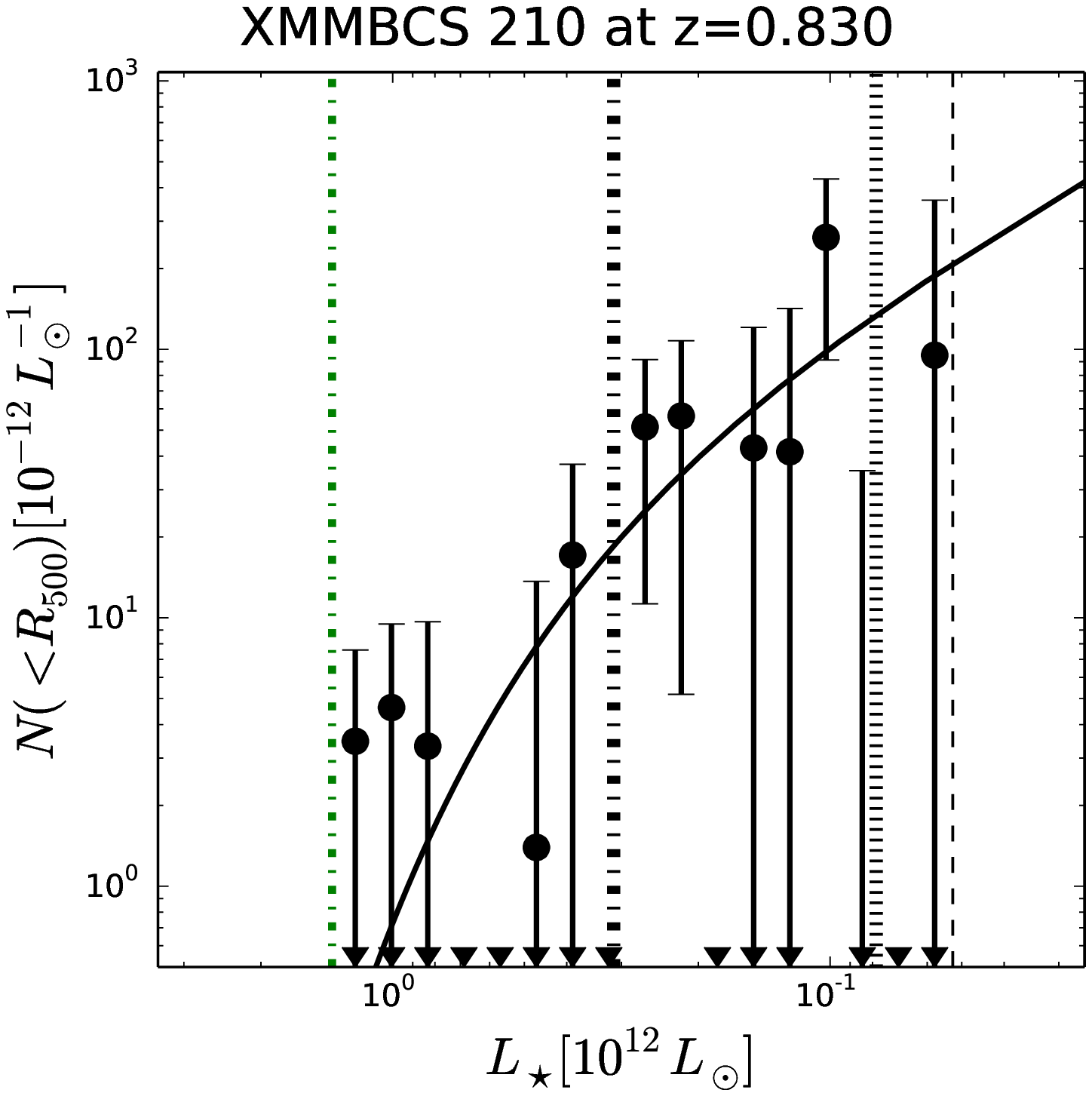} \end{subfigure}    

    \begin{subfigure}[b]{0.20\textwidth} \includegraphics[width=\textwidth]{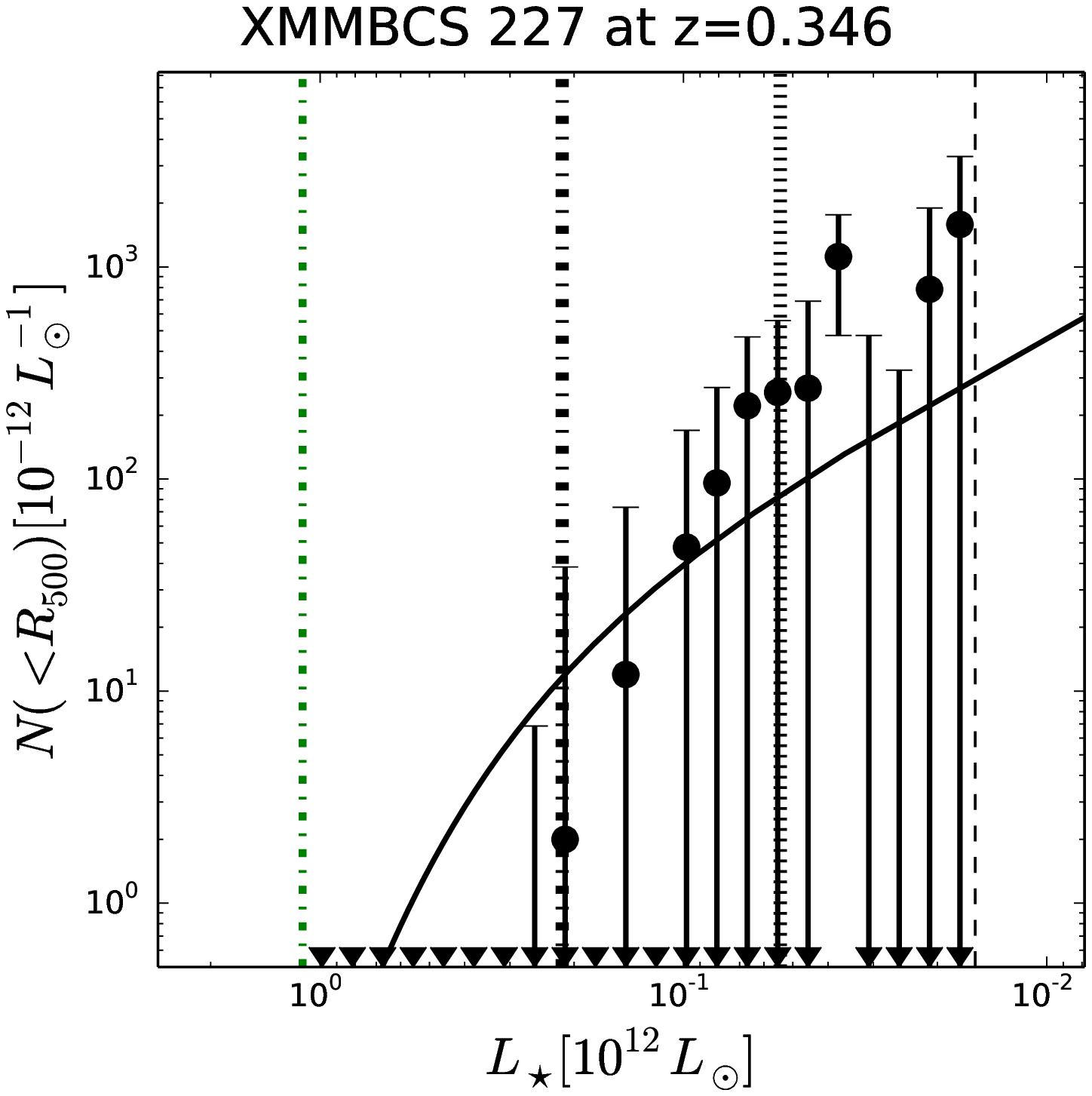} \end{subfigure}    
    \begin{subfigure}[b]{0.20\textwidth} \includegraphics[width=\textwidth]{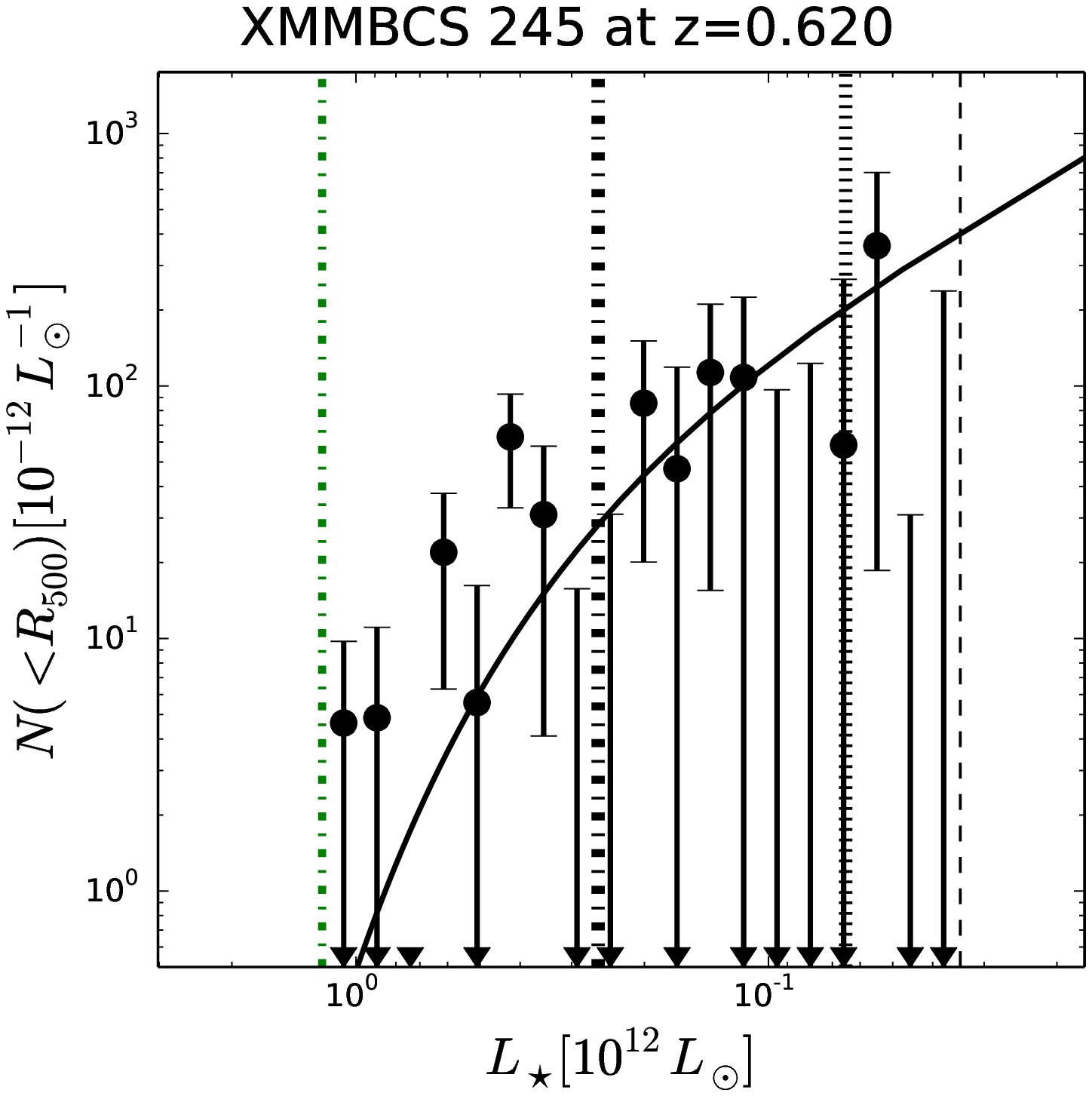} \end{subfigure}      
    \begin{subfigure}[b]{0.20\textwidth} \includegraphics[width=\textwidth]{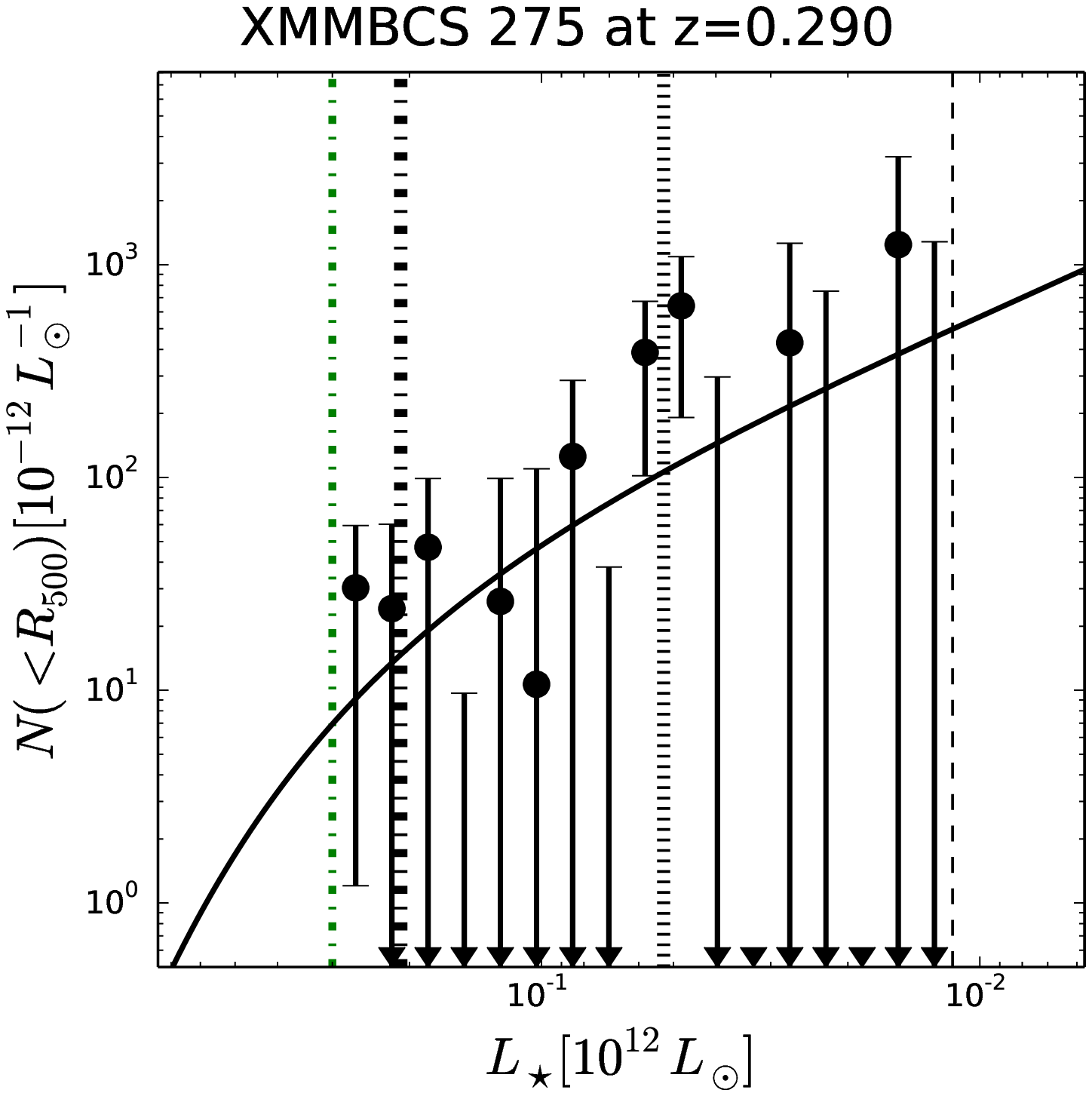} \end{subfigure}    
    \begin{subfigure}[b]{0.20\textwidth} \includegraphics[width=\textwidth]{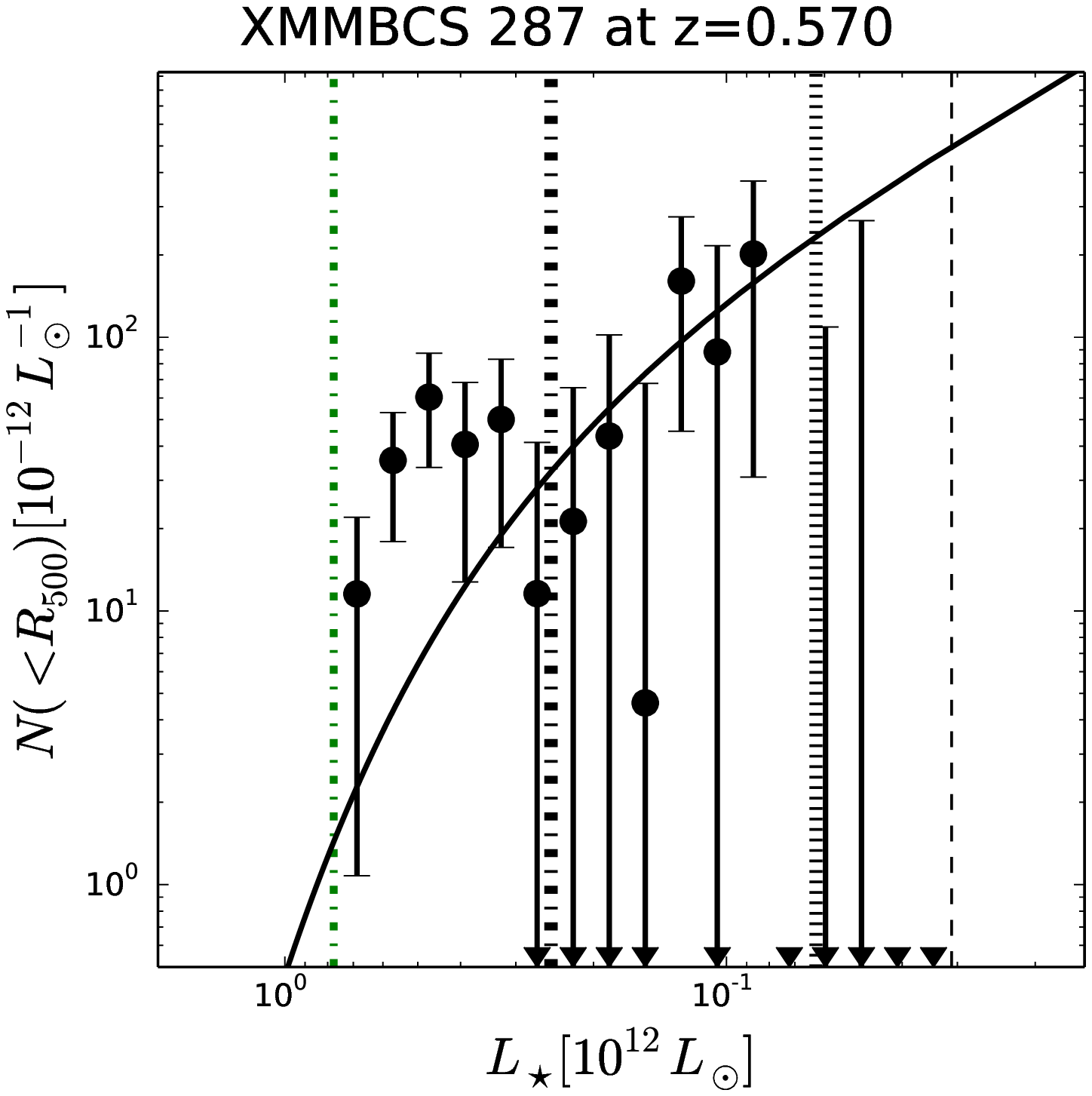} \end{subfigure}    
    
    \begin{subfigure}[b]{0.20\textwidth} \includegraphics[width=\textwidth]{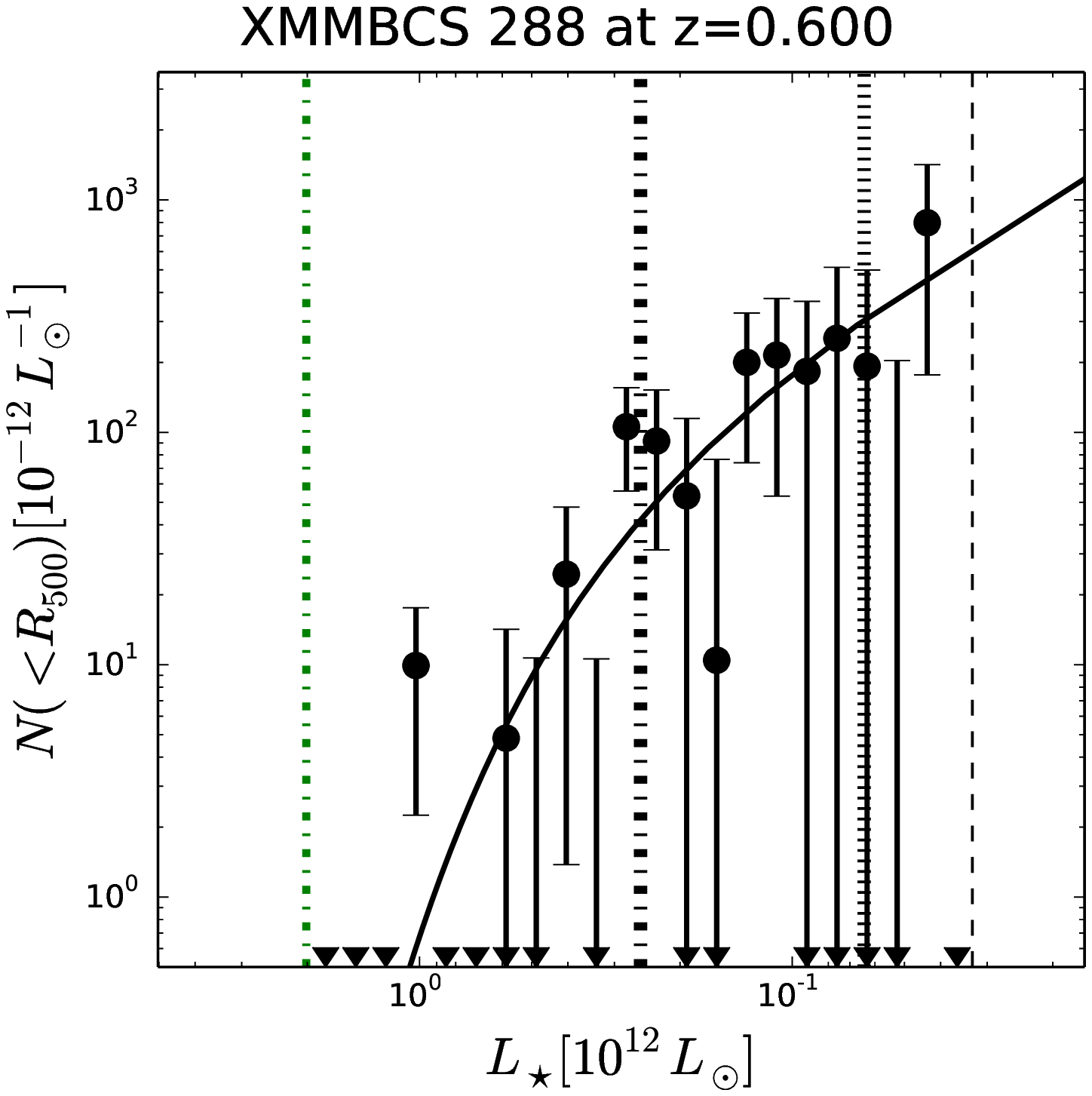} \end{subfigure}      
    \begin{subfigure}[b]{0.20\textwidth} \includegraphics[width=\textwidth]{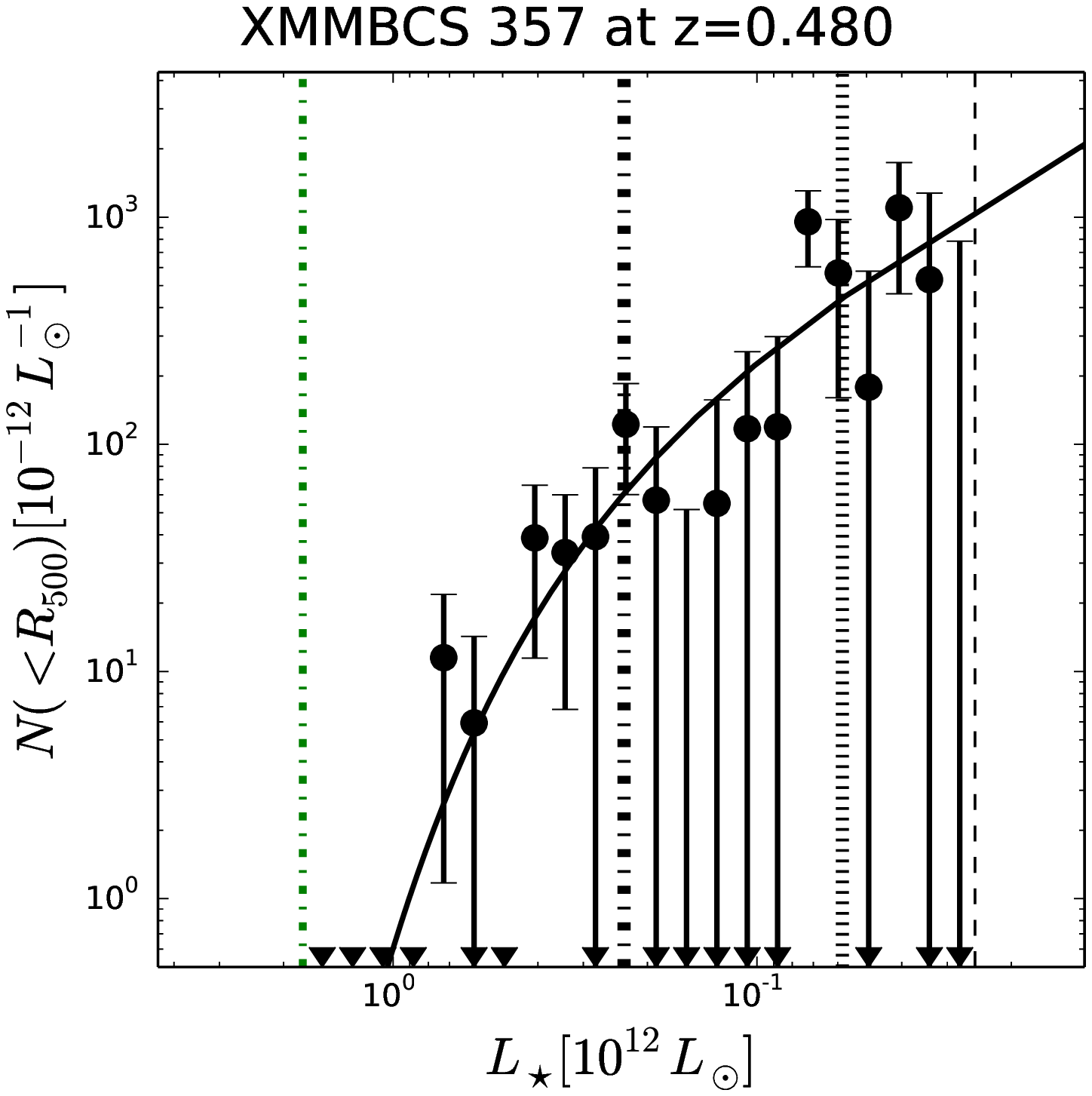} \end{subfigure}    
    \begin{subfigure}[b]{0.20\textwidth} \includegraphics[width=\textwidth]{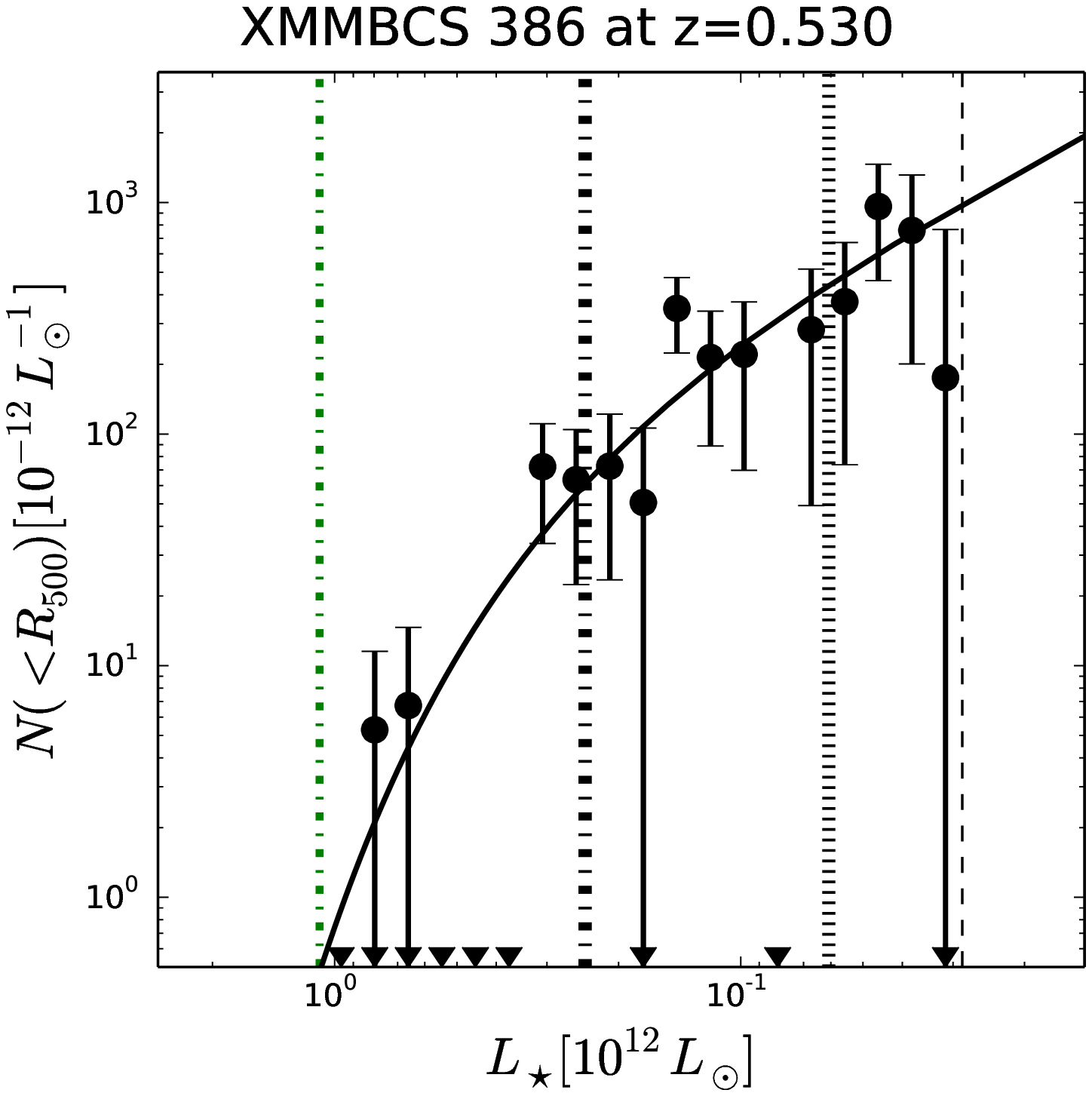} \end{subfigure}    
    \begin{subfigure}[b]{0.20\textwidth} \includegraphics[width=\textwidth]{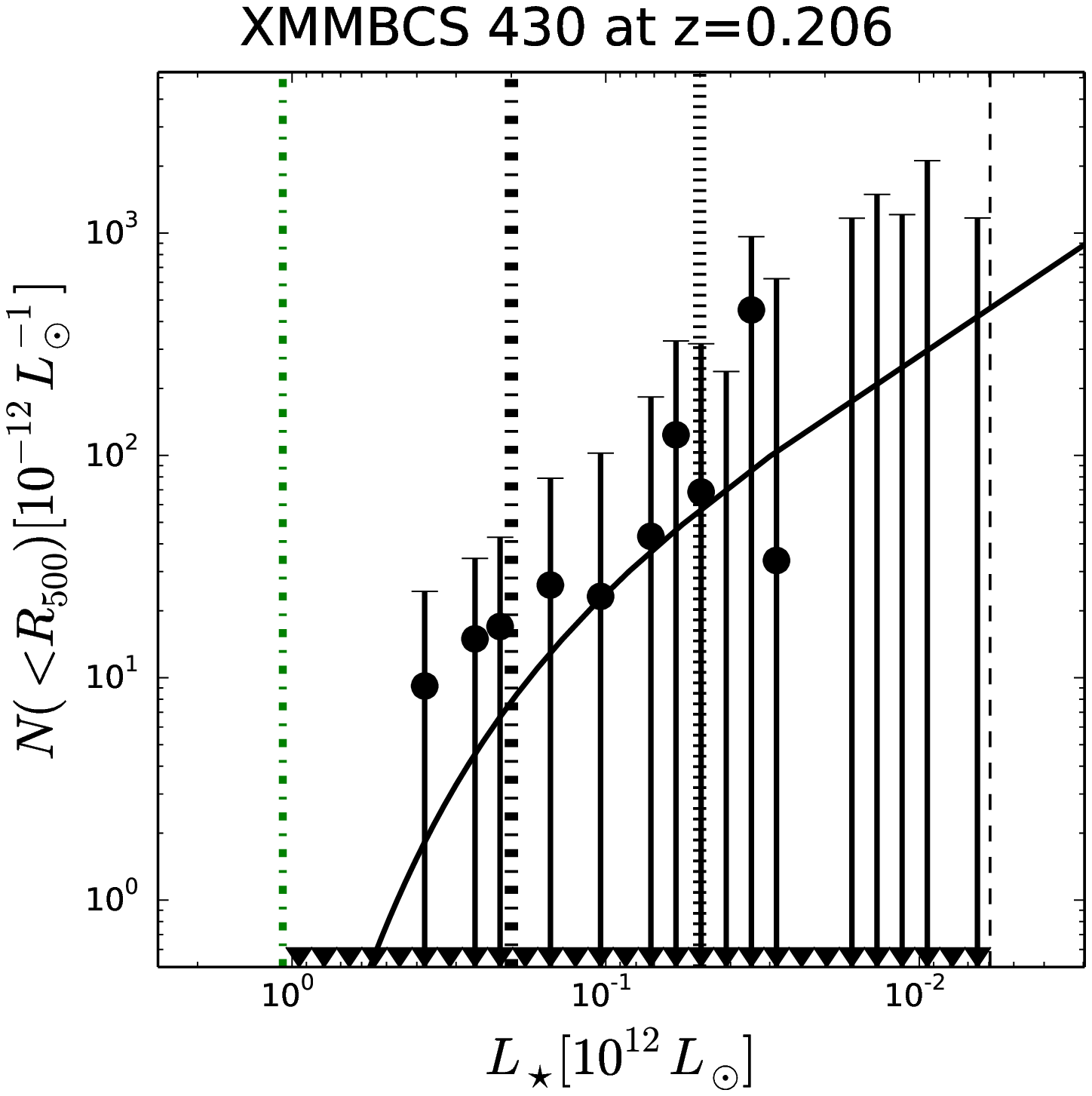} \end{subfigure}      
  
    \begin{subfigure}[b]{0.20\textwidth} \includegraphics[width=\textwidth]{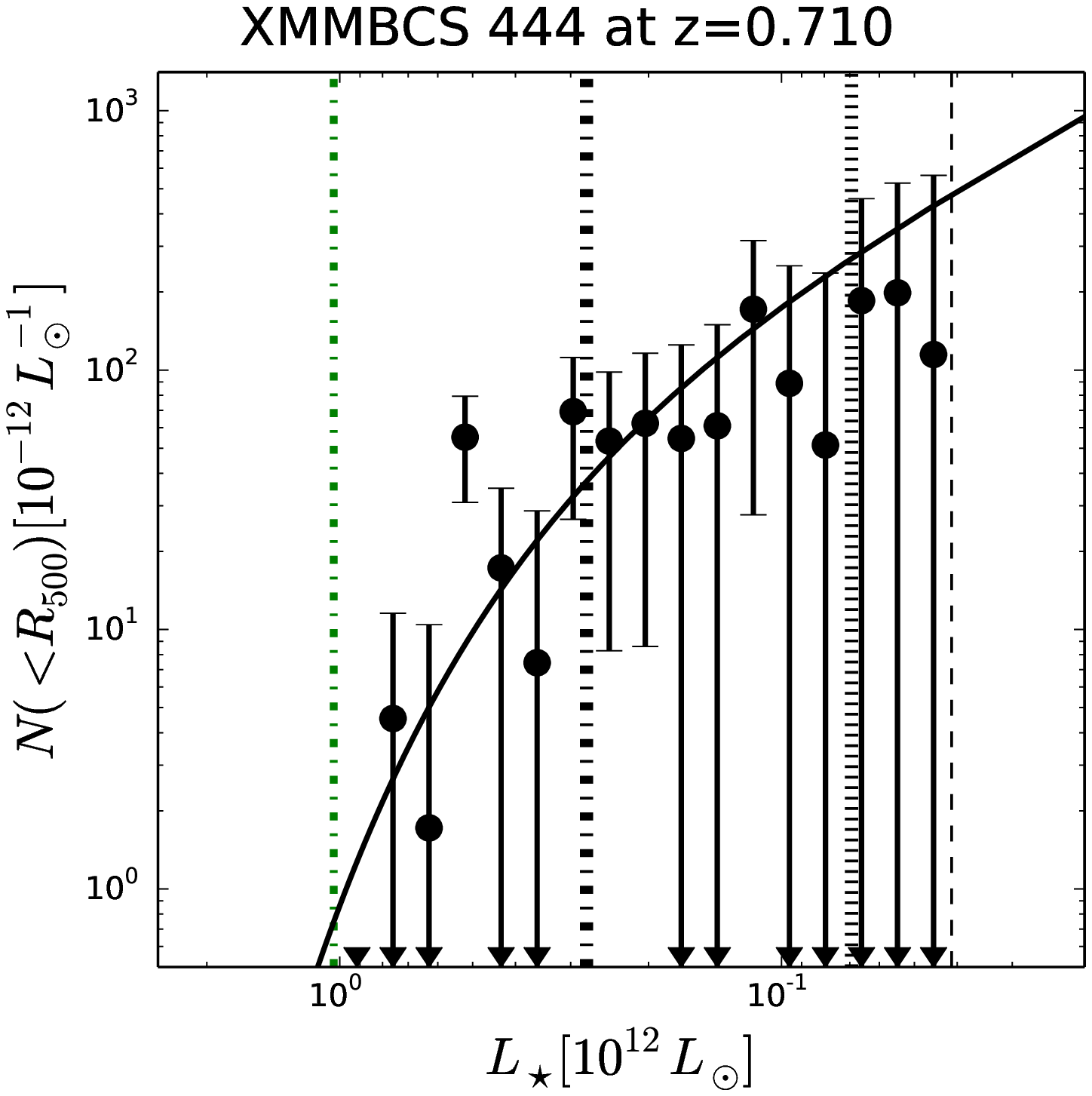} \end{subfigure}    
    \begin{subfigure}[b]{0.20\textwidth} \includegraphics[width=\textwidth]{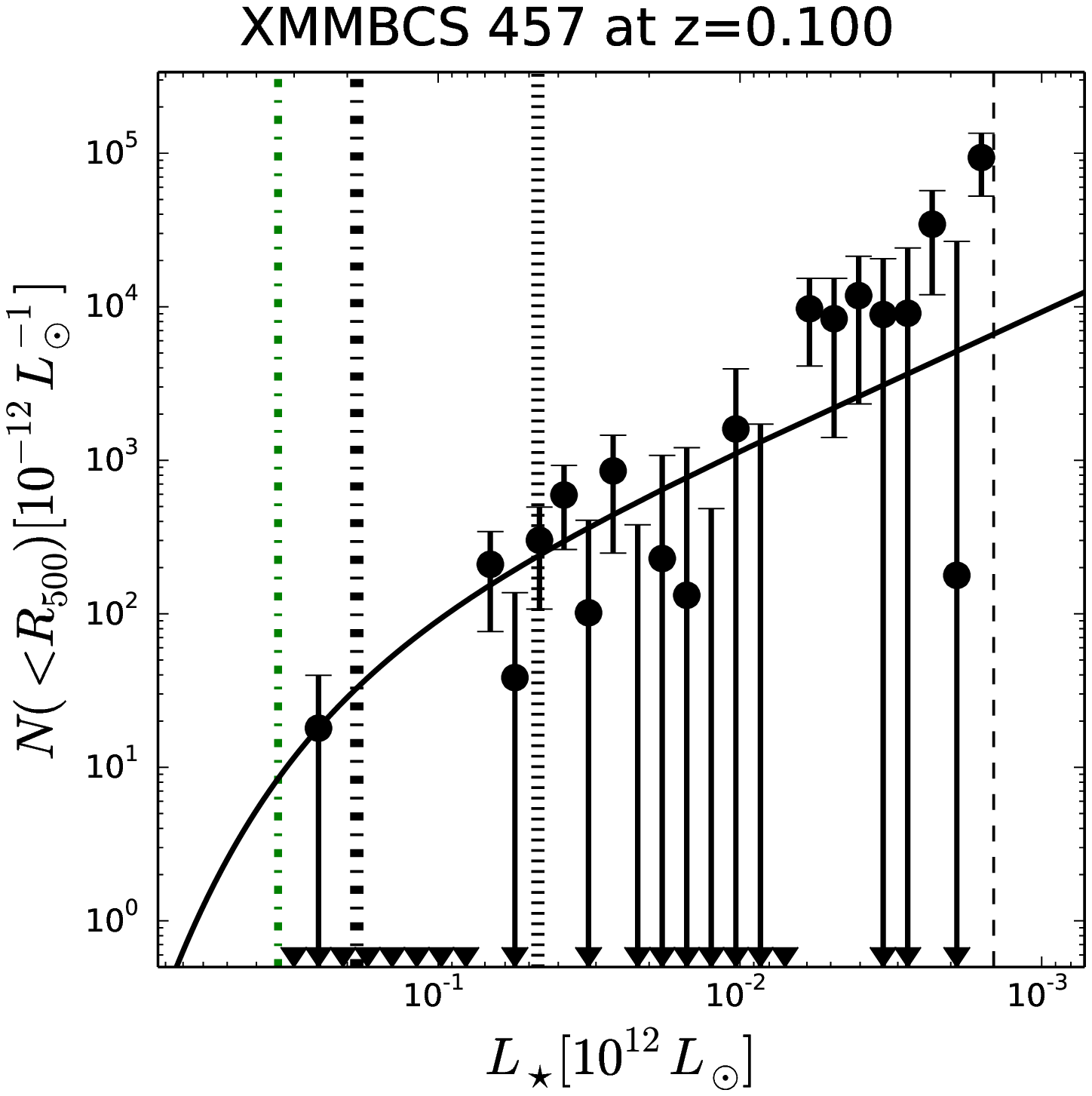} \end{subfigure}    
    \begin{subfigure}[b]{0.20\textwidth} \includegraphics[width=\textwidth]{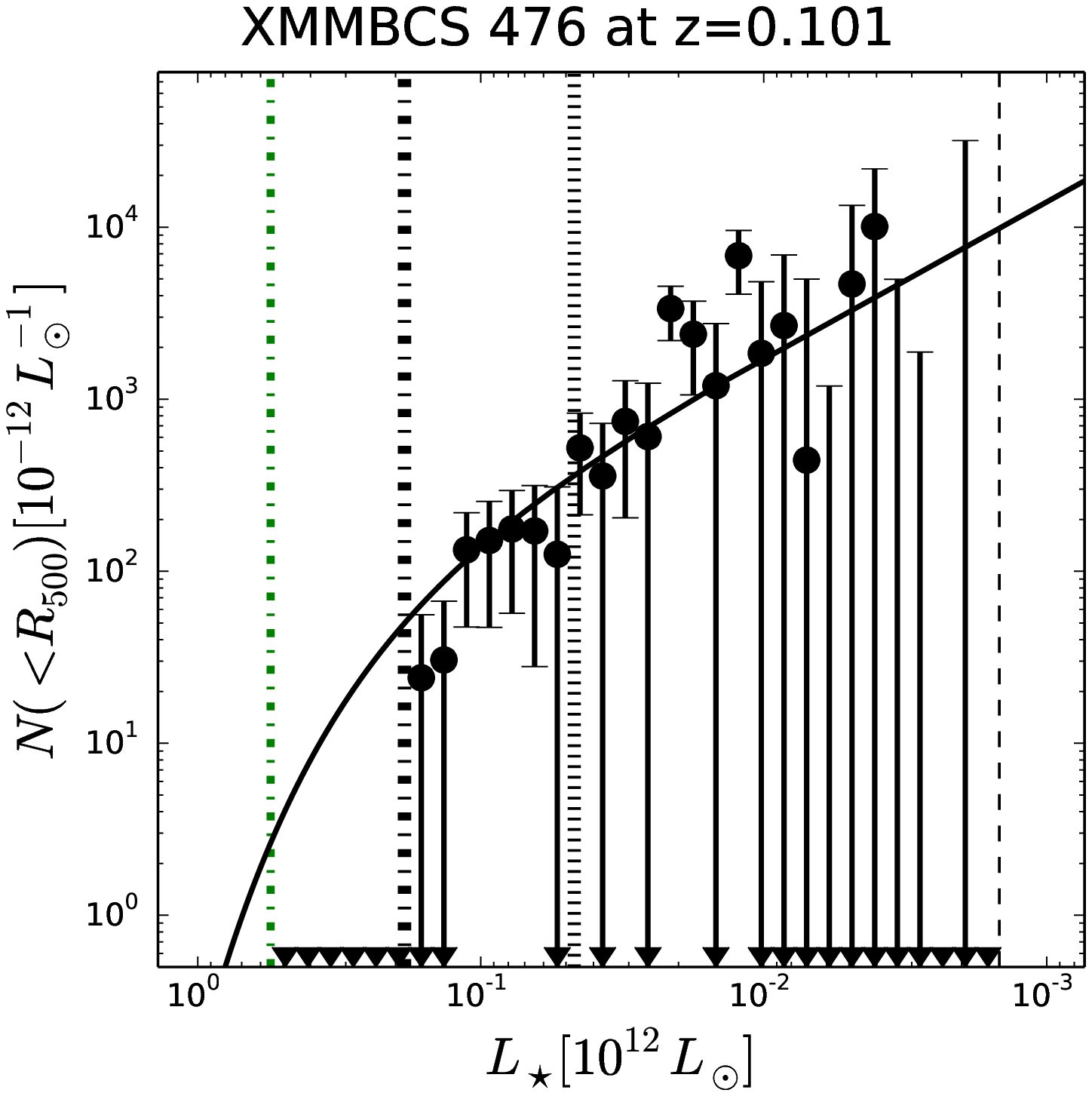} \end{subfigure}      
    \begin{subfigure}[b]{0.20\textwidth} \includegraphics[width=\textwidth]{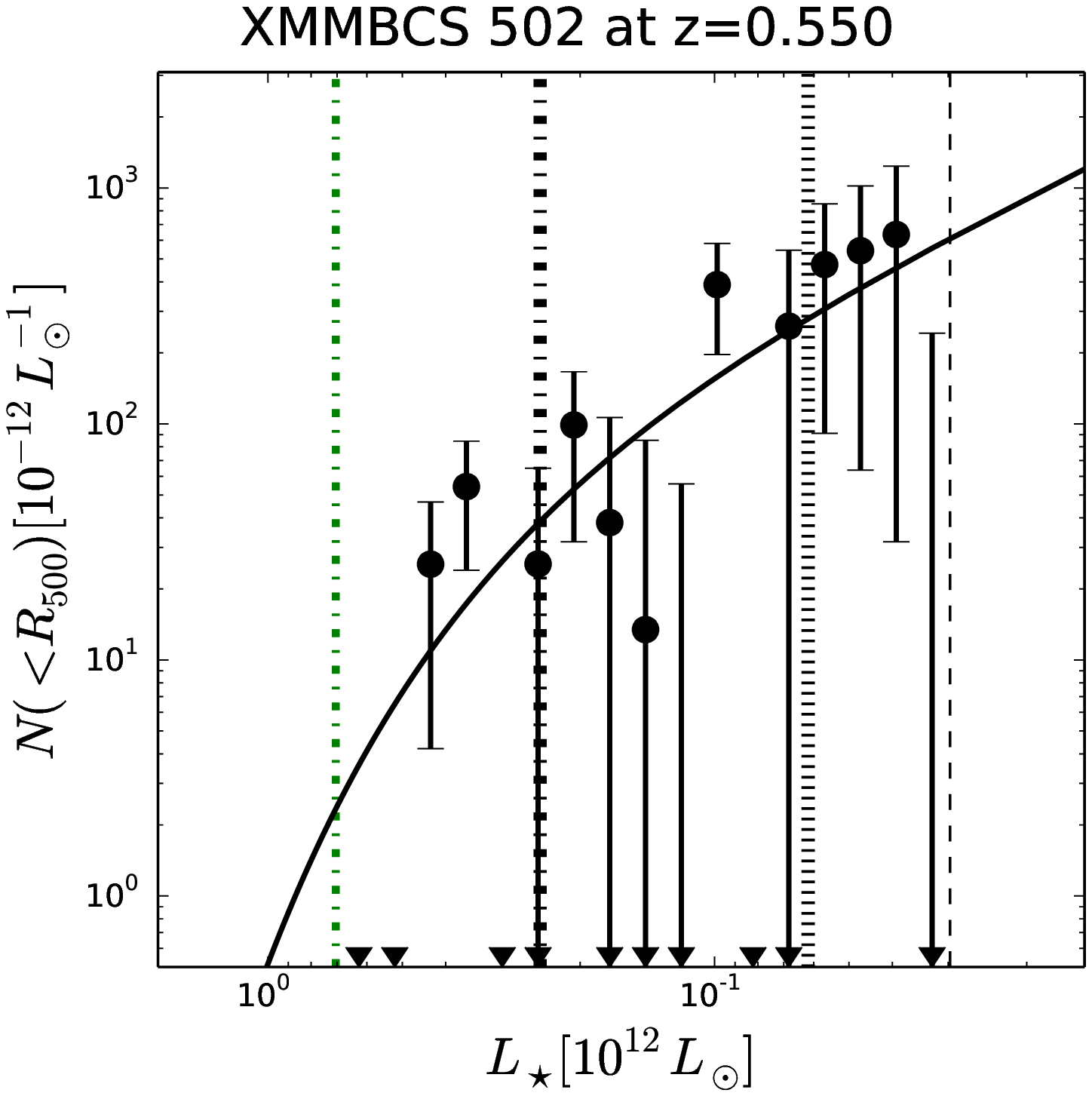} \end{subfigure}    

    \begin{subfigure}[b]{0.20\textwidth} \includegraphics[width=\textwidth]{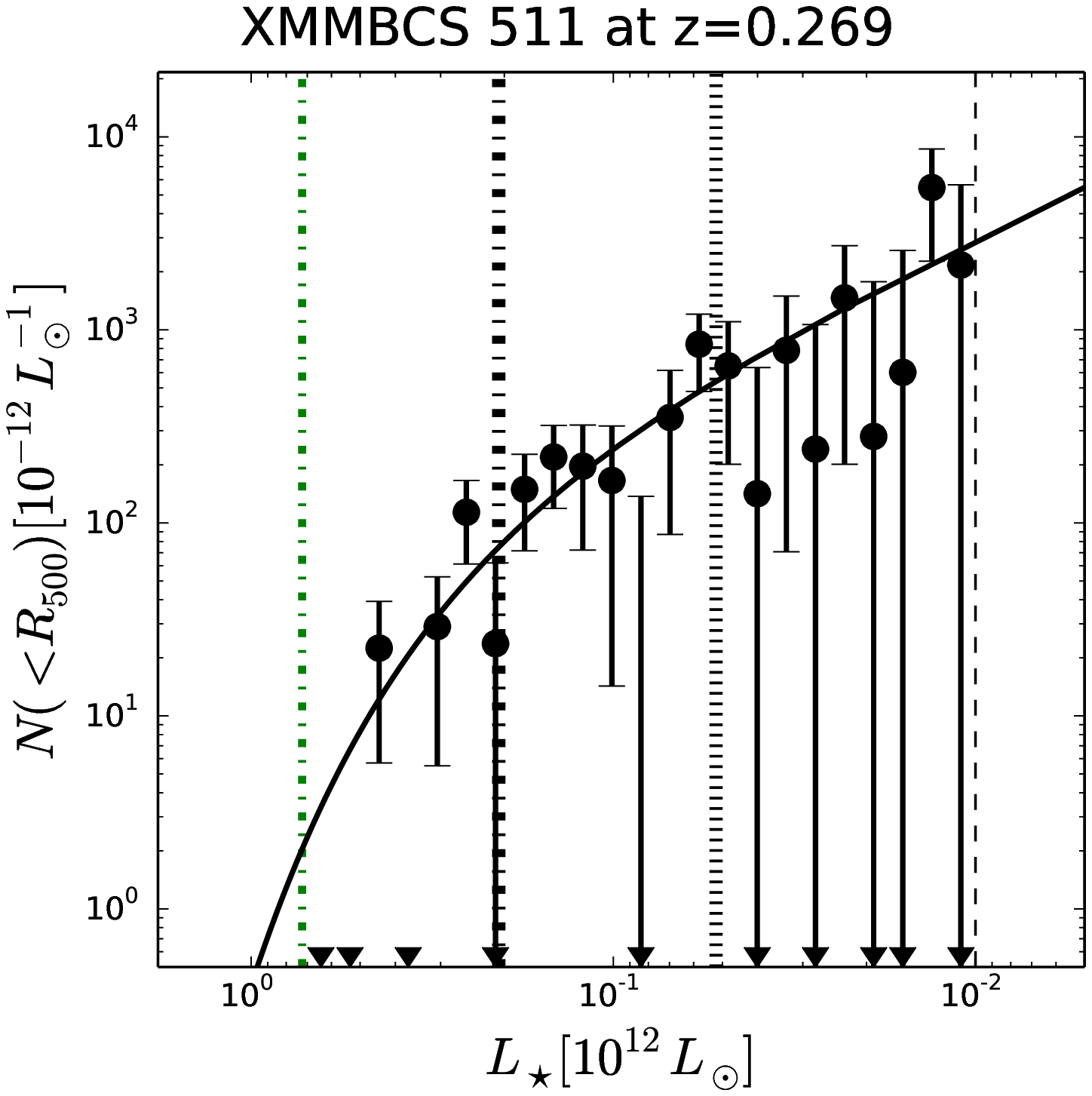} \end{subfigure}    
    \begin{subfigure}[b]{0.20\textwidth} \includegraphics[width=\textwidth]{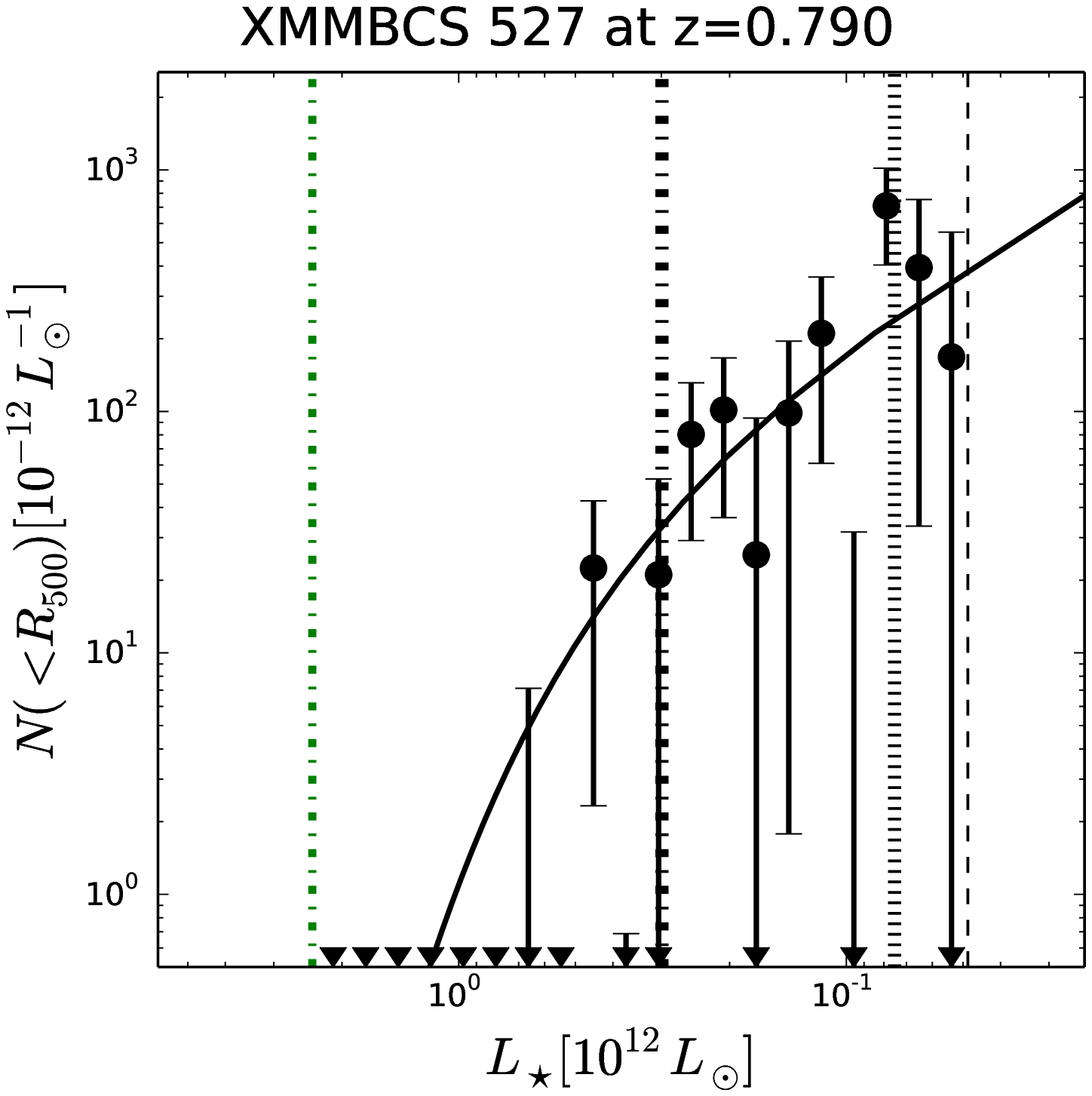} \end{subfigure} 
    \begin{subfigure}[b]{0.20\textwidth} \includegraphics[width=\textwidth]{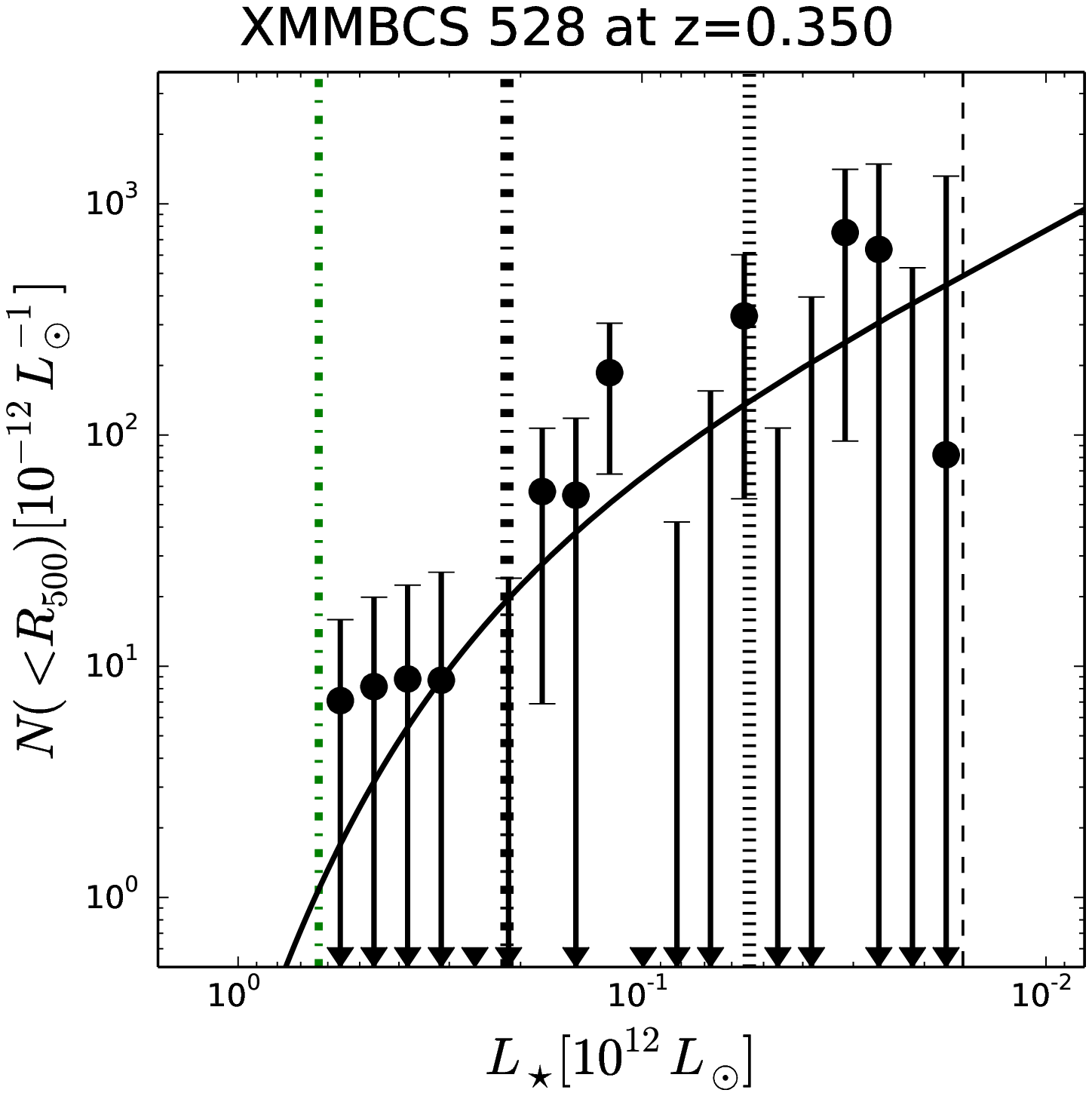} \end{subfigure}    
    \begin{subfigure}[b]{0.20\textwidth} \includegraphics[width=\textwidth]{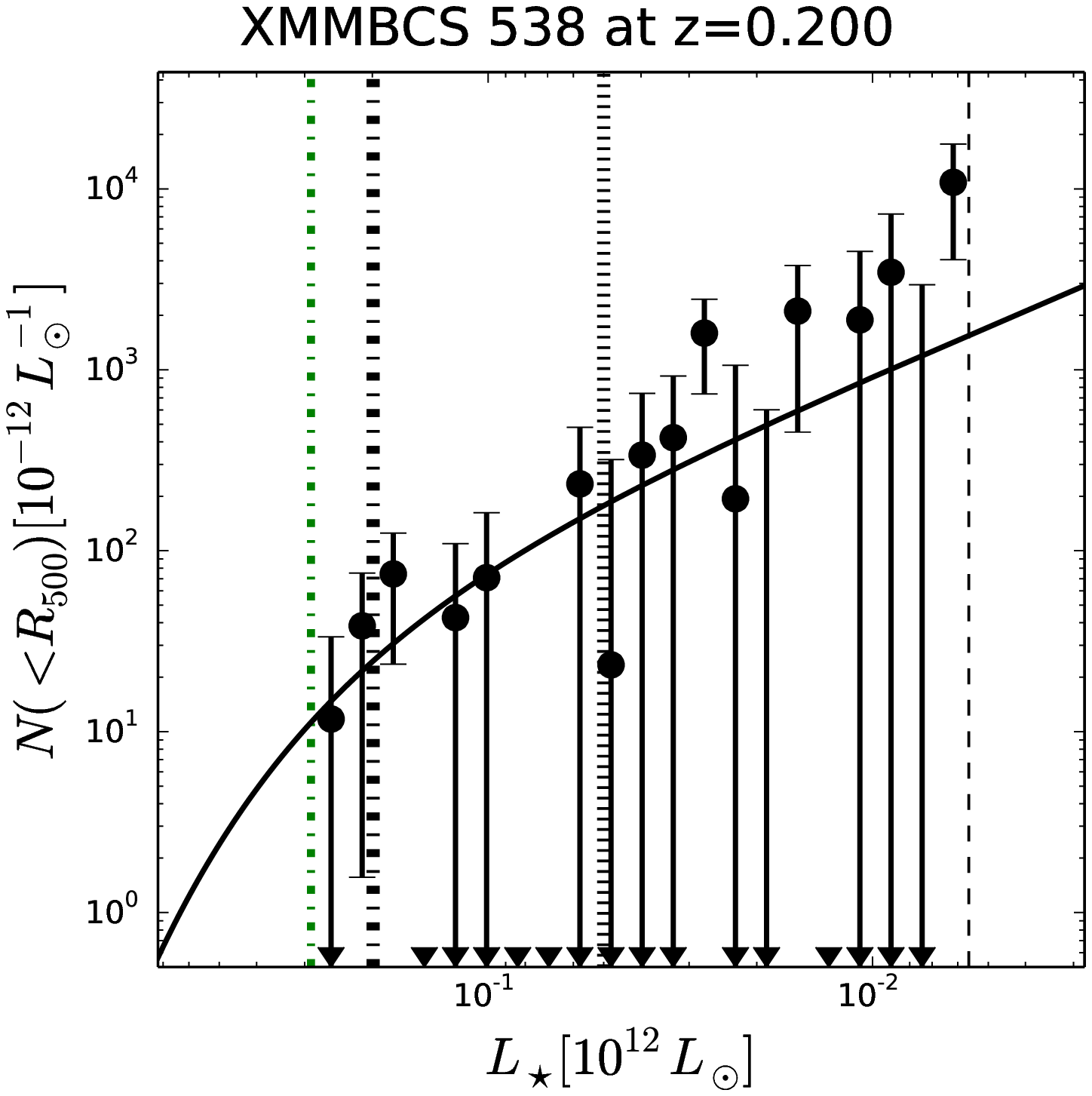} \end{subfigure}    

    \begin{subfigure}[b]{0.20\textwidth} \includegraphics[width=\textwidth]{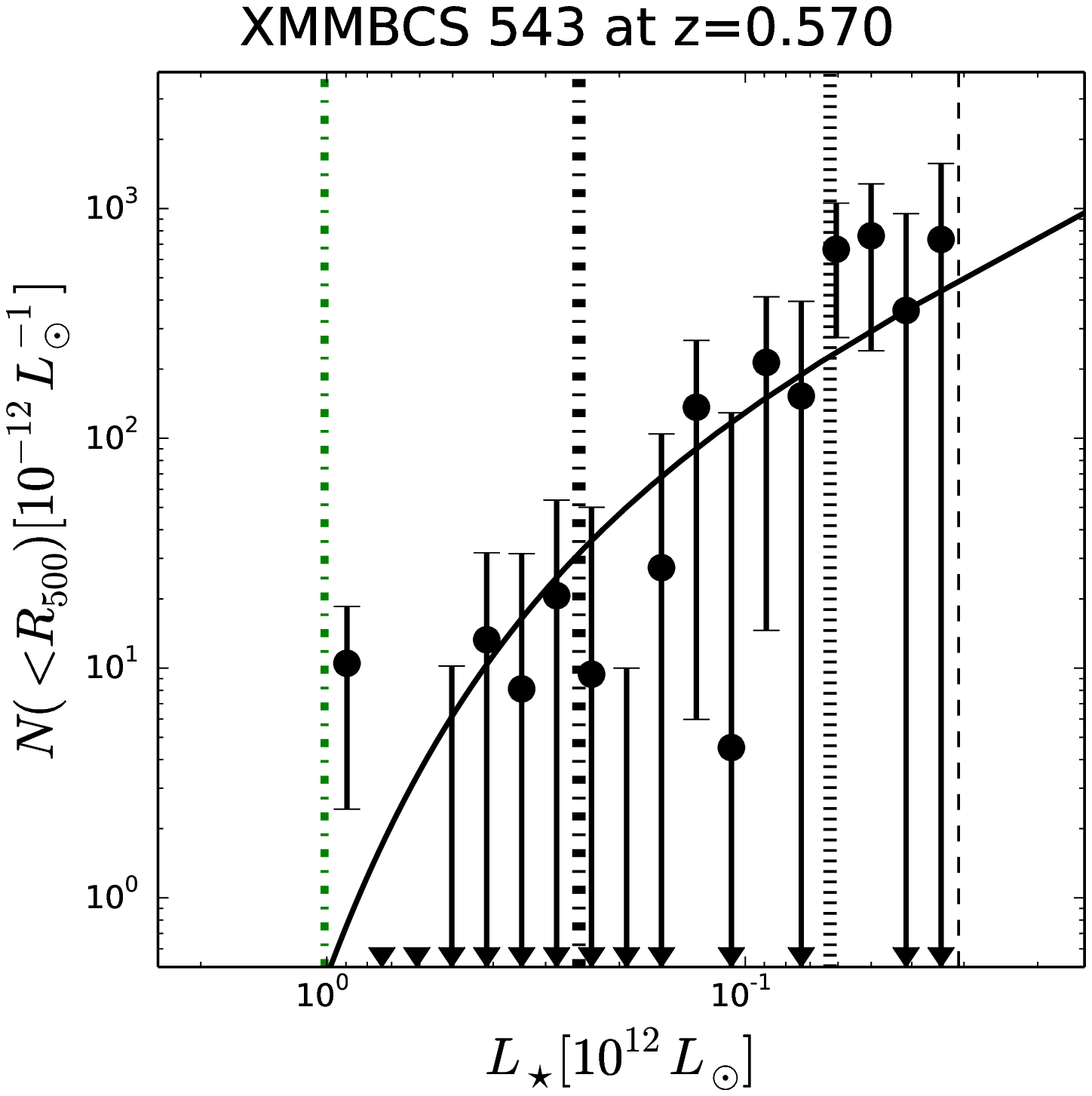} \end{subfigure}      
    \begin{subfigure}[b]{0.20\textwidth} \includegraphics[width=\textwidth]{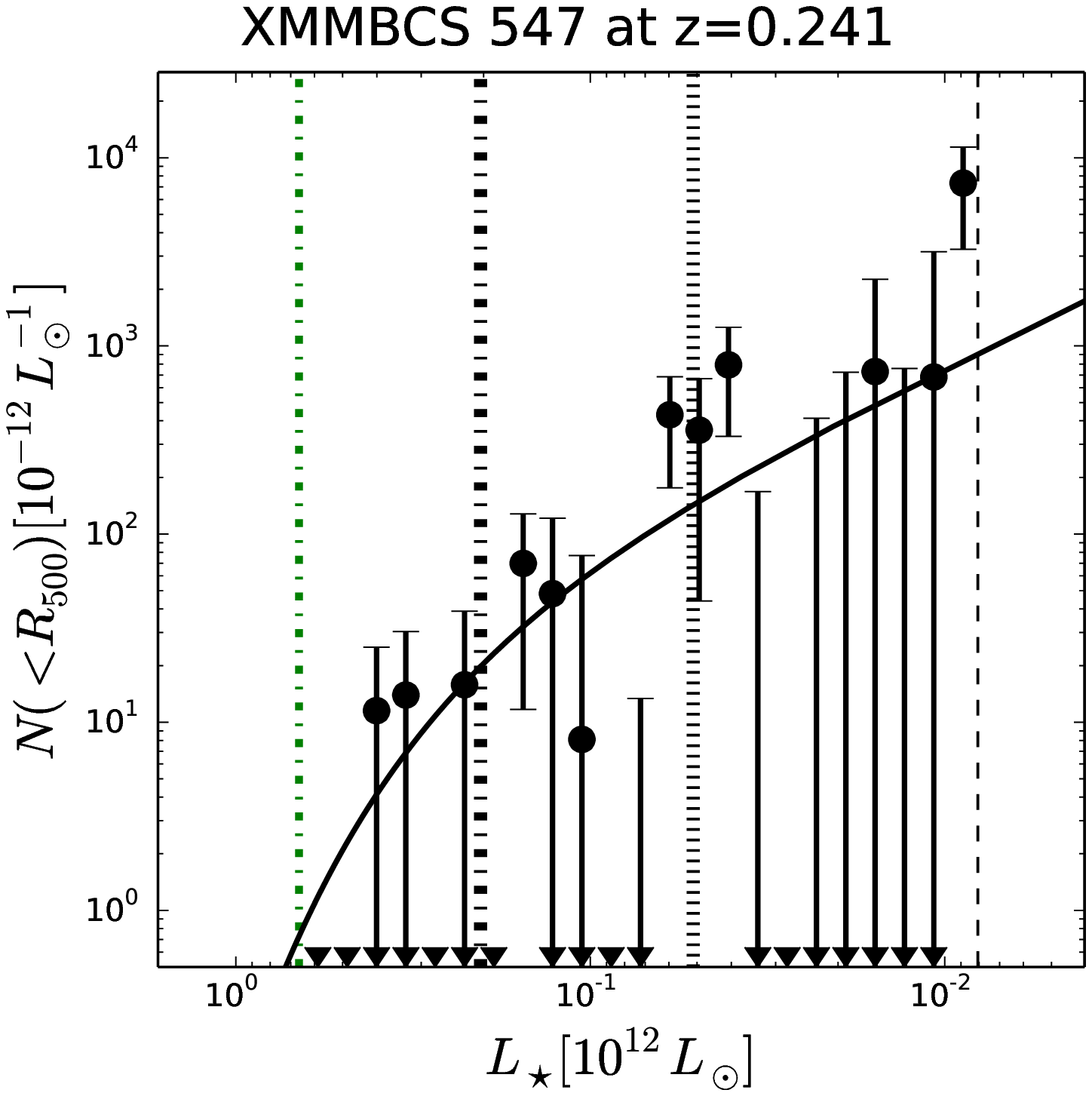} \end{subfigure}
    \caption{
    See caption in Figure~\ref{fig:lf_cm_1}.
    }
    \label{fig:lf_cm_2}    
\end{figure*}

\section{RESULTS}
\label{sec:result_discussion}

We present the estimated stellar masses of 46 \XMMBCS\ clusters in Table~\ref{tab:measurement}.  In addition to the stellar mass estimates, Table~\ref{tab:measurement} contains the cluster ID and redshift, the LF normalization $\phi_0$, the $p$-value that the model LF and observations are drawn from the same parent distribution and the measured blue fraction \fblue.  

We find that the LF provides a good description for the data for most of the clusters.  The observed LFs and the best-fit models of 46 \XMMBCS\ clusters are shown in Figure~\ref{fig:lf_cm_1} and Figure~\ref{fig:lf_cm_2}.  However, we find two clusters with $p$-values that indicate inconsistency between model and data at the $\approx3\sigma$ level.  The two outliers are \XMMBCS287 at $\redshift=0.57\pm0.04$ and \XMMBCS457 with a spectroscopic redshift $\redshift=0.1$. There are five very bright stars in the center of \XMMBCS287 contaminating the photometry of the cluster galaxies and causing the poor fit in the LF.  On the other hand, only 7 galaxies brighter than $\IRAConestar+1.5$ are detected within the cluster \Rfiveoo\ in \XMMBCS457, explaining the high \Cstat\ values and low $p$-values in the fit.   We include both of these systems while deriving the scaling relation.  In addition, we fail to detect the cluster galaxy population for \XMMBCS152 at $\redshift=0.139$; there are only two galaxies brighter than $\IRAConestar+1.5$ (before statistical background subtraction) that lie projected within the cluster \Rfiveoo\ and  (corresponding to $\IRACone\approx18$~mag), this results in $\Lstarsat=0~\Lsun$ in the LF fitting. Nevertheless, we also include \XMMBCS152 in deriving the scaling relation parameters because its BCG is clearly detected although the \Lstarsat\ is statistically consistent with zero. We discuss the systematics caused by these three clusters in Section~\ref{sec:sys_badfit}.

The best-fit parameters of the resulting scaling relations-- both total stellar mass \Mstar\ and satellite galaxy stellar mass \Mstarsat\ versus binding mass \Mfiveoo-- are presented in Table~\ref{tab:prior_and_fit}.  Neither scaling relation shows statistically significant redshift evolution, a point we will discuss more below.  The fully marginalized and joint parameter constraints for the total stellar mass \Mstar--\Mfiveoo\ relation are shown in Figure~\ref{fig:triangle_full}.  There is a noticeable covariance between the mass power law index \Bstar\ and the redshift power law index \Cstar, which is presumably driven by the characteristics of our flux limited sample that leads to the lowest mass systems being found only at low redshift.

We show the observed \Mstar\ and its best-fit scaling relation as a function of cluster mass and redshift in Figure~\ref{fig:sr_mz}.   The observed \Mstar\ for each cluster is corrected to the pivot redshift ($\ZPIV=0.47$; left panel) and the pivot mass ($\MPIV=8\times10^{13}M\odot$; right panel)  using the best-fit scaling relation $\Mstar = \Mstar(\Mfiveoo, \redshift)$.  

The \Mstar\ inside cluster \Rfiveoo\ increases with cluster masses ($\Mstar\propto{\Mfiveoo}^{0.69}$) and no significant redshift trend is observed. The \Cstar\ is statistically consistent with zero, suggesting the stellar contents inside the cluster \Rfiveoo\ sphere with respect to the halo mass of $\MPIV=8\times10^{13}\Msun$ is not evolving out to $\redshift\approx1$.  The intrinsic log-normal scatter of $\Mstar = \Mstar(\Mfiveoo, \redshift)$ is $\Dstarcom=0.36$, corresponding to $e^{\Dstar} - 1\approx(43\pm10)\percent$ scatter in \Mstar\ at fixed halo mass.  Together with the scaling relation power law index in mass, this implies a binding mass scatter for a given \Mstar\ of $\Delta\ln\Mfiveoo|\Mstar= \Dstar/\Bstar \approx (0.52\pm0.09)$, corresponding to $e^{\Dstar/\Bstar} - 1\approx (68\pm16)\percent$. 

The scaling relation where the BCG stellar mass is excluded is similar except with a lower \Astar\ and a larger \Dstar\ than the values inferred from including the BCG.  The amplitude of the total stellar mass scaling relation is $\approx36$\percent\ higher than the relation that only includes the light from the satellite galaxies, indicating that the BCG characteristically contains $\approx27$\percent\ of the stellar mass in this sample of low mass clusters and groups.

To actually use \Mstar\ as a mass indicator one would typically have only a cluster redshift and no knowledge of the virial radius \Rfiveoo.  Thus, we examine also the \Mstar--\Mfiveoo\ scaling relation when the \Mstar\ is extracted within a fixed metric radius of $0.5~$Mpc.  The analysis yields the parameters $(\Astar / 10^{12}\Msun, \Bstar, \Cstar, \Dstar)$ 
\[
(
1.77^{+0.11}_{-0.11}, 
0.50^{+0.13}_{-0.13},
0.15^{+0.43}_{-0.43},
0.33^{+0.06}_{-0.05}
) \, ,
\]
which indicates an \Mfiveoo\ scatter of $e^{\Dstar/\Bstar} - 1\approx (93\pm11)~\percent$ at fixed \Mstar(0.5~Mpc, $z$).

We compare \XMMBCS\ results with two other high-\redshift\ cluster samples from the literature.  The first one is the Gemini CLuster Astrophysics Spectroscopic Survey \citep[GCLASS;][hereafter vdB14]{burg14} sample for which they studied the 10 low mass clusters or groups selected from a NIR survey at redshift between 0.86 and 1.34, while the second one is the SPT-selected sample \citep[][hereafter Chiu16]{chiu16} consisting of 14 massive clusters with $\Mfiveoo\gtrsim3\times10^{14}\Msun$ at redshift between 0.56 and 1.32.

As demonstrated in Chiu16, the systematics, such as the different mass calibrators (velocity dispersion $\sigma_{\nu}$ or X-ray luminosity \Lx) or the different initial mass functions, could lead to spurious mass or redshift trends if no homogenization is applied to the different samples.  For example, the \Mfiveoo\ inferred from the X-ray mass proxies (e.g., \Lx) and velocity dispersion $\sigma_{\nu}$ could be underestimated by $\approx44\percent$ and $\approx23\percent$, respectively, as compared to the SZE-inferred masses including the CMB cosmological constraint \citep{bocquet14}. 

We therefore homogenize the vdB14 and Chiu16 samples before the comparison by bringing all three mass scales to the mass scale of the \XMMBCS\ sample, whose masses are determined from the X-ray \Lx\ calibrated using X-ray hydrostatic masses.  Specifically, the \Mfiveoo\ of vdB14 and Chiu16 samples are multiplied by $1.23/1.44$ and $1/1.44$ to reach the mass floor inferred by the X-ray mass proxy (\Lx) used in this work. Following the same procedure in Chiu16, we multiply the BCG-excluded stellar masses by the factors $0.95$ and $0.88$ in the vdB14 and Chiu16 samples, respectively, which corrects for the change in \Mstar\ caused by the reduction in \Rfiveoo\ that comes from the lower \Mfiveoo.  The same initial mass function \cite{chabrier03} has been used in all three studies. 

The comparison is shown in Figure~\ref{fig:sr_mz}. The \XMMBCS\ clusters are in good agreement with the GCLASS and the SPT samples, with the GCLASS and the SPT clusters serving as extensions of the high redshift and high mass ends of the \XMMBCS\ sample. The scaling relation $\Mstar(\Mfiveoo, \redshift)$ of \XMMBCS\ results in a mass trend $\Bstar=0.69\pm0.15$ which is statistically consistent with the GCLASS ($\Bstar=0.62\pm0.12$) and SPT ($\Bstar=0.63\pm0.09$) clusters. Interestingly, our measured mass trend is also in good agreement with results extracted from cluster and group samples in different redshift ranges and using a variety of techniques \citep{lin03b,giodini09,lin12,ziparo15}. The consistent values of \Bstar\ affirm that a similar relationship between the stellar and halo masses exists for all systems above the mass scale $\Mfiveoo\approx  2\times10^{13}\Msun$. 

The redshift trends of the SPT ($\Mstar\propto(1+z)^{0.26\pm0.18}$) and \XMMBCS\ ($\Mstar\propto(1+z)^{-0.04\pm0.47}$) samples are both statistically consistent with zero, suggesting that the stellar masses of galaxies within \Rfiveoo\ do not evolve out to redshift $\redshift\approx1.35$ over the full mass range of groups and clusters. This result is in excellent agreement with that from a sample of 94 massive clusters at $0\le\redshift\le0.6$ \citep{lin12}, and  the same picture is also implied by studies of X-ray or optically selected groups, some of which extend to redshifts $\redshift\approx1.0-1.6$ \citep{giodini09, connelly12, leauthaud12b, ziparo13}. Combining the GCLASS/SPT sample at $0.6\lesssim\redshift\lesssim1.35$ and our \XMMBCS\ clusters extending to the low mass end between redshift $\approx0.1$ and $\approx1$, we conclude that the stellar masses inside the cluster \Rfiveoo\ sphere are well established for halo masses $\Mfiveoo\gtrsim2\times10^{13}\Msun$ since $\redshift\approx1.35$.

\section{SYSTEMATICS}
\label{sec:sys}

We discuss the potential systematics due to the problematic clusters, the LF fitting, the mass-to-light ratio \ML, the blending in the \Spitzer\ imaging and the mass estimates below. 

\subsection{The problematic clusters}
\label{sec:sys_badfit}

We find that the LF modeling of two clusters (\XMMBCS287 and \XMMBCS457) shows inconsistency between the model and observed data at the $\approx3\sigma$ level;
in addition, we do not detect the non-BCG galaxy population for \XMMBCS152.
To quantify the systematics caused by these problematic clusters, we repeat the whole likelihood maximization excluding these clusters and compare the scaling relation parameters.
The parameters $(\Astar / 10^{12}\Msun, \Bstar, \Cstar, \Dstar)$ for the scaling relation $\Mstar=\Mstar(\Mfiveoo,\redshift)$ excluding these clusters are 
\[
(
1.89^{+0.13}_{-0.12}, 
0.67^{+0.15}_{-0.15},
-0.12^{+0.46}_{-0.46},
0.34^{+0.07}_{-0.06}
) \, ,
\] which are statistically consistent with the values including these clusters (see Table~\ref{tab:prior_and_fit}).
Therefore, our result is not biased by these clusters.

\subsection{The LF fitting}
\label{sec:sys_lffit}

To quantify the systematics raised from the LF fitting, 
we estimate the stellar masses of all \XMMBCS\ clusters by fixing ${\mstar}_{,~\mathrm{stacked}}$ and $\alpha$ to the values, which are shifted by $1\sigma$  from the best-fit values determined by fitting to the full stacked sample, and repeat the whole analysis to obtain the resulting scaling relation parameters.   Specifically, we use the two extreme cases, the $1\sigma$ shift from the best-fit values of ${\mstar}_{,~\mathrm{stacked}}$ and $\alpha$ (the black circle in the right panel of Figure~\ref{fig:stacked_lf}) toward the upper right ($1\sigma_{\mathrm{right}}$) and lower left ($1\sigma_{\mathrm{left}}$) along the direction of the parameter degeneracy.  Accordingly, the $1\sigma_{\mathrm{left}}$ ($1\sigma_{\mathrm{right}}$) shift implies that the characteristic magnitude \mstar\ predicted by our CSP model is fainter by $\approx-0.51$~mag (brighter by $\approx0.37$~mag) and $\alpha=-1.25$ ($\alpha=-0.43$).
The resulting parameters $(\Astar / 10^{12}\Msun, \Bstar, \Cstar, \Dstar)$ for scaling relation $\Mstar=\Mstar(\Mfiveoo,\redshift)$ are
\[ 
( 
1.80^{+0.13}_{-0.12}, 
0.72^{+0.15}_{-0.15},
0.23^{+0.48}_{-0.49},
0.36^{+0.07}_{-0.06}
) \, ,
\] 
and
\[ 
( 
1.98^{+0.13}_{-0.13}, 
0.65^{+0.15}_{-0.15},
-0.16^{+0.47}_{-0.47},
0.36^{+0.06}_{-0.06}
) \,
\] 
for $1\sigma_{\mathrm{left}}$ and $1\sigma_{\mathrm{right}}$ shift, respectively.
The resulting parameters of the scaling relations are all statistically consistent (within $1\sigma$) with the values obtained using ${\mstar}_{,~\mathrm{stacked}} = 0$ and $\alpha=-0.89$ 
as in Table~\ref{tab:prior_and_fit}; therefore we conclude that the systematics associated with adopting the best fit LF parameters from the cluster stack when fitting the LF in individual clusters are not dominant.

\subsection{The mass-to-light ratio \ML}
\label{sec:sys_m2l}

The mass-to-light ratio ${\ML}_{,\mathrm{blue}}$ of the blue population is estimated assuming the synthetic galaxy population with the star formation history ($\tau=10$~Gyr) and one solar metallicity at formation redshift $z_{\mathrm{f}} = 3$.  
Using ${\ML}_{,\mathrm{blue}}$ derived from $\tau=5$~Gyr raises the stellar mass estimates by $1.3\percent$, $2.1\percent$ and $2.5\percent$ for the cases of $\fblue=0.2$, $\fblue=0.3$ and $\fblue=0.4$, respectively. 
Using ${\ML}_{,\mathrm{blue}}$ derived from $\tau=15$~Gyr lowers the stellar mass estimations by $0.4\percent$, $0.7\percent$ and $1.2\percent$ for the cases of $\fblue=0.2$, $\fblue=0.3$ and $\fblue=0.4$, respectively. 
Changing the metallicity or increasing the formation redshift to $z_{\mathrm{f}}=5$ in deriving ${\ML}_{,\mathrm{blue}}$ has only negligible impact on the ${\ML}_{,\mathrm{blue}}$ derived from $\tau=10$~Gyr and $\redshift_{\mathrm{f}}=3$ model. That is, the systematic uncertainty raised from the blue population is dominated by the large scatter of the estimated \fblue\ rather than the assumed ${\ML}_{,\mathrm{blue}}$.

If we re-run the likelihood maximization using $\ML = {\ML}_{,\mathrm{CSP}}$ for all cluster galaxies, the resulting parameters $(\Astar / 10^{12}\Msun, \Bstar, \Cstar, \Dstar) $ of the scaling relation $\Mstar=\Mstar(\Mfiveoo,\redshift)$ are
\[ 
( 
2.03^{+0.14}_{-0.13}, 
0.65^{+0.15}_{-0.15},
-0.04^{+0.48}_{-0.48},
0.37^{+0.07}_{-0.06}
) \, .
\]
This result is effectively assuming there is no blue galaxy population in our sample and thus represents an upper limit to the \Mstar\ estimation for each system.  We show the best-fit model assuming $\ML = {\ML}_{,\mathrm{CSP}}$ as the purple region in Figure~\ref{fig:sr_mz}.  As seen in  Figure~\ref{fig:sr_mz}, including the \fblue\ correction causes a steeper \Bstar\ at the $\approx0.2\sigma$ level and has no significant effect on the redshift trends. However, assuming $\ML = {\ML}_{,\mathrm{CSP}}$ leads to an \Mstar\ estimate, which is biased high by $\approx8.5\percent$, which is equivalent to a $\approx0.86\sigma$ shift, for a cluster of mass $\Mfiveoo=0.8\times10^{14}\Msun$ at $\redshift=0.47$. Our best-fit \fblue\ model of \XMMBCS\ clusters suggests that the $\fblue\approx18\pm10\percent$ for a cluster with $\Tx=3$~keV at $\redshift=1$, while the mean of the \fblue\ estimates for the entire \XMMBCS\ sample is $\approx(31\pm4)\percent$.   

\subsection{Blending}
\label{sec:sys_blending}

For the \IRACone\ band used in this work, the FWHM is $\approx1.8\arcsec$, making it challenging to deblend the fluxes from the multiple neighboring objects without introducing external information on the source distribution.  However, the blending among cluster galaxies does not affect the total luminosity estimated in this work, because integrating the best-fit LF is equivalent to estimating the excess light of the cluster galaxy population.  Blending of cluster and non-cluster galaxies, on the other hand could bias the cluster light, and this would be most likely in denser cluster core around the BCG.

We quantify the systematic effects of BCG blending with non-cluster galaxies as follows.  We first calculate the probability $P_{\mathrm{blend}}(m)$ of BCG blending with the fore/background galaxies with magnitude $m$.  Specifically, the $P_{\mathrm{blend}}(m)$ is derived by re-normalizing the background magnitude distribution of the background aperture (i.e, \Rfiveoo) to the angular area of the BCG.  The angular area of the BCG is approximated by the aperture with the radius of $2\times\mathtt{FLUX\_RADIUS}$, which is a parameter derived by  \texttt{SExtractor}.  Assuming a Poisson distribution for $P_{\mathrm{blend}}(m)$ for a given magnitude $m$, we find by sampling 1000 realizations that the probability of BCG blending with the fore- and background galaxies is $\approx25\pm14\percent$.  Note that the majority of the non-cluster members blending with the BCG takes place in the fainter magnitude range because of the more abundant faint galaxy population.

Second, we estimate the total flux blending with the BCG from 1000 realizations where blending takes place.  The resulting mean excess of the flux from the non-cluster members (i.e., the fore- and background) contributes on average an additional $\approx11\pm6\percent$ to the BCG flux if blending takes place.  
In the end, we calculate the expected excess flux due to blending by weighing the extra blended flux by the probability of the BCG being blended.  As a result, we find that the expected excess of the blended flux from the fore- and background over the whole \XMMBCS\ sample results in a bias in the BCG flux at the level of $\approx 2.3\pm1.5\percent$, which is well below the statistical uncertainties in our analysis.  Moreover, the excess flux due to blending estimated from our 1000 realizations  shows no trends in mass and redshift over the \XMMBCS\ sample.  If the flux of the BCG is catastrophically overestimated by a factor of two due to blending with the non-cluster galaxies, then the total stellar mass estimation \Mstar\ would be overestimated by $\approx10\percent$ for \XMMBCS\ sample.  However, a factor of two overestimation of the BCG fluxes is a rare occurrence.  Moreover, our calculation shows that this systematic would not introduce biases into the mass or redshift trends of the scaling relation.

\subsection{Cluster Binding Masses}
\label{sec:sys_aperture}

We further quantify systematics associated with the binding mass \Mfiveoo, which is inferred from the X-ray luminosity.  For \XMMBCS\ clusters, the uncertainty of \Mfiveoo\ ($\approx30\percent$) results in an uncertainty of \Rfiveoo\ at the level of $\approx10\percent$, and this radius is used to define the region from which the \Mstar\ is extracted.  The resulting uncertainty in \Mstar\ is at the level of $\approx8\percent$ given the NFW distribution of the galaxies.  Given that our measurement uncertainties for \Mstar\ are already larger than this, this additional scatter does not impact our analysis.

However, there are also systematic uncertainties in \Mfiveoo.  We quantify the systematics between the \Mfiveoo\ estimates inferred from the mass proxies of the SZE-signatures and X-ray luminosities following the work of \citet{bocquet15}, where measurements of SZE-inferred cluster binding masses calibrated using X-ray data, velocity dispersions and a cosmological analysis with information from external probes were compared.  In our baseline analysis we adopt the X-ray luminosity-mass relation as calibrated using X-ray hydrostatic masses \citep{pratt09}.  If we instead adopt the SZE-inferred masses for our analysis, then the resulting  \Mfiveoo\ and \Mstar\ would be higher by $44\percent$ and $13.6\percent$, respectively, following the procedure described in Section~\ref{sec:result_discussion}.  Assuming this systematic offset of the mass proxies has no dependence on the cluster mass and redshift (i.e., it only affects the absolute scale of cluster mass), adopting the SZE-inferred mass would lead to the normalization \Astar\ of the scaling relation dropping by $1 - 1.136/1.44^{\Bstar} \approx 11.5\percent$.

We also estimate the impact of adopting the weak lensing based luminosity-mass relation from a recent study of 70 clusters and groups at $0.1\le \redshift \le 0.83$ with masses ranging from $\approx2\times10^{13}\Msun$ to $\approx2\times10^{15}\Msun$ \citep{kettula15}.  We first estimate the mass scale offset between the  \cite{kettula15} and \cite{pratt09} luminosity-mass relations.   Specifically, we derive the X-ray masses for the 70 cluster sample using the core-extracted \Lx--\Mfiveoo\ relation from \cite{pratt09} and then compare those X-ray derived masses to the lensing mass measurements.  We find that the mean of the ratio of the lensing masses to the X-ray masses is $1.22\pm0.10$, and there is no significant mass dependence.  Thus, the gravitational lensing measurements prefer masses that are $\approx22\percent$ higher than the X-ray masses.  If we scale up our X-ray masses by $22\percent$, then the resulting stellar masses would increase by $7.2\percent$ due to the increasing radius \Rfiveoo.  The $22\percent$ increment in \Mfiveoo\ and $7.2\percent$ increment in \Mstar\ would lead to the normalization \Astar\ of the scaling relation dropping by $1 - 1.072/1.22^{\Bstar} \approx 6.5\percent$.

%
%

\section{CONCLUSIONS}
\label{sec:conclusion}

We use IRAC \IRACone\ band photometry from the wide field \SSDF\ survey \citep{ashby13a} together with blue fraction \fblue\ measurements relying on $griz$ photometry from the \BCS\ survey \citep{desai12} to estimate the stellar masses of 46 X-ray selected low mass clusters and groups from the \XMMBCS\ survey \citep{suhada12}.  This sample has masses in the range $2\times10^{13}\Msun\lesssim\Mfiveoo\lesssim2.5\times10^{14}\Msun$ (median mass $8\times10^{13}\Msun$) and redshifts in the range $0.1\le\redshift\le1.02$ (median redshift $0.47$).  The stellar masses of the full population and the BCG-excluded population are estimated for each cluster.  

We employ a Bayesian likelihood developed in a previous analysis \citep{liu15b} that leverages an existing X-ray luminosity mass relation \citep{pratt09} to constrain the stellar mass-halo mass scaling relations for this sample.  The form of the scaling relation is a power law in mass and redshift with log-normal intrinsic scatter.  The normalization, the power law indices in mass and redshift, and the intrinsic scatter of the stellar mass at fixed halo mass are fully quantified in this work.  The best-fit stellar mass-halo mass scaling relation is
\[
\frac{\Mstar}{10^{12}\Msun} = 1.87^{+0.13}_{-0.12}
\left( \frac{\Mfiveoo}{8\times10^{13}\Msun  } \right)^{0.69\pm0.15} 
\left(\frac{1+z}{1.47} \right)^{-0.04\pm0.47} \, ,
\]
with log-normal intrinsic scatter $\Dstarcom=0.36^{+0.07}_{-0.06}$.

The best-fit scaling relation of \XMMBCS\ clusters behaves as $\Mstar\propto \Mfiveoo^{0.69\pm0.15}$, indicating a strong mass dependence of the stellar mass fraction within \Rfiveoo. The intrinsic log-normal scatter $\Dstar=0.36^{+0.07}_{-0.06}$ of \Mstar\ at a given cluster mass is comparable to the scatter in the X-ray luminosity-halo mass scaling relation $(\Dx=0.38\pm0.06)$. No significant redshift trend of stellar mass is seen; the best-fit scaling relation that describes the total stellar mass \Mstar\ evolves as $\Mstar\propto\left(1+\redshift\right)^{-0.04\pm0.47}$.  Thus, our analysis provides no evidence for redshift evolution of the stellar mass fraction within \Rfiveoo\ of low mass clusters and groups out to $\redshift\approx1$.

We compare \XMMBCS\ clusters with the SPT massive clusters ($\Mfiveoo\approx6\times10^{14}\Msun$; Chiu16) and GCLASS low mass clusters and groups ($\Mfiveoo\approx1\times10^{14}\Msun$; vdB14) at redshift $0.6\lesssim\redshift\lesssim1.3$.  After correcting for the systematics of different mass calibrators, we find that there is good agreement among the \XMMBCS, GCLASS and SPT clusters.  The mass trend \Bstar\ of \XMMBCS\ clusters is statistically consistent with the results of the GCLASS and the SPT samples.  Together with the results of the GCLASS and the SPT samples extending to the high redshift and the high mass regimes, the \XMMBCS\ sample provides no evidence for a redshift trend in the stellar mass fraction of the galaxy populations in clusters with masses $\Mfiveoo\gtrsim2\times10^{13}\Msun$ out to redshift $\redshift\approx1.3$.  Larger samples with uniform selection and mass estimation would allow for a more precise study of the redshift trend.

We investigate the systematic effects raised from (1) the clusters which have problematic LF fitting, (2) the LF modeling, (3) the blue population in clusters, (4) the blending in the imaging and (5) the cluster mass uncertainty.  The systematics raised from the problematic clusters and the LF modeling are smaller than the statistical uncertainties.  We find that the blending is more severe in the cluster core, but that the expected bias of the BCG flux due to blending with non-cluster members is at the level of $\lesssim2.5\percent$, which is too small to be important in this work.  We find that the most important systematic effect is from the mass-to-light ratio \ML, making it important to include blue fraction \fblue\ measurements to avoid biasing scaling relation parameters.  We estimate the blue fraction \fblue\ using the \BCS\ optical catalog and statistically apply the correction to the \Mstar\ estimations using the measured redshift and X-ray temperature of each system.  The mean \fblue\ of the \XMMBCS\ sample is $31\pm4\percent$ with a tendency for \fblue\ to be higher in the low mass systems.  On the other hand, no significant redshift trend is seen for \fblue\ in the \XMMBCS\ sample.  The absence of a redshift trend is really applicable only to the clusters and higher mass groups in our sample, because due to the X-ray flux limited selection, our sample has low mass groups only at low redshift.   Assuming $\fblue=0$ has no statistically significant impact on the mass and redshift trends (\Bstar\ and \Cstar); however, the normalization \Astar\ is biased high by $\approx8.5\percent$ (corresponding to a $\approx0.86\sigma$ shift) at a characteristic mass  $\Mfiveoo=8\times10^{13}\Msun$ and redshift $\redshift=0.47$.

We also examine whether \Mstar\ could be a promising mass proxy.  Based on this work we conclude that the stellar mass enclosed by the projected radius $r=0.5~$Mpc provides a mass proxy with an intrinsic scatter of $\approx93\percent$ (1$\sigma$ in mass) for the low mass clusters and groups out to redshift $\redshift\approx1$.  This scatter is larger than the mass scatter one sees at fixed $K$-band luminosity in a group and cluster sample at $\redshift\lesssim0.05$ \citep{lin04a} and larger than the optical richness in a sample of massive, SZE selected clusters extending to $\redshift\approx0.8$  \citep{saro15}.  While this scatter is high compared to some other mass proxies such as the X-ray $Y_{\mathrm{X}}$ \citep[e.g.][]{kravtsov06a,vikhlinin09,arnaud10}, X-ray temperature \citep[e.g.][]{arnaud05}, the ICM mass \citep[e.g.][]{okabe10} and the SZE signal to noise \citep[e.g.][]{benson13,bocquet15}, the stellar mass could still be useful as a mass proxy for low mass and high redshift systems where other proxies are typically in short supply. 

Our work suggests that the stellar mass enclosed within \Rfiveoo\ for clusters or groups of a particular binding mass \Mfiveoo\ exhibits large scatter (43\percent$\pm$10\percent, 1$\sigma$ log-normal) but with a characteristic value that is approximately the same at any point in the last $\approx9$~Gyr of evolution.  This is a remarkable result at first glance, given the sharp trend for decreasing stellar mass fraction with halo mass that exists over this same timespan.  One is driven to ask how massive halos could exhibit different stellar mass fractions than their building blocks, which include the lower mass halos.  However, as has been explored with structure formation simulations \citep{mcgee09}, massive clusters accrete material not only in the form of lower mass clusters and groups but also directly from the surrounding field.  As discussed previously in Chiu16, one possible scenario is that as halos accrete and become more massive the material from lower mass halos with higher stellar mass fractions is roughly balanced by accretion of material from the field that tends to have  lower stellar mass fraction \citep[for $z\approx1$ measurement see, e.g.,][]{vdB13}.  Within such a scenario, the large scatter in \Mstar\ for systems of similar binding mass \Mfiveoo\ would reflect differences in assembly histories.  Other processes such as the stripping of stellar material from infalling galaxies, which would remove those stars from our galaxy based stellar mass measurements, must also play some role \citep{lin04b}.  

Our current analysis invites a more careful comparison to structure formation simulations that include galaxy formation.  In addition, new studies are needed to enable a more precise characterization of the mass and the redshift trends in the cluster galaxy populations;  these will require larger samples of clusters and groups that (1) have been uniformly selected over the full redshift range in a manner that does not rely on their galaxy population and (2) have low scatter mass proxies where the connection to halo mass is well understood over the full redshift range.  

%
%

\section*{Acknowledgements}
\label{sec:acknowledgements}
We express thanks to Drs. B.~Hoyle, J.~Song and M.~Klein for discussions that led to improvements in this paper.  We acknowledge the support by the DFG Cluster of Excellence ``Origin and Structure of the Universe'', the DLR award 50 OR 1205 that supported I. Chiu during his PhD project, and the Transregio program TR33 ``The Dark Universe''. 
This work is based in part on archival data obtained with the \Spitzer\ Space Telescope, which is operated by the Jet Propulsion Laboratory, California Institute of Technology under a contract with NASA.
This work is based on observations obtained with \XMMNEWTON, an ESA science mission with instruments and contributions directly funded by ESA Member States and the USA (NASA).
This paper includes data gathered with the Blanco 4~m telescope, located at the Cerro Tololo Inter-American Observatory in Chile, which is part of the U.S. National Optical Astronomy Observatory, which is operated by the Association of Universities for Research in Astronomy (AURA), under contract with the NSF.

%
%

\bibliographystyle{mn2e}
\bibliography{spt}

%
%

\appendix

\begin{figure*}
\centering
\includegraphics[scale=0.7]{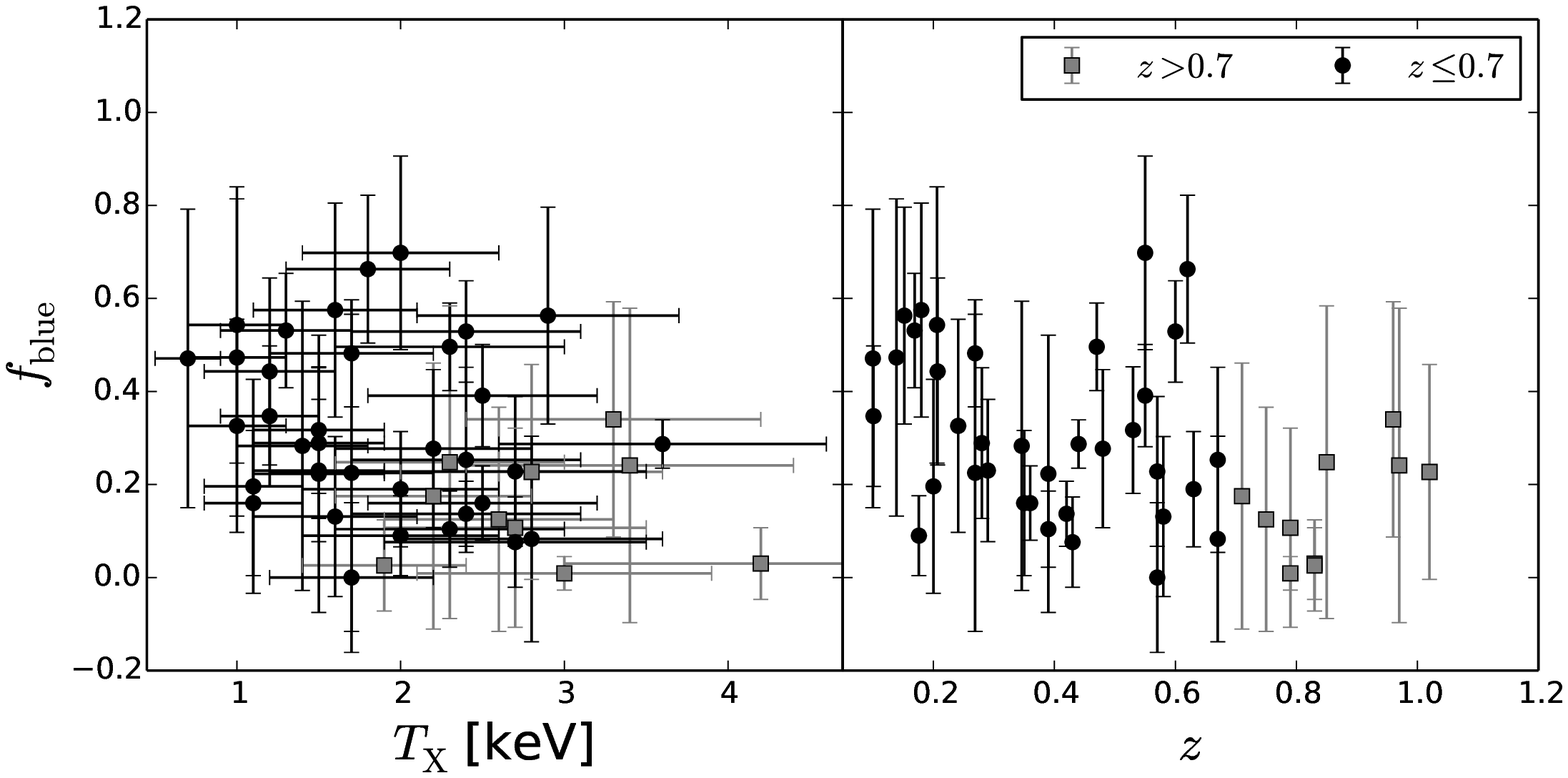}
\caption{
The blue fraction \fblue\ of \XMMBCS\ clusters as a function of cluster X-ray temperature \Tx\ (left) and redshift \redshift\ (right).  The \XMMBCS\ clusters with $\redshift\le0.7$ and $\redshift>0.7$ are shown with black circles and grey squares, respectively.
}
\label{fig:fblue_tz}
\end{figure*}
\begin{figure*}
\centering
\includegraphics[scale=0.7]{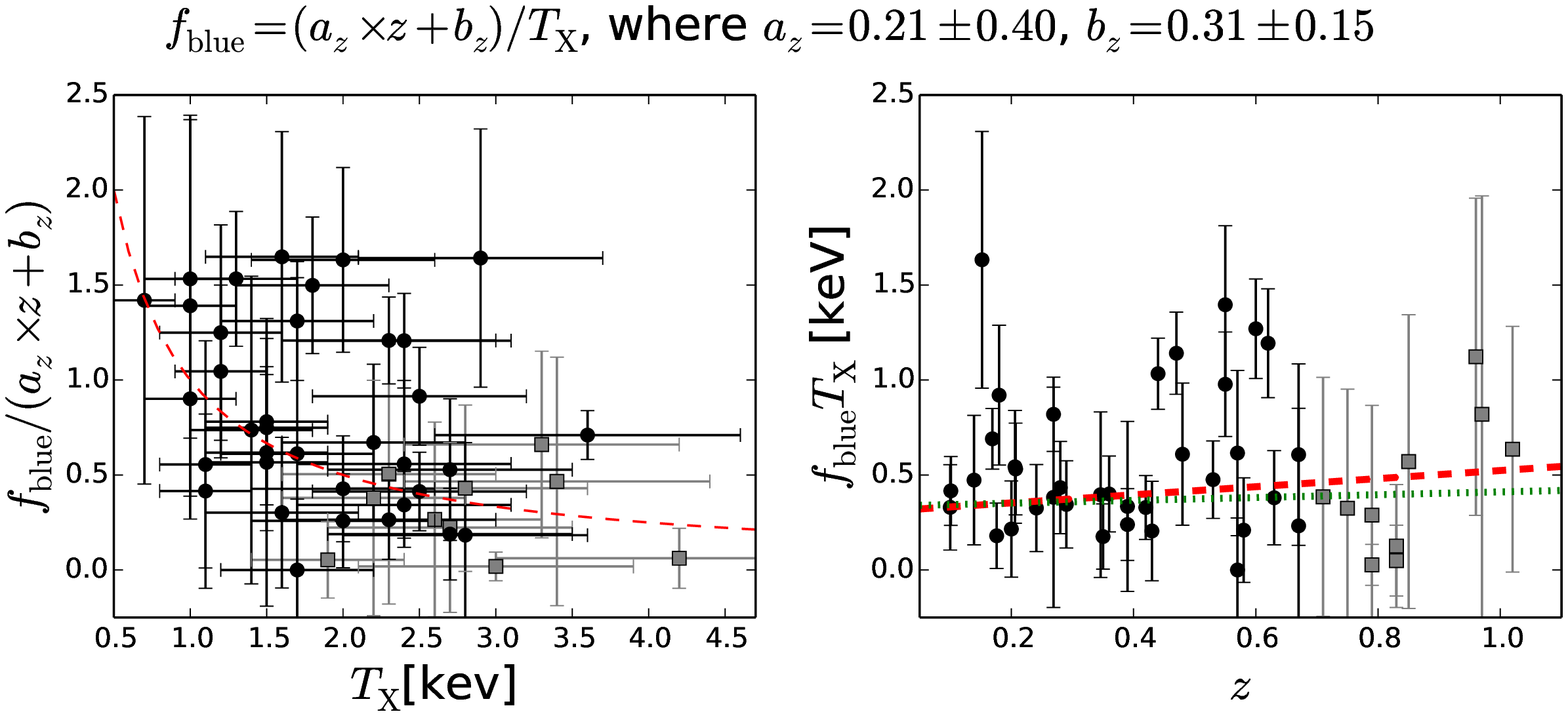}
\caption{
The blue fraction \fblue\ of \XMMBCS\ clusters after correcting for the best-fit mass and redshift trends as a function of X-ray temperature \Tx\ after correcting for the best-fit redshift trend $a_{z}\times\redshift + b_{z}$ (left) and as a function of redshift after correcting for the mass trend by with \Tx.  The \XMMBCS\ clusters with $\redshift\le0.7$ and $\redshift>0.7$ are in the black circles and grey squares, respectively.
The red dashed line indicates the best-fit $\fblue(\Tx,\redshift)$ relation for the \XMMBCS\ cluster with $\redshift\le0.7$, while the green dotted line is the best-fit model fitting to the full sample.
}
\label{fig:fblue_trends}
\end{figure*}

\section{Blue fractions}
\label{app:fblue_xmmbcs}

To enable a more accurate \ML\ for the galaxy populations we measure the \fblue\ of the \XMMBCS\ clusters using the \BCS\ optical catalog \citep{desai12}.  The \BCS\ catalog contains the calibrated photometry of the optical bands $griz$ and the derived photometric redshift estimates of the sources identified in the \BCS\ survey.  The resulting $10\sigma$ depths of the galaxies (point sources) in $griz$ are $23.3$ (23.9), 23.4 (24.0), 23.0 (23.6) and 21.3 (22.1)~mag, respectively.   The data reduction, the source extraction, the photometry calibration and the photometric properties are fully described elsewhere \citep{desai12}.  We describe the estimation of the blue fraction \fblue\ here.

First we perform star/galaxy separation in $i$ band by selecting the galaxies with $\mathtt{spread\_model\_i}\geq2\times10^{-3}$ and restrict the catalog to the central $6\deg^{2}$ region of \XMMBCS\ survey (see Section~\ref{sec:stellarmass_estimation}). The blank sky used for the statistical background subtraction is defined by the central $6\deg^{2}$ region of \XMMBCS\ survey excluding the cluster fields (see Section~\ref{sec:stellarmass_estimation}).  Second, we estimate the completeness of the \BCS\ survey by comparing the source count-magnitude relation between the \BCS\ tiles and the COSMOS \citep{ilbert2008} survey in the same way conducted in \cite{zenteno11}.  Specifically, we fit a power law, which is assumed to be the complete source count-magnitude relation with the slope fixed to the value derived from COSMOS field, to the observed source count-magnitude in \BCS\ survey in the magnitude range between 18 and 20.  The completeness function $\fcom(m)$ as a function of magnitude $m$ is then obtained by fitting an error function to the ratio of the observed source counts to the complete source counts predicted by the best-fit power law.  The completeness functions are separately derived for four bands $griz$ in all \BCS\ tiles overlapping the \XMMBCS\ survey because the depth variation is large among the tiles and the filters.  As a result, we find that the median of the 50\percent\ (90\percent)  completeness of the tiles overlapping the \XMMBCS\ central region is  $23.97$ ($23.18$),  $23.41$ ($22.62$), $22.76$ ($21.89$) and $21.22$ ($19.83$)~mag for $griz$, respectively.

We estimate the blue fractions of the galaxy population projected within the \Rfiveoo\ of 46 \XMMBCS\ clusters by separating the galaxies according to their colors.
In the similar fashion of estimating the NIR luminosity function (Section~\ref{sec:stellarmass_estimation}), we remove the non-cluster members by statistically subtracting the background galaxy counts from the galaxy counts of the cluster field in the color-magnitude space.  For simplicity, we denote the galaxy counts as the position in the color-magnitude space by CMC (color-magnitude-counts) hereafter.  The color and the magnitude of the filter used in deriving the CMC are defined (in the rest frame) by the bands straddling the $4000\angstrom$ and the band redder than $4000\angstrom$, respectively.
Precisely, the color and the magnitude used in CMC are ($g-r$, $r$), ($r-i$, $i$) and ($i-z$, $z$) for the clusters with redshift $\zd\le0.33$, $0.33<\zd\le0.70$ and $\zd>0.70$, respectively.
The steps of $0.05$~mag ($0.25$~mag) are used in binning in color (magnitude) to derive the CMC.

We also construct the completeness map in the observed color-magnitude space by propagating the completeness function of the bands used in the CMC.  We discard the galaxies which lie outside the cluster redshift \zd\ at $3\sigma$ level if the reliable photometric redshift estimates are available, i.e. we discard the galaxies with $\| \mathtt{z\_photo} - \zd \| \geq 3 \times \delta_{\redshift}\times(1+\zd)$ and $\mathtt{z\_photo\_flag} = 1$, where $\delta_{\redshift} = 0.061$ is the scatter of the photometric redshift performance in \BCS\ survey.  To estimate the color distribution of the selected galaxies in the cluster field we project all the galaxies, which lie projected within \Rfiveoo\ and are brighter (fainter) than $\mstar+1.5$ (the BCG), 
to the line perpendicular to the RS tilt predicted by the CSP model at cluster redshift \zd.  
The characteristic magnitude $\mstar(z)$ is defined using the CSP model described in Section~\ref{sec:xmmbcs_catalog}.
Additionally, we take the completeness correction into account by inversely weighting the galaxy counts by the completeness as a function of position in the color-magnitude space.

On the other hand, we construct the mean background CMC (with the same magnitude cut) by extracting the mean value of the CMC of 200 apertures, which are randomly drawn from the blank sky with the same radii of cluster's \Rfiveoo. The completeness correction of each randomly-drawn background aperture is taken into account when we construct the mean background CMC.  We then project the resulting mean background CMC to the line perpendicular to the RS tilt of CSP model in the same way of the cluster field.

In the end, the projected color distribution of the cluster galaxy population along the RS tilt predicted by our CSP model is derived by statistically subtracting the projected CMC of the mean background from the cluster field.  Similar to \cite{zenteno11}, we define the blue and the RS populations by the galaxies with $\Delta C <0.2$~mag and $-0.2\le\Delta C\le0.2$, respectively, where $\Delta C \equiv C_{\mathrm{gal}} - C_{\mathrm{RS}}$ is the difference of the projected colors between the galaxy ($C_{\mathrm{gal}}$) and the RS ($C_{\mathrm{RS}}$) predicted by the CSP model at cluster's redshift.  The blue fraction \fblue\ is then calculated as the ratio of the number of the blue galaxies to the sum of the blue and RS galaxies.
The uncertainty of the \fblue\ of each cluster is derived as the standard deviation of 10000 realizations, where each realization is generated from the observed numbers of the blue and RS galaxies assuming Poisson distribution.

The estimated \fblue\ of 46 \XMMBCS\ clusters are presented in Table~\ref{tab:measurement}, and we show the estimated \fblue\ of 46 \XMMBCS\ clusters in Figure~\ref{fig:fblue_tz}.  The scatter of \fblue\ is large and the mean of \fblue\ is $(31\pm4)\percent$ with the median $25\percent$.  Motivated by \cite{urquhart10}, we extract the mass and redshift trends of the \fblue\ by fitting a parametrized function, which is a function of X-ray temperature \Tx\ and cluster redshift, to the estimated \fblue.  We assume that the ensemble of \fblue\ increases linearly as the cluster redshift increases (the BO effect), which is parametrized by two parameters $a_{z}$ and $b_{z}$, while the \fblue\ at given redshift is inversely proportional to the cluster mass, which is linked to the X-ray temperature \Tx.
I.e., 
\begin{equation}
\label{eq:fblue_func}
\fblue(\Tx, \redshift) = \frac{(a_{z}\times\redshift+b_{z})}{\Tx} \, .
\end{equation}
To estimate the best-fit parameters, we fit the model to the synthetic data sets of 10000 realizations, where each realization consists of $\{{\fblue}_{,i}\}$  ($i$ runs over 46 \XMMBCS\ clusters) generated from the estimated \fblue\ of each cluster.  The best-fit and the $1\sigma$ uncertainty of the parameters are estimated as the mean and the standard deviation of the best-fit parameters of these 10000 realizations.

The measurements of \fblue\ of the \XMMBCS\ cluster sample are in Figure~\ref{fig:fblue_tz}.
We present the best-fit \fblue\ relation of \XMMBCS\ clusters ($\redshift\le0.7$) with the best-fit parameters $(a_{z}, b_{z}) = (0.21, 0.31)$  in Figure~\ref{fig:fblue_trends}.  To plot the mass and redshift trends of the estimated \fblue\ of each cluster  in Figure~\ref{fig:fblue_trends}, we correct the redshift and mass trends with respect to the obtained best-fit trends (i.e., eq~(\ref{eq:fblue_func}) with $(a_{z}, b_{z}) = (0.21, 0.31)$).  Specifically, the estimated \fblue\ of each cluster is divided by the redshift trend ($a_{z}\times\redshift + b_{z}$) and weighted by the temperature \Tx\ in the left and right panels of Figure~\ref{fig:fblue_trends}, respectively.
Although we discard the high redshift clusters ($\redshift>0.7$) in the fit, their \fblue\ behavior are consistent with the best-fit relation estimated from the low redshift clusters ($\redshift\le0.7$) alone. 

\end{document}